%% file: main.tex
\documentclass[
aps,
prd, 
reprint,
superscriptaddress,
nofootinbib,
]{revtex4-2}

\usepackage{graphicx}
\usepackage[usenames,dvipsnames,table,xcdraw]{xcolor}
\usepackage{xspace}
\usepackage{amsmath,amsfonts,amssymb,amsthm,bm}
\usepackage[varg]{txfonts} 
\usepackage{mathtools}
\usepackage[utf8]{inputenc}
\usepackage[normalem]{ulem}
\usepackage{makecell}
\usepackage{multirow}
\usepackage{float}
\usepackage{adjustbox}
\usepackage[caption=false]{subfig}
\usepackage{rotating}
\usepackage{dcolumn}
\usepackage{enumitem}
\usepackage{natbib}
\usepackage[htt]{hyphenat}
\xdefinecolor{mylinkcolor}{rgb}{0,0,0.7}
\usepackage[
	bookmarksnumbered, bookmarksopen, bookmarksopenlevel=2,
	breaklinks=true,
	colorlinks=true, filecolor=mylinkcolor, citecolor=mylinkcolor,
	linkcolor=mylinkcolor, urlcolor=mylinkcolor, menucolor=mylinkcolor,
]{hyperref}

\usepackage{setspace}
\usepackage[hang]{footmisc} 
\allowdisplaybreaks

\def\mr{\mathrm}

\def\mF{\mathcal{F}}

\def\solarmass{M_\odot}

\def\NEVENTS{27\xspace}

\graphicspath{ {fig/} }

\newcommand{\AEI}{\affiliation{Max Planck Institute for Gravitational Physics (Albert Einstein Institute), Am M\"uhlenberg 1, Potsdam 14476, Germany}}
\newcommand{\Maryland}{\affiliation{Department of Physics, University of Maryland, College Park, MD 20742, USA}}

\newcommand{\UoN}{\affiliation{Nottingham Centre of Gravity \& School of Mathematical Sciences, University of Nottingham, University Park, Nottingham NG7 2RD, United Kingdom}}

\begin{document}

\title{
Eccentric and unbound compact binaries in the LIGO--Virgo--KAGRA catalog:\\ parameter estimation and waveform systematics with \texttt{SEOBNRv6EHM}
}

\author{Lorenzo Pompili}
\email{lorenzo.pompili@nottingham.ac.uk}
\UoN
\AEI

\author{Aldo Gamboa}
\email{aldo.gamboa@aei.mpg.de}
\AEI

\author{Alessandra Buonanno}
\email{alessandra.buonanno@aei.mpg.de}
\AEI
\Maryland


\begin{abstract}
Orbital eccentricity encodes key information about compact-binary formation channels and astrophysical environments, making it a critical target for gravitational-wave (GW) inference.
We present parameter-estimation (PE) analyses of GWs from eccentric compact binaries with the \texttt{SEOBNRv6EHM} waveform model.
Using long, highly eccentric numerical-relativity waveforms as synthetic signals, we compare parameter recovery across state-of-the-art eccentric models. We find that \texttt{SEOBNRv5EHM} and \texttt{TEOBResumS-Dal\'i} can yield biased estimates of eccentricity, masses, and spins in the most challenging configurations, while \texttt{SEOBNRv6EHM} significantly reduces these biases.
Applying \texttt{SEOBNRv6EHM} to \NEVENTS GW events from the O1--O4 LIGO--Virgo--KAGRA observing runs---including binary black hole, neutron-star--black-hole, and binary neutron-star mergers---we identify five events with mild support for eccentricity over the quasi-circular precessing-spin hypothesis, with Bayes factors $\log_{10} \mathcal{B}^{\text{EAS}}_{\text{QCP}} > 0.5$.
Since \texttt{SEOBNRv6EHM} is applicable to generic planar binaries, we reanalyze five high-mass events allowing for unbound initial conditions. For three of them---including GW190521, previously claimed to originate from a dynamical capture---a direct-capture configuration is comparable to, or marginally favored over, both the eccentric aligned-spin and quasi-circular precessing-spin hypotheses ($\log_{10}\mathcal{B}^{\rm unbound}_{\rm QCP} \approx 0.2$--$0.6$ for GW190521). 
The recovered configurations are, however, astrophysically unrealistic and cannot be confidently discriminated from highly eccentric bound orbits, so these results do not, by themselves, support an unbound origin for these events.
\texttt{SEOBNRv6EHM} is three to five times faster in PE analyses than \texttt{SEOBNRv5EHM}, while improving waveform accuracy, enabling efficient, large-scale GW inference with eccentric waveforms.
\end{abstract}

\date{\today}

\maketitle


\section{Introduction}
\label{sec:intro}

Gravitational waves (GWs) from compact-binary coalescences, detected by the LIGO--Virgo--KAGRA (LVK) interferometers~\cite{TheLIGOScientific:2014jea, VIRGO:2014yos, KAGRA:2020tym}, have opened a new window on the astrophysics of black holes (BHs) and neutron stars (NSs)~\cite{LIGOScientific:2016aoc, LIGOScientific:2018mvr, LIGOScientific:2020ibl, LIGOScientific:2021usb, Kagra:2021vkt, LIGOScientific:2025slb, LIGOScientific:2025pvj}.
Among the source properties encoded in these signals, orbital eccentricity stands out as a powerful diagnostic of binary formation channels. Binaries formed through isolated evolution~\cite{Bethe:1998bn, Belczynski:2001uc, Dominik:2012kk, Stevenson:2017tfq} are expected to circularize efficiently through GW emission long before entering the sensitive band of ground-based detectors~\cite{Peters:1963ux,Peters:1964zz,Hinder:2007qu,Sperhake:2007gu}.
In contrast, dynamically assembled binaries---through capture or multi-body interactions in dense star clusters~\cite{Miller:2001ez, Miller:2002pg, OLeary:2005vqo, OLeary:2008myb, Samsing:2013kua, Zevin:2018kzq, Chattopadhyay:2023pil}, or von Zeipel--Kozai--Lidov oscillations in hierarchical triple systems~\cite{1910AN....183..345V, kozai1962secular, lidov1962evolution, Wen:2002km}---can retain measurable eccentricity in the frequency range to which detectors are sensitive.
Numerical simulations quantify this expectation: approximately $5\%$ of dynamically assembled mergers in globular clusters retain $e > 0.1$ at $10\,$Hz~\cite{Samsing:2017xmd, Samsing:2017rat, Rodriguez:2018pss}, while estimates from galactic nuclei can reach up to $70\%$~\cite{Samsing:2020tda, Tagawa:2020jnc, Gondan:2020svr}.
The detection, or absence, of eccentricity in GW observations therefore provides a direct handle on the fraction of compact binaries originating from dynamical formation channels~\cite{Zevin:2021rtf}.

While the masses and spins of compact binaries have been routinely measured since the first detection~\cite{LIGOScientific:2016aoc}, eccentricity has only recently come within reach as an observational target.
This has been made possible by sustained progress in modeling GWs from binaries in eccentric orbits~\cite{Hinder:2008kv, Mroue:2010re, Hinder:2017sxy, Hinderer:2017jcs, Cao:2017ndf, Ramos-Buades:2019uvh, Chiaramello:2020ehz, Nagar:2020xsk, Nagar:2021gss, Ramos-Buades:2021adz, Ramos-Buades:2022lgf, Healy:2022wdn, Albanesi:2023bgi, Liu:2023ldr, Nagar:2024oyk, Faggioli:2024ugn, Gamboa:2024hli, Gamboa:2024imd, Gamba:2024cvy, Paul:2024ujx, Morras:2025nlp, Planas:2025feq, Islam:2025llx, Nee:2025zdy, Nee:2025nmh, Maurya:2025shc, Albanesi:2025txj, Ramos-Buades:2026kbq, Morras:2026fho, Faggioli:2026alx, Albanesi:2026qtx}, which provide waveform templates to search for eccentricity signatures via Bayesian parameter estimation (PE).
Building on these models, several inference studies have investigated the presence of eccentricity in LVK data~\cite{Romero-Shaw:2019itr, Romero-Shaw:2020thy, Wu:2020zwr,Lenon:2020oza, Gayathri:2020coq, OShea:2021faf, Gamba:2021gap, Iglesias:2022xfc, Romero-Shaw:2022xko, Bonino:2022hkj, Ramos-Buades:2023yhy, Gupte:2024jfe, Fei:2024ruj, Morras:2025xfu, Planas:2025plq, Planas:2025jny, Jan:2025fps, Kacanja:2025kpr, Chiaramello:2025bhi, LIGOScientific:2025brd, Gupte:2026whi}. However, the unambiguous identification of an eccentric binary remains challenging.

The first large-scale PE studies targeting orbital eccentricity were carried out in Refs.~\cite{Romero-Shaw:2019itr, Romero-Shaw:2020thy, Romero-Shaw:2022xko} using likelihood reweighting and the \texttt{SEOBNRE} waveform model~\cite{Cao:2017ndf} on events from the first three LVK observing runs (O1--O3), finding potential signatures in GW190521, GW190620\_030421, GW191109\_010717, and GW200208\_222617.

Using the \texttt{SEOBNRv4EHM} model~\cite{Ramos-Buades:2022lgf} with the machine-learning inference code \texttt{DINGO}~\cite{Dax:2021tsq, Dax:2022pxd}, Ref.~\cite{Gupte:2024jfe} analyzed 57 binary black hole (BBH) events from O1--O3 and identified three events with Bayes factors favoring nonzero orbital eccentricity---GW200129\_065458, GW190701\_203306, and GW200208\_222617---as well as others with mild support.
The analysis was extended to 85 BBHs from the first part of the fourth observing run (O4a)~\cite{Gupte:2026whi}, using the \texttt{SEOBNRv5EHM} model~\cite{Gamboa:2024hli} with both \texttt{DINGO} and \texttt{Bilby}~\cite{Ashton:2018jfp, Romero-shaw:2020owr}, finding no additional candidates with significant support.
Both studies highlighted the critical role of glitch mitigation: the inferred support for eccentricity in GW200129\_065458 and GW190701\_203306---both affected by glitches in the LIGO Livingston detector---ranges from modest to strong depending on the subtraction method used.
An independent analysis of BBH events from the first four LVK observing runs with \texttt{RIFT}~\cite{Lange:2018pyp}, using the \texttt{SEOBNRv5EHM} and \texttt{TEOBResumS-Dal\'i}~\cite{Nagar:2024oyk} models, was performed in Ref.~\cite{Malagon:2026uev}, also investigating population-level inference~\cite{Zeeshan:2026pga}. This study finds evidence for eccentricity in GW200129\_065458, although the result is sensitive to analysis choices, but not in GW190701\_203306, in contrast with Refs.~\cite{Gupte:2024jfe, Gupte:2026whi, Planas:2025jny}.

Among these candidates, GW200208\_222617 has been identified as the potentially least ambiguous~\cite{Romero-Shaw:2025vbc}, as it is not affected by data-quality issues and the moderate mass of its source makes the eccentricity--precession degeneracy less severe~\cite{Romero-shaw:2022fbf, CalderonBustillo:2020xms}; however, it has a high false-alarm rate~\cite{Kagra:2021vkt}, even when using an eccentric template bank~\cite{Wang:2025yac}, and a comparatively weaker Bayes factor.

Complementary analyses using the \texttt{IMRPhenomTEHM} model~\cite{Planas:2025feq} confirmed the evidence for eccentricity in GW200129\_065458 and GW200208\_222617 from a sample of 17 BBH events from O1--O3~\cite{Planas:2025jny}, with milder support for the high-mass events GW190701\_203306 and GW190929\_012149.
The same model was also used to analyze the first four neutron-star--black-hole (NSBH) events, finding support for eccentricity in GW200105\_162426~\cite{Planas:2025plq}, consistent with earlier analyses~\cite{Fei:2024ruj, Morras:2025xfu}; this event, however, reaches more modest Bayes factors compared to the most significant BBH candidates, and the evidence is sensitive to the choice of prior~\cite{Morras:2025xfu, Planas:2025plq, Kacanja:2025kpr, Jan:2025fps, Clarke:2026cuw}.
The \texttt{IMRPhenomTEHM} analysis was extended to O4a events~\cite{Xu:2025ajj}, finding only some candidates with mild support that partially overlap with those of Ref.~\cite{Gupte:2026whi}, though with some differences.
GW190521, which showed early signs of eccentricity~\cite{Romero-Shaw:2020thy, Gayathri:2020coq}, is not recovered as eccentric by more recent waveform models~\cite{Iglesias:2022xfc, Ramos-Buades:2023yhy, Gamboa:2024hli}.
However, Ref.~\cite{Gamba:2021gap} found evidence that GW190521 is better described as a dynamical capture of two nonspinning BHs on initially unbound orbits, using an earlier version of \texttt{TEOBResumS-Dal\'i}~\cite{Chiaramello:2020ehz, Nagar:2020xsk}.

Together, these studies reveal a recurring set of candidates with moderate support for eccentricity, but also expose persistent challenges: the degeneracy between eccentricity and spin-precession effects~\cite{Romero-shaw:2022fbf, CalderonBustillo:2020xms}, the sensitivity of the results to the choice of eccentricity prior~\cite{Gupte:2024jfe, Morras:2025xfu, Planas:2025jny, Clarke:2026cuw}, the impact of non-Gaussian noise artifacts on the inferred eccentricity~\cite{Gupte:2024jfe, Gupte:2026whi}, and systematic differences arising from waveform modeling uncertainties~\cite{Gupte:2024jfe, Jan:2025zcm, Divyajyoti:2025cwq}.
Addressing waveform systematics is one of the main goals of this work.

In this work, we employ the newly developed \texttt{SEOBNRv6EHM} waveform model~\cite{Gamboa:2026jht} for aligned-spin BBHs in generic planar-orbits to perform Bayesian PE analyses of \NEVENTS GW events from the O1--O4 LVK observing runs.
The \texttt{SEOBNRv6EHM} model incorporates several key improvements over its predecessor \texttt{SEOBNRv5EHM}~\cite{Gamboa:2024hli}, including a new factorization and resummation of the eccentricity corrections in the radiation-reaction force and the waveform modes, as well as a recalibration of the effective-one-body (EOB) Hamiltonian to quasi-circular (QC) aligned-spin numerical-relativity (NR) simulations.
Crucially, the model achieves a significant reduction in computational cost---by factors of $\sim 2 - 8$ per waveform evaluation compared to other state-of-the-art EOB eccentric models---while simultaneously improving waveform accuracy, making it the most faithful aligned-spin eccentric model to NR simulations to date~\cite{Gamboa:2026jht}.
Against 319 eccentric NR simulations, the median of the maximum mismatches over the total-mass range $[20, 200]\,\solarmass$ is ${\sim}\,0.02\%$, and mismatches remain below or close to $2\%$ for eccentricities up to $e_{\rm gw} \sim 0.9$---an improvement of about an order of magnitude over \texttt{SEOBNRv5EHM} at high eccentricities, and over \texttt{TEOBResumS-Dal\'i} across the entire eccentricity range~\cite{Gamboa:2026jht}.
This speedup brings the cost of eccentric PE closer to that of QC analyses, enabling systematic studies of large event catalogs, and the analysis of long-duration signals such as those from NSBH and binary neutron-star (BNS) mergers.

We perform a series of zero-noise injection-recovery analyses of synthetic NR signals with increasing eccentricity, using the \texttt{SEOBNRv6EHM}, \texttt{SEOBNRv5EHM}, and \texttt{TEOBResumS-Dal\'i} models to assess waveform systematics.
For the most challenging configurations, such as the long, highly eccentric simulation \texttt{SXS:BBH:2527}, the latter two models yield biased eccentricity and intrinsic-parameter estimates, while \texttt{SEOBNRv6EHM} significantly reduces these biases.
To our knowledge, this is the first study to investigate biases in PE using NR signals (particularly, long, highly eccentric ones) and multiple state-of-the-art inspiral--merger--ringdown eccentric waveform models.
Within the same NR injection study, we also examine how the waveform-generation starting frequency affects parameter recovery for highly eccentric signals---an aspect that has not received systematic treatment in the literature. The issue arises from the mismatch between the \textit{orbit-averaged} template starting frequency and the higher \emph{instantaneous} frequencies attained at periastron passages; we discuss practical strategies to recover the missing signal power without prohibitive computational cost.

We then apply \texttt{SEOBNRv6EHM} to a diverse sample of compact-binary mergers, including BBHs, NSBH, and BNS events. We focus on signals previously identified as potential eccentric candidates, as well as events with extreme parameters (e.g., high mass ratio, high spins, long duration) to further validate the robustness of the model across a broad region of parameter space.
We identify five events with mild to moderate support for eccentricity, consistent with the findings of previous analyses in the literature, specifically, GW200129\_065458, GW200208\_222617, GW231223\_032836, GW190701\_203306, and GW200105\_162426, listed in decreasing order of Bayes factor relative to the \texttt{SEOBNRv5PHM} model~\cite{Ramos-Buades:2023ehm,Estelles:2025zah}.

Additionally, exploiting the applicability of \texttt{SEOBNRv6EHM} to generic planar orbits, we reanalyze five high-mass events allowing for unbound initial conditions. For four of them---including GW231123\_135430---the unbound model gives a comparable or better fit than the bound eccentric aligned-spin model. For three of these---GW190521 (for which a dynamical-capture origin was previously suggested~\cite{Gamba:2021gap}), GW191109\_010717, and GW231221\_135041---a direct-capture configuration fits the data comparably well, or is even marginally preferred, when compared against the quasi-circular precessing-spin (QCP) hypothesis ($\log_{10}\mathcal{B}^{\rm unbound}_{\rm QCP} \approx 0.2$--$1.6$, depending on the event and reference precessing model).

The remainder of this paper is organized as follows.
In Sec.~\ref{sec:methods}, we briefly summarize the \texttt{SEOBNRv6EHM} model and its key improvements, and describe the Bayesian inference framework and analysis settings adopted throughout.
We then present validation results from synthetic NR injections in Sec.~\ref{sec:pe_injections}, and report the analysis of real GW events in Sec.~\ref{sec:pe_events}.
In Sec.~\ref{sec:pe_unbound}, we extend the analysis to a selection of high-mass events under an unbound-orbit hypothesis.
We conclude in Sec.~\ref{sec:conclusions} with a summary of our findings and an outlook on future work.
Four appendices collect additional technical material: Appendix~\ref{app:f_start_and_t_start} examines the impact of the waveform-generation starting frequency on parameter recovery; Appendix~\ref{app:dali_benchmarks} reports timing comparisons against \texttt{TEOBResumS-Dal\'i}; Appendix~\ref{app:unbound_injections} tests, through injection-recovery analyses of GW190521-like signals, how well the unbound and QCP hypotheses can be distinguished; and Appendix~\ref{app:GW231123_unbound} provides additional details on the unbound analysis of GW231123\_135430.

\section{Methods}
\label{sec:methods}

We briefly describe the \texttt{SEOBNRv6EHM} waveform model and its key improvements over its predecessor \texttt{SEOBNRv5EHM}, followed by the Bayesian PE framework, prior choices, and analysis settings adopted throughout this work.

\subsection{The \texttt{SEOBNRv6EHM} waveform model}
\label{sec:model}

The \texttt{SEOBNRv6EHM} waveform model~\cite{Gamboa:2026jht} generates inspiral--merger--ringdown waveforms for BBHs in generic planar orbits within the EOB formalism~\cite{Buonanno:1998gg, Buonanno:2000ef, Damour:2000we, Damour:2001tu, Buonanno:2005xu}, and includes the $(\ell, |m|) \in \{ (2,2),\, (2,1),\, (3,3),\, (3,2),\, (4,4),\, (4,3) \}$ spherical-harmonic modes. 
Here, we briefly summarize the key improvements of \texttt{SEOBNRv6EHM} over its predecessor \texttt{SEOBNRv5EHM}~\cite{Gamboa:2024hli}; a comprehensive description is provided in Ref.~\cite{Gamboa:2026jht}.
For the purposes of this work, the most relevant features are that the model applies to generic planar orbits (including initially unbound ones), that it is more accurate than its predecessor at high eccentricities, and that it is faster; the rest of this subsection outlines how these are achieved.

In the EOB formalism, the binary dynamics is governed by Hamilton's equations for the phase-space variables $(r, \phi, p_{r_*}, p_\phi)$, supplemented by a radiation-reaction (RR) force that accounts for energy and angular-momentum losses due to GW emission.
The \texttt{SEOBNRv5EHM} model parametrizes the eccentricity corrections to the RR force and the waveform modes in terms of the orbit-averaged Keplerian parameters $(x, e, \zeta)$---with $e$ the Keplerian eccentricity and $\zeta$ the relativistic anomaly, defined through the Keplerian parametrization $r/M = p/(1 + e \cos \zeta)$ with $p$ the semi-latus rectum~\cite{Darwin1959gravity, Gamboa:2024imd}--- obtained by solving a system of auxiliary post-Newtonian (PN) equations coupled to the EOB equations of motion. While this approach yields accurate waveforms, it introduces two limitations:
(i)~the overdeterminacy of the equations of motion leads to a desynchronization of the dynamics in challenging configurations (e.g., combinations of high eccentricities, high spins, and/or long waveforms), leading to unphysical features in the waveforms, and
(ii)~the reliance on Keplerian parameters restricts the model to bound orbits.
The \texttt{SEOBNRv6EHM} model overcomes both limitations by expressing all eccentricity corrections directly in terms of the EOB phase-space variables $(r, \dot r, \dot p_{r_*})$, eliminating the need for auxiliary Keplerian equations. This reparametrization makes the equations of motion applicable to generic equatorial orbits, including unbound (scattering) trajectories and dynamical captures, which are also described within the \texttt{TEOBResumS-Dal\'i} model~\cite{Nagar:2020xsk, Nagar:2024oyk}.

The RR force is factored into components that depend on the QC waveform modes ($\mF_{\phi}^{\mathrm{modes}}$, $\mF_{r}^{\mathrm{modes}}$) and eccentricity correction factors ($\mF_{\phi}^{\mathrm{ecc}}$, $\mF_{r}^{\mathrm{ecc}}$).
The latter are obtained through a \emph{sigmoid resummation}, which ensures a controlled behavior at all times: the corrections reduce to unity in the circular-orbit and low-velocity limits, and remain bounded during strong periastron passages and the plunge.
The part of the RR force that depends on QC waveform modes includes results at high PN orders, while the PN-expanded arguments of the sigmoid resummation are accurate up to 1PN order and fully consistent with the known eccentric RR force at that order~\cite{Khalil:2021txt, Gamboa:2024imd}; the explicit expressions are given in Eqs.~(11) of Ref.~\cite{Gamboa:2026jht}.

For the waveform modes, a new factorization is adopted in which the Newtonian prefactors are generalized to generic orbits (expressed in terms of EOB variables), and multiplicative eccentricity corrections $h_{\ell m}^{\mathrm{ecc}}$ are computed at relative 1PN order for all the six modes  returned by the model.

The EOB Hamiltonian is based on that of the \texttt{SEOBNRv5} models~\cite{Khalil:2023kep}, and it incorporates the full 4PN and partial 5PN nonspinning information. However, it is recalibrated to QC aligned-spin NR simulations due to the modified RR force.
The calibration follows a hierarchical approach similar to \texttt{SEOBNRv5HM}~\cite{Pompili:2023tna}: a nonspinning 5PN parameter $a_6$ is fitted to nonspinning NR waveforms, and a 4.5PN spin-orbit parameter $\hat{d}_{\mathrm{SO}}$ is fitted to spinning NR waveforms.
The merger-ringdown attachment time is also recalibrated to QC NR simulations through the parameter $\Delta t^{22}_{\mathrm{ISCO}}$.

Additionally, the $\rho_{\ell m}$ amplitude functions entering the QC part of the RR force include calibration parameters $\Delta\rho_{\ell m}^{(1)}$, which represent the linear-in-$\nu$ contributions at high PN orders, with $\nu=m_1 m_2/(m_1 + m_2)^2$ being the symmetric mass ratio.
For all modes except $(\ell, |m|) = (2,2)$, these are fitted to numerical second-order gravitational self-force (2GSF) flux data as recalibrated in Ref.~\cite{Leather:2025nhu}, following a correction to the original data~\cite{Warburton:2021kwk} used in the \texttt{SEOBNRv5} models~\cite{Vandemeent:2023ols}.
Since the leading-order Newtonian RR force in \texttt{SEOBNRv6EHM} differs from that of \texttt{SEOBNRv5}, this constitutes an approximation, as detailed in Ref.~\cite{Gamboa:2026jht}.
Instead, $\Delta\rho_{22}^{(1)}$ is calibrated to an equal-mass, nonspinning QC NR waveform, as this yields a more regular calibration structure for spinning configurations.

The \texttt{SEOBNRv6EHM} model is validated against 592 QC, 319 eccentric, one dynamical-capture, and two scattering \texttt{SXS} NR waveforms~\cite{Scheel:2025jct}, and through scattering-angle comparisons against 61 \texttt{SXS} NR simulations~\cite{Long:2025nmj}, achieving comparable or improved waveform accuracy with respect to \texttt{SEOBNRv5EHM} across the parameter space, with the largest improvements for the most eccentric configurations~\cite{Gamboa:2026jht}.
Furthermore, the new parametrization, the elimination of the auxiliary Keplerian equations, and numerous implementation optimizations in the \texttt{pySEOBNR} code~\cite{Mihaylov:2023bkc} make \texttt{SEOBNRv6EHM} faster than other state-of-the-art EOB eccentric models by factors of $\sim 2-8$ in single waveform evaluations~\cite{Gamboa:2026jht}, with the QC limit improving on the cost of the dedicated QC model \texttt{SEOBNRv5HM}~\cite{Pompili:2023tna}.

\subsection{Bayesian inference framework}
\label{sec:pe_method}

Extracting source parameters from GW data requires comparing the observed strain $d(t)$ against template waveforms within a Bayesian framework.
Given observed data $d(t) = h(\boldsymbol{\vartheta}, t) + n(t)$, where $h(\boldsymbol{\vartheta}, t)$ is the GW signal described by a set of parameters $\boldsymbol{\vartheta}$ and $n(t)$ is detector noise, the posterior probability distribution for the parameters is 
$p(\boldsymbol{\vartheta} \mid d) = \mathcal{L}(d \mid \boldsymbol{\vartheta}) \, p(\boldsymbol{\vartheta}) / \mathcal{Z}(d)$
where $\mathcal{L}(d \mid \boldsymbol{\vartheta})$ is the likelihood, $p(\boldsymbol{\vartheta})$ is the prior probability, and $\mathcal{Z}(d) = \int \mathcal{L}(d \mid \boldsymbol{\vartheta}) \, p(\boldsymbol{\vartheta}) \, \mr{d}\boldsymbol{\vartheta}$ is the evidence.
Assuming stationary, Gaussian noise, the log-likelihood takes the standard form~\cite{Veitch:2014wba}
\begin{equation}
\label{eq:loglikelihood}
\ln \mathcal{L}(d \mid \boldsymbol{\vartheta}) \propto -\frac{1}{2} \sum_k \left( d_k - h_k(\boldsymbol{\vartheta}) \mid d_k - h_k(\boldsymbol{\vartheta}) \right),
\end{equation}
where the sum runs over the detectors in the network, and the noise-weighted inner product is defined as
\begin{equation}
\label{eq:overlap}
( h_1 \mid h_2 ) \equiv 4 \,\Re \int^{f_{\text{max}}}_{f_{\text{min}}} \frac{\tilde{h}_1(f) \,\tilde{h}^*_2(f)}{S_\text{n}(f)} \, \text{d}f,
\end{equation}
where the tilde $ \, \tilde{} \, $ denotes the Fourier transform, the star $ ^{*} $ denotes the complex conjugate, and $S_\text{n}$ is the one-sided power-spectral density (PSD) of the detector's noise.

Among the source parameters $\boldsymbol{\vartheta}$, the orbit of eccentric binaries is additionally characterized by its eccentricity and a radial phase angle (or \textit{anomaly}---the fraction of the orbital period elapsed since the last periastron passage), both of which are time-dependent quantities typically specified at a given orbit-averaged reference frequency $\langle f_{\text{ref}} \rangle$~\cite{Ramos-Buades:2023yhy, Shaikh:2023ypz}. The \texttt{SEOBNRv6EHM} model parametrizes the orbit in terms of the Keplerian eccentricity $e$ and relativistic anomaly $\zeta$~\cite{Darwin1959gravity}. For aligned-spin, eccentric binaries, this yields six intrinsic parameters: the component masses $m_1$ and $m_2$ (with $m_1 \geq m_2$), the dimensionless spin projections $\chi_1$ and $\chi_2$ along the orbital angular momentum, the Keplerian eccentricity $e$ and relativistic anomaly $\zeta$ at the reference frequency.
Throughout this work we report results in terms of the following derived mass and spin parameters: the total mass $M = m_1 + m_2$, the chirp mass $\mathcal{M} = (m_1 m_2)^{3/5}/(m_1 + m_2)^{1/5}$, the mass ratio $q = m_1/m_2 \geq 1$, and the binary's effective spin parameter $\chi_{\text{eff}} = (m_1 \chi_1 + m_2 \chi_2)/(m_1 + m_2)$.
The standard extrinsic parameters include the luminosity distance $d_L$, the right ascension and declination, the inclination angle $\iota$ between the total angular momentum and the line of sight, the polarization angle $\psi$, the reference phase $\varphi$, and the coalescence time $t_c$.

For the eccentric parameters, we employ a uniform prior on $e \in [0, 0.8]$ and a uniform prior on $\zeta \in [0, 2\pi]$ unless otherwise specified. This is not an astrophysically motivated prior~\cite{Rozner:2026jtj}, but it allows us to identify events for which the data have high likelihood support for eccentricity. Astrophysical information on the relative rates of eccentric and QC mergers can then be incorporated \textit{a posteriori} through odds ratios or population-level analyses~\cite{Gupte:2024jfe,Gupte:2026whi, Zeeshan:2026pga}.
The remaining priors follow standard LVK conventions~\cite{LIGOScientific:2025yae}: uniform in component masses (in the detector frame), isotropic in sky location and spin orientation, uniform in comoving volume and source-frame time, and uniform in reference phase and polarization angle.

Comparing eccentric parameters across different waveform models requires additional postprocessing, as eccentricity and anomaly are gauge-dependent and not uniquely defined in general relativity. To facilitate direct comparisons with results from other waveform models, one can map the model-specific eccentricity to waveform-based definitions such as the eccentricity $e_{\mathrm{gw}}$ and mean anomaly $l_{\mathrm{gw}}$ implemented in the \texttt{gw\_eccentricity} Python package~\cite{Shaikh:2023ypz, Shaikh:2025tae, Ramos-Buades:2022lgf}.
For the NR injection study, we adopt these definitions and report results in terms of $e_{\mathrm{gw}}$ and $l_{\mathrm{gw}}$ at a dimensionless orbit-averaged reference frequency $M \langle f_{\text{ref}} \rangle = 0.01$, where $M = (1+z)\,M_{\text{src}}$ is the detector-frame total mass, as this value exceeds the starting frequency of all NR simulations considered and works robustly for all waveform models in our comparison. We stress that this choice is purely illustrative, since the eccentricity posterior samples obtained at one reference frequency can be mapped to any other. For the NR injection study, we also reconstruct the full eccentricity evolution $e_{\text{gw}}(M \langle f_{\text{ref}} \rangle)$ from the posterior samples, which provides a more complete picture of model accuracy across the inspiral.

For real GW events, however, some systems have posterior support for very high eccentricities with only few periastron passages occurring even when integrating backwards in time; in such cases the conversion to $e_{\mathrm{gw}}$ becomes unreliable~\cite{Shaikh:2023ypz}. We therefore report the Keplerian eccentricity $e$ from the \texttt{SEOBNRv6EHM} model, evaluated at the orbit-averaged starting frequency of waveform generation $\langle f_{\text{start}} \rangle$ (see Sec.~\ref{sec:pe_events} for details). 

We perform all PE analyses using the \texttt{Bilby} inference library~\cite{Ashton:2018jfp, Romero-shaw:2020owr} and the \texttt{dynesty} sampler~\cite{Speagle:2019ivv}, which constitute the standard tools adopted in recent LVK analyses~\cite{LIGOScientific:2025yae}. In some cases, we employ \texttt{parallel Bilby} (\texttt{pBilby})~\cite{Smith:2019ucc}, a parallelized implementation designed to accelerate inference on multiple computing nodes. Specifically, we use \texttt{Bilby} on a single computing node with 64 CPU cores for all BBH signals, while for some of the longer-duration signals from NSBH and BNS mergers we use \texttt{pBilby} across 16 nodes with 32 cores each.
For the sampler settings, we adopt the \texttt{acceptance-walk} stepping method, with average number of accepted steps per Markov chain Monte Carlo (MCMC) chain $\texttt{naccept}=60$ and a total number of live points $\texttt{nlive}=1000$, keeping the remaining sampler settings to their default values, unless otherwise specified. To reduce the dimensionality of the parameter space and improve sampler convergence we apply distance marginalization in all analyses. These choices are consistent with those used in the latest LVK analyses~\cite{LIGOScientific:2025yae}.

Regarding frequency settings, we generally start integrating the likelihood at $f_{\text{min}} = 20~\mathrm{Hz}$, except in cases where this was changed in LVK analyses due to data-quality issues, and generate waveforms starting from an orbit-averaged $(2,2)$ mode frequency $\langle f_{\text{start}} \rangle = 10~\mathrm{Hz}$. The lower starting frequency ensures that, at least for QC binaries, higher-order modes---which for time-domain models start at $f_{\ell m} \approx (m/2)~f_{22}$---are fully captured in the frequency range of interest.
This is most relevant for high-mass, asymmetric binaries~\cite{Ursell:2025ufb}. For BNS and NSBH events, where the impact of missing higher-order-mode power is expected to be negligible, we start waveform generation at the same frequency as the likelihood to reduce computational cost.

The choice of a starting frequency of $10~\mathrm{Hz}$ is more conservative than that usually adopted in QC LVK analyses, which typically begin waveform generation at $\sim 13.33~\mathrm{Hz}$ to capture the $(3,3)$ mode starting at the minimum likelihood frequency of $20~\mathrm{Hz}$~\cite{LIGOScientific:2025yae}. For eccentric binaries, however, the \emph{instantaneous} mode frequency during periastron passages is higher than the \emph{orbit-averaged} one, so a lower starting frequency is needed to capture the power radiated in periastron passages whose instantaneous frequency crosses $20~\mathrm{Hz}$.
The impact of the starting frequency on the inferred parameters, as well as potential alternatives to reducing the starting frequency, are assessed in detail for some of the simulated NR signals in Appendix~\ref{app:f_start_and_t_start}.

\section{Analysis of numerical-relativity synthetic signals}
\label{sec:pe_injections}

We validate the \texttt{SEOBNRv6EHM} model through synthetic injections of NR waveforms from the SXS catalog~\cite{Scheel:2025jct}.
Specifically, we consider five equal-mass, nonspinning BBH configurations with different initial eccentricities, corresponding to the SXS IDs \texttt{SXS:BBH:1355}, \texttt{SXS:BBH:1359}, \texttt{SXS:BBH:1362}, \texttt{SXS:BBH:2525}, and \texttt{SXS:BBH:2527}.
The mass ratio, the spins and the inclination are deliberately held fixed, so that eccentricity is the only parameter varied across the sequence and the differences between models can be attributed to their modeling of the eccentric dynamics rather than to other effects.
Two of these simulations, \texttt{SXS:BBH:1355} and \texttt{SXS:BBH:1359}, correspond to binaries with eccentricities of $e_{\rm gw} \approx 0.05$ and $e_{\rm gw} \approx 0.09$ at a dimensionless reference frequency of $M\langle f_{\rm ref}\rangle = 0.01$, and were included in the validation of the \texttt{SEOBNRv5EHM} model~\cite{Gamboa:2024hli}. 
The other three, \texttt{SXS:BBH:1362}, \texttt{SXS:BBH:2525}, and \texttt{SXS:BBH:2527}, are analyzed here for the first time, and have eccentricities of $e_{\rm gw} \approx 0.17$, $e_{\rm gw} \approx 0.24$, and $e_{\rm gw} \approx 0.34$ at the same reference frequency, respectively.
Notably, \texttt{SXS:BBH:2527} is the simulation in the public SXS catalog with the highest initial eccentricity of $e_{\rm gw}\approx 0.9$ at $M\langle f_{\rm ref}\rangle \approx 0.001$, and 14 periastron passages over a total duration of $\sim 16000\,M$.
We do not analyze the simulation \texttt{SXS:BBH:1363}, used in Ref.~\cite{Gamboa:2024hli}, as it has since been deprecated by the SXS collaboration.

\begin{figure}[t]
    \centering
    \includegraphics[width=\columnwidth]{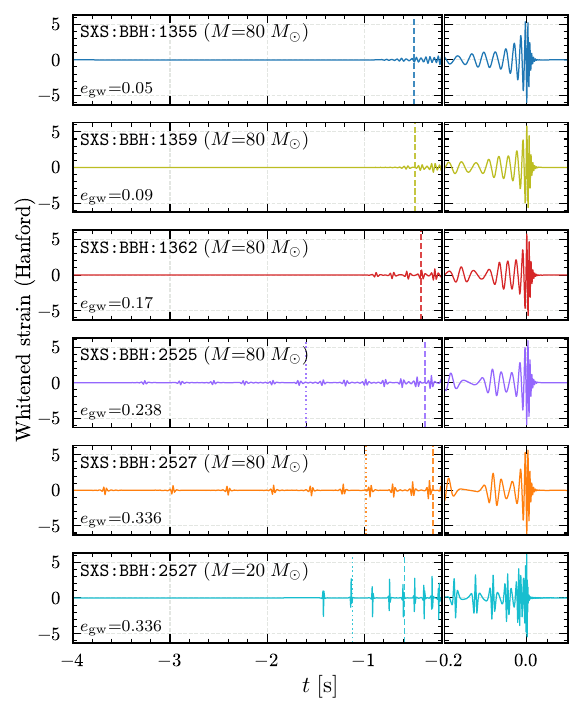}
    \caption{
	Whitened-strain waveforms in the LIGO Hanford detector for the six NR synthetic-signal injections, ordered by increasing eccentricity. Dotted and dashed vertical lines mark the times at which the orbit-averaged $(2,2)$-mode frequency reaches $\langle f_{22}\rangle = 10$ and $20$~Hz, respectively, shown only when the corresponding crossing lies after the start of the NR simulation. The SXS simulation ID, total mass, and injected GW eccentricity $e_{\rm gw}$ at the reference frequency $M\langle f_{\rm ref} \rangle = 0.01$ are indicated in each panel.}
    \label{fig:NR_injections_waveforms}
\end{figure}

The injections are performed in zero noise for the Hanford and Livingston detectors, assuming the A+ design sensitivity PSD~\cite{Ligo:noisecurves}.
We choose extrinsic parameters similarly to the previous injection study done for the \texttt{SEOBNRv5EHM} model~\cite{Gamboa:2024hli}, see Table~\ref{tab:peinjections}.
Still, our injection setup differs from the one used in the \texttt{SEOBNRv5EHM} validation~\cite{Gamboa:2024hli} in several respects:
\begin{itemize}
	\item We interface directly with the \texttt{sxs} Python package~\cite{SXSPackage_v2025.0.16}, which provides access to waveforms in the latest SXS format and applies standardized preprocessing steps---including tapering, transition to a constant value at late times, padding, and line subtraction---that allow for clean Fourier transforms even in the presence of GW memory (included in the latest simulations)~\cite{Chen:2024ieh}.
	\item The total mass is set to $80\,M_\odot$ (compared to $70\,M_\odot$ in the previous study), chosen to ensure that the minimum frequency of the likelihood integration $f_{\rm min}$, fixed at $20\,$Hz, lies above the orbit-averaged $(2,2)$ mode frequency at the end of the tapering window even for the shortest simulations, to minimize the impact of the windowing on the analysis.  The NR waveforms are windowed using the default SXS tapering prescription, applied until $500\,M$ after the NR reference time. 
	\item We assume the A+ design-sensitivity PSD~\cite{Ligo:noisecurves} (expected design sensitivity for the fifth observing run O5) rather than the zero-detuned Advanced LIGO PSD, yielding slightly higher signal-to-noise ratios (SNRs). The five injections at $80\,M_\odot$ have network optimal SNRs of $36.1$, $35.9$, $36.8$, $38.3$, and $40.3$ for \texttt{SXS:BBH:1355}, \texttt{SXS:BBH:1359}, \texttt{SXS:BBH:1362}, \texttt{SXS:BBH:2525}, and \texttt{SXS:BBH:2527}, respectively, compared to $20.0$ in the previous study.
	\item While the previous study recovered injections using only the dominant $(\ell, m) = (2,2)$ mode, we include all available modes of the \texttt{SEOBNRv6EHM} model, i.e., $(\ell, |m|) \in \{(2,2),\, (2,1),\, (3,3),\, (3,2),\, (4,4),\, (4,3)\}$, and inject the NR signals with all modes up to $\ell \leq 8$, including GW memory contributions.
\end{itemize}
For these reasons, we do not expect exact quantitative agreement with the results of Ref.~\cite{Gamboa:2024hli} for the two simulations that are common to both studies.
For the long, high-eccentricity simulation \texttt{SXS:BBH:2527}, we additionally consider a total mass of $20\,M_\odot$, which provides a considerably more demanding test: at lower mass, a larger portion of the NR inspiral falls within the detector's sensitive band, where eccentricity modulations are more pronounced and waveform accuracy is tested over many more orbital cycles and periastron passages. For the \texttt{SXS:BBH:2527} injection at $20\,M_\odot$, we reduce the luminosity distance from $2\, {\rm Gpc}$ to $500\, {\rm Mpc}$ to yield a comparable SNR of $43.5$.

The chosen detector-frame total masses ($20\,M_\odot$ and $80\,M_\odot$) probe astrophysically relevant regions of the source-frame BBH mass distribution inferred from GWTC-4.0~\cite{LIGOScientific:2025pvj}. At the injected luminosity distances ($d_L = 500\,\mathrm{Mpc}$ and $2\,\mathrm{Gpc}$, i.e.\ $z \approx 0.10$ and $z \approx 0.36$), the $M = 20\,M_\odot$ and $M = 80\,M_\odot$ configurations correspond to source-frame component masses of $\approx (9 + 9) M_\odot$ and $\approx (29 + 29) M_\odot$, sitting near the prominent $m_1 \approx 10\,M_\odot$ peak and the $30$--$40\,M_\odot$ feature, respectively. 
The eccentricity range spanned by our injections broadly overlaps the $90\%$ range $e_{10\,\mathrm{Hz}} \approx 0.1$--$0.5$ predicted for detectable eccentric BBHs in globular-cluster simulations~\cite{Singh:2025ojp}; the \texttt{SXS:BBH:2527} configuration at $M = 20\,M_\odot$ reaches somewhat higher $e_{10\,\mathrm{Hz}}$, but remains astrophysically plausible given that lower mass binaries have a higher eccentricity at a fixed dimensionful reference frequency compared to higher mass ones. The injected SNRs lie near the upper edge of the $90\%$ range ($13$--$57$, with median $\sim 20$) expected for measurably eccentric binaries with an O4-like detector network~\cite{Singh:2025ojp}, and would fall closer to the bulk at O5-like sensitivity.

\begin{figure*}[t]
    \centering
    \includegraphics[width=\textwidth]{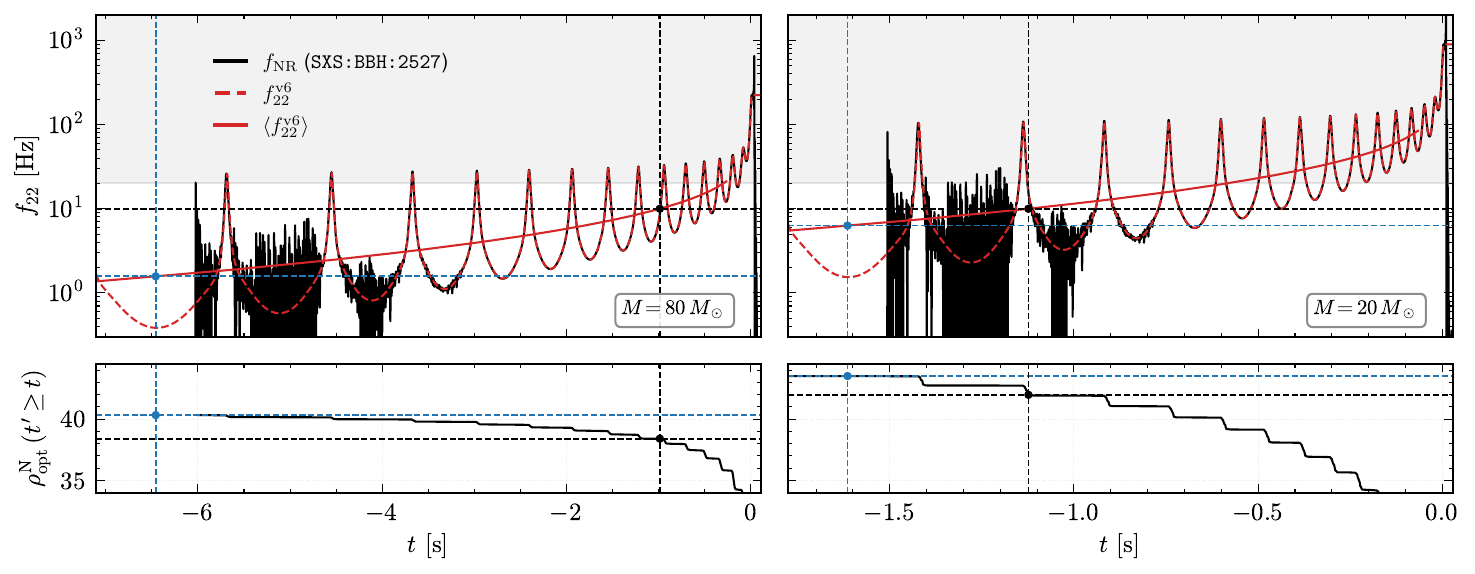}
    \caption{
	Instantaneous and orbit-averaged frequency of the $(2,2)$ mode for the NR simulation \texttt{SXS:BBH:2527} at total mass $M = 80\,M_\odot$ (left) and $M = 20\,M_\odot$ (right). The black curve shows the instantaneous frequency $f_{22}$ from the NR data, the red dashed curve is the instantaneous frequency of the best-fitting \texttt{SEOBNRv6EHM} waveform, and the solid red curve is its orbit-averaged counterpart $\langle f_{22}^{\mathrm{v6}} \rangle$.
	The grey shaded region marks the detector sensitivity band $[20,\, 2048]\,$Hz. The black dashed line indicates the default starting frequency $\langle f_{\text{start}} \rangle = 10\,$Hz used for the recovery templates, and the blue dashed line marks the orbit-averaged frequency which would be needed to capture the full NR frequency content starting the waveform at apastron. In both cases, several early periastron passages have instantaneous frequencies above $20\,$Hz but occur before $\langle f_{22}^{\mathrm{v6}} \rangle$ reaches $10\,$Hz, and are therefore missed by the recovery templates in our default analyses.
    The bottom panels show the optimal network SNR in the segment $t^{\prime}>t$: it asymptotes to $\rho^{\mathrm{N}}_{\mathrm{opt}}$ when the full NR signal is captured, and drops by $1$--$2$ units when the signal starts at $\langle f_{22}\rangle = 10$~Hz.
    }
    \label{fig:NR_orbit_averaged_frequencies}
\end{figure*}

Figure~\ref{fig:NR_injections_waveforms} shows the whitened injected NR waveforms in the LIGO Hanford detector for the six configurations considered, ordered by increasing eccentricity. The impact of eccentricity is clearly visible in the time-domain morphology: as the eccentricity increases, the waveform develops increasingly pronounced amplitude and frequency modulations during the inspiral, with sharp bursts of radiation at each periastron passage separated by quiescent apastron phases. For the most eccentric cases (\texttt{SXS:BBH:2525} and \texttt{SXS:BBH:2527}), these bursts dominate the early inspiral and the signal morphology departs from the smooth chirp characteristic of QC mergers.
Note that, because the figure shows whitened strain, the morphology of a given NR simulation depends on the total mass: rescaling $M$ shifts the waveform's frequency content, and the PSD weighting then suppresses portions of the signal that fall in the low-frequency region of the noise curve. As a result, the apastron arcs between consecutive periastron bursts appear either as low-amplitude oscillations when their frequency content lies above $20\,$Hz, or as flat, near-zero stretches when it falls below it. For \texttt{SXS:BBH:2527}, this is why only the last apastron passage is resolved in whitened strain at $M = 80\,M_\odot$---consistent with Fig.~\ref{fig:NR_orbit_averaged_frequencies}, where only the final apastron has instantaneous frequency above $20\,$Hz---whereas at $M = 20\,M_\odot$ the zoomed-in panel resolves five to six early apastron passages, all of which exceed $20\,$Hz in instantaneous frequency.
For the same reasons, the early periastron passages ($ t \lesssim -0.2 \,$s) for \texttt{SXS:BBH:2527} at $M = 80\,M_\odot$ look less pronounced than the ones at $M = 20\,M_\odot$, despite the binary being highly eccentric.

Beyond the setup changes above, care is taken to ensure consistent frequency content between the injected NR signals and the recovery templates. 
Specifically, for each simulation we determine the orbit-averaged $(2,2)$ mode frequency at the NR reference time from the best-fitting \texttt{SEOBNRv6EHM} waveform (i.e., the waveform whose eccentricity and anomaly parameters yield the lowest mismatch against the NR simulation, as obtained in Ref.~\cite{Gamboa:2026jht}) and we use this value as the starting frequency for waveform generation.
This approach is preferred over reading the frequency directly from the NR data, as numerical noise can prevent an accurate measurement of the orbit-averaged frequency at early times.
When the resulting frequency falls below $10\,$Hz, we start the template at $10\,$Hz instead, for computational efficiency and consistency with the real-data analyses.
For \texttt{SXS:BBH:1355}, \texttt{SXS:BBH:1359}, and \texttt{SXS:BBH:1362}, the orbit-averaged frequency at the reference time is above $10\,$Hz, so the template and injection have matching frequency content.
For \texttt{SXS:BBH:2525} and \texttt{SXS:BBH:2527}, the simulations are longer and the orbit-averaged frequency at the reference time falls below $10\,$Hz (starting the waveform at apastron, a starting frequency of about $6\,$Hz is needed to capture the full NR content for \texttt{SXS:BBH:2525} with $M = 80\,M_\odot$ and \texttt{SXS:BBH:2527} with $M = 20\,M_\odot$, and $1.5\,$Hz for \texttt{SXS:BBH:2527} with $M = 80\,M_\odot$); the templates therefore start at an orbit-averaged frequency of $10\,$Hz and miss the earliest periastron passages of the NR signal whose instantaneous frequency exceeds $20\,$Hz. This is illustrated in Fig.~\ref{fig:NR_orbit_averaged_frequencies} for \texttt{SXS:BBH:2527} at both total masses. 
The bottom panels show that this leads to a reduction of $\approx 1-2$ in the optimal network SNR. The curves showing the optimal network SNR as function of time decrease in steps rather than smoothly: each jump marks a periastron passage leaving the segment as $t$ increases.
We assess the impact of this approximation in Appendix~\ref{app:f_start_and_t_start}.

\subsection{Parameter recovery with \texttt{SEOBNRv6EHM}}
\label{sec:pe_injections_v6}

\begin{table*}
\centering
\small
\input{tab/NR_injection_table.tex}
\caption{
Injected values and recovered posterior medians with $90\%$ credible intervals for the six NR synthetic-signal injections, corresponding to equal-mass, nonspinning BBH configurations with increasing eccentricity, recovered with \texttt{SEOBNRv6EHM}.
The reported parameters are: total mass $M$, chirp mass $\mathcal{M}$, inverse mass ratio $1/q$, effective spin $\chi_{\text{eff}}$, inclination angle $\iota$, luminosity distance $d_L$, GW eccentricity $e_{\text{gw}}$ and mean anomaly $l_{\text{gw}}$ evaluated at a dimensionless orbit-averaged frequency $M \langle f_{\text{ref}} \rangle = 0.01$ (corresponding to $\approx 25\,$Hz for a total mass of $80\,M_\odot$), and network matched-filter SNR $\rho^{\text{N}}_{\text{mf}}$. The numbers in parentheses correspond to the configuration with $M = 20\,M_\odot$.
}
\label{tab:peinjections}
\end{table*}

\begin{figure*}
    \centering
    \includegraphics[width=0.8\textwidth]{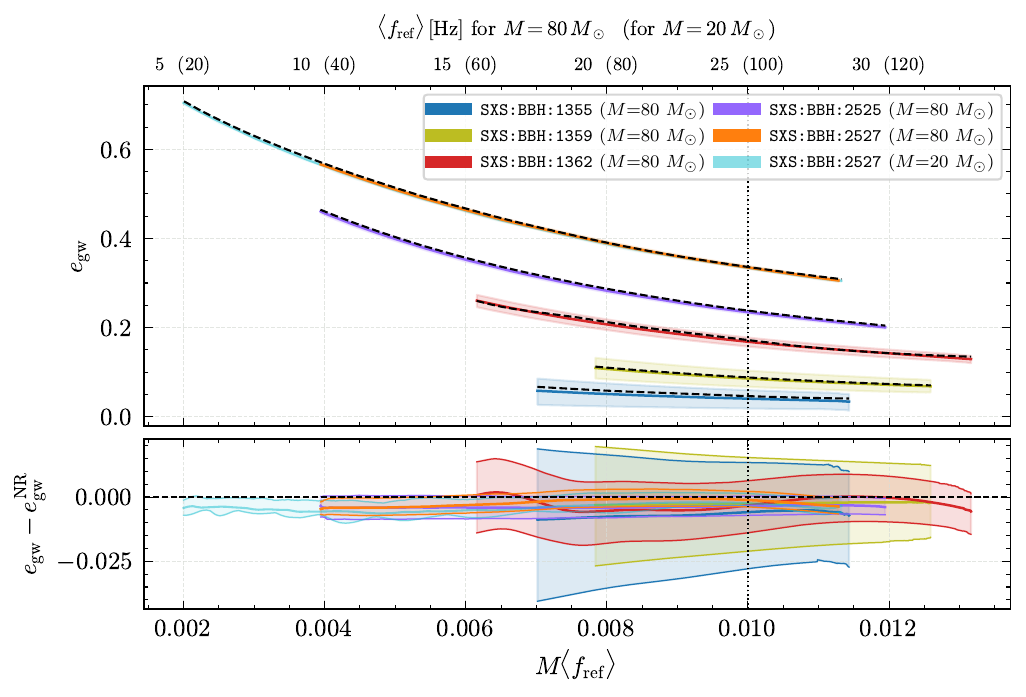}
    \caption{
    Reconstruction of the gravitational-wave eccentricity $e_{\rm gw}$ as a function of the dimensionless orbit-averaged frequency $M\langle f_{\mathrm{ref}} \rangle$ for the six NR injections recovered with \texttt{SEOBNRv6EHM}. Colored bands show the 90\% credible interval from the posterior samples, with solid lines indicating the median. Black dashed curves show the eccentricity measured from the SXS NR waveforms. The lower panel shows the residuals $e_{\rm gw} - e_{\rm gw}^{\rm NR}$. The top axis indicates the corresponding GW frequency in Hz for $M = 80\,M_\odot$, with values for $M = 20\,M_\odot$ in parentheses. The curves extend until the maximum frequency at which we are able to reliably measure the eccentricity from both the NR waveforms and the model using the \texttt{gw\_eccentricity} package, which is typically a few cycles before the merger. The vertical dotted line marks the reference frequency $M\langle f_{\rm ref} \rangle = 0.01$ at which the eccentricity values in Table~\ref{tab:peinjections} are reported.
    }
    \label{fig:NR_injections_eccentricity}
\end{figure*}

The results of the \texttt{SEOBNRv6EHM} recovery are summarized in Table~\ref{tab:peinjections} and Fig.~\ref{fig:NR_injections_eccentricity}.
The GW eccentricity $e_{\text{gw}}$ and mean anomaly $l_{\text{gw}}$ at the reference frequency $M \langle f_{\text{ref}} \rangle = 0.01$ are accurately recovered for all six configurations, with the injected values lying within the $90\%$ credible intervals. Both quantities are measured with increasing precision at higher eccentricities: for instance, the $90\%$ credible interval on $e_{\text{gw}}$ shrinks from $\pm 0.02$ for \texttt{SXS:BBH:1355} ($e_{\text{gw}} \approx 0.05$) to $\pm 0.003$ for \texttt{SXS:BBH:2527} ($e_{\text{gw}} \approx 0.34$).
The intrinsic parameters (total mass, chirp mass, mass ratio, and effective spin) are also very well recovered. The inclination angle $\iota$ shows a mild bias away from its face-on injected value of $\iota = 0$, which is expected since the prior on the binary orientation ($p(\iota) \propto \sin\iota$) has zero weight at $\iota = 0$; this prior-dominated effect also leads to a slight underestimation of the luminosity distance, as the two parameters are strongly correlated.

The network matched-filter SNR is perfectly recovered for \texttt{SXS:BBH:1355}, \texttt{SXS:BBH:1359}, and \texttt{SXS:BBH:1362}, for which the template and injection frequency content are matched. For \texttt{SXS:BBH:2525} and \texttt{SXS:BBH:2527} at $80\,M_\odot$, the recovered SNRs ($37.3$ and $38.3$, respectively) are measurably lower than the injected values ($38.3$ and $40.3$), reflecting the loss of signal power in the early periastron passages that the templates do not capture.
For \texttt{SXS:BBH:2527} at $20\,M_\odot$, the SNR loss is smaller ($42.7$ recovered versus $43.5$ injected) because the lower total mass places the orbit-averaged starting frequency of $10\,$Hz at an earlier point in the inspiral, capturing a larger fraction of the NR signal.
Despite this SNR loss, the recovered parameters are largely unbiased.
This can be understood geometrically: a waveform difference $\delta h$ can be decomposed into a component parallel to the waveform manifold, which can be absorbed by shifts in the model parameters and therefore leads to biases, and a component perpendicular to it, which results in an irreducible loss of SNR~\cite{Flanagan:1997kp,Lindblom:2008cm}.
The difference between waveforms generated with a lower and higher starting frequency $\langle f_{\text{start}} \rangle$ is predominantly perpendicular to the manifold: intuitively, the additional early periastron cycles cannot be reproduced by adjusting the masses, eccentricity, or any other parameter without explicitly lowering $\langle f_{\text{start}} \rangle$.
For the analyses presented here, the near-absence of parameter biases despite the SNR loss confirms that a fixed orbit-averaged starting frequency of $10\,$Hz is adequate for current detector sensitivities and SNRs.
Still, recovering the full SNR of the signal is desirable as it could maximize the precision of parameter measurements; potential strategies to achieve this---such as lowering the starting frequency or integrating the EOB equations of motion backward in time for a fixed duration---are explored in Appendix~\ref{app:f_start_and_t_start}.

While Table~\ref{tab:peinjections} reports eccentricity values at the single reference frequency $M \langle f_{\text{ref}} \rangle = 0.01$, the posterior samples can be used to reconstruct the full eccentricity evolution $e_{\text{gw}}(M\langle f_{\text{ref}}\rangle)$ by re-evaluating the \texttt{gw\_eccentricity} definitions at different reference frequencies for each sample.
Figure~\ref{fig:NR_injections_eccentricity} shows the result of this procedure for all six configurations. The recovered curves track the NR eccentricity across the full inspiral, with residuals remaining within the $90\%$ credible intervals for all configurations at $80\,M_\odot$, demonstrating that the model accurately captures the eccentricity decay throughout the inspiral and not just at a single reference point.
For the most demanding configuration (\texttt{SXS:BBH:2527} at $20\,M_\odot$), the NR eccentricity falls marginally outside the $90\%$ credible interval in the early inspiral, where the eccentricity is also higher, reflecting the tighter posteriors afforded by the higher SNR and longer duration; the bias nonetheless remains small in absolute terms, and none of the other parameters are affected;
the median residuals $e_{\rm gw} - e_{\rm gw}^{\rm NR}$ are below $\pm 0.01$ across the frequency range for all configurations.
These results confirm that \texttt{SEOBNRv6EHM} can reliably measure eccentricity across a wide range, even at the higher SNRs considered here compared to the previous \texttt{SEOBNRv5EHM} validation study~\cite{Gamboa:2024hli}.

\subsection{Waveform systematics}
\label{sec:pe_injections_systematics}

Having established that \texttt{SEOBNRv6EHM} accurately recovers all injected parameters, we now assess the sensitivity of the results to the choice of waveform model.
To do so, we also recover all six configurations with \texttt{SEOBNRv5EHM}~\cite{Gamboa:2024hli} and \texttt{TEOBResumS-Dal\'i}~\cite{Nagar:2024oyk}%
\footnote{We use the version of \texttt{TEOBResumS-Dal\'i} in the branch \texttt{LVK-Dali} of the public repository at \url{https://bitbucket.org/teobresums/teobresums}, which corresponds to the version reviewed within the LVK collaboration.},
to assess waveform systematics by comparing parameter recovery across state-of-the-art eccentric models.
For all configurations, we use priors and settings identical to those used for \texttt{SEOBNRv6EHM}, but we limit the eccentricity prior for \texttt{TEOBResumS-Dal\'i} to $e \leq 0.6$  for the systems with $M = 80\,M_\odot$, to avoid rare waveform generation failures at higher eccentricities.
This bound does not truncate the posteriors: even for the most eccentric configuration at this mass, \texttt{SXS:BBH:2527}, the eccentricity at the $10\,$Hz starting frequency is $e_{\rm gw} \approx 0.47$ (Fig.~\ref{fig:NR_injections_eccentricity}).
For the \texttt{SXS:BBH:2527} injection at $M = 20\,M_\odot$, the orbit-averaged starting frequency of $10\,$Hz corresponds to an earlier point in the inspiral where the eccentricity is higher; we therefore extend the upper bound of the eccentricity prior to $e \leq 0.9$ for all models to ensure the prior covers the full posterior support.
Additionally, for \texttt{TEOBResumS-Dal\'i} we sample uniformly in eccentricity and true anomaly, rather than the default uniform-in-mean-anomaly parameterization, for consistency with the uniform-in-relativistic-anomaly prior used by the \texttt{SEOBNR} models; this is a more consistent choice because the relativistic and true anomaly coincide in the Newtonian limit.

Since all injected configurations are equal-mass and observed face-on, higher-order modes (whose content differs across the three models) contribute negligibly, except for the $(\ell,|m|)=(3,2)$ mode. Biases from neglecting higher-order modes in eccentric inference grow with mass ratio and inclination~\cite{Tang:2026jvl} (see also Fig.~14 of Ref.~\cite{Gamboa:2026jht}); the systematic differences reported below therefore mostly reflect the differing modeling of the eccentric orbital dynamics.
For the same reason, the biases quoted below are established for equal-mass, nonspinning, face-on binaries; how they scale with mass ratio, spin and inclination would require a dedicated study, which we leave to future work.

\begin{figure*}
    \centering
    \includegraphics[width=0.85\textwidth]{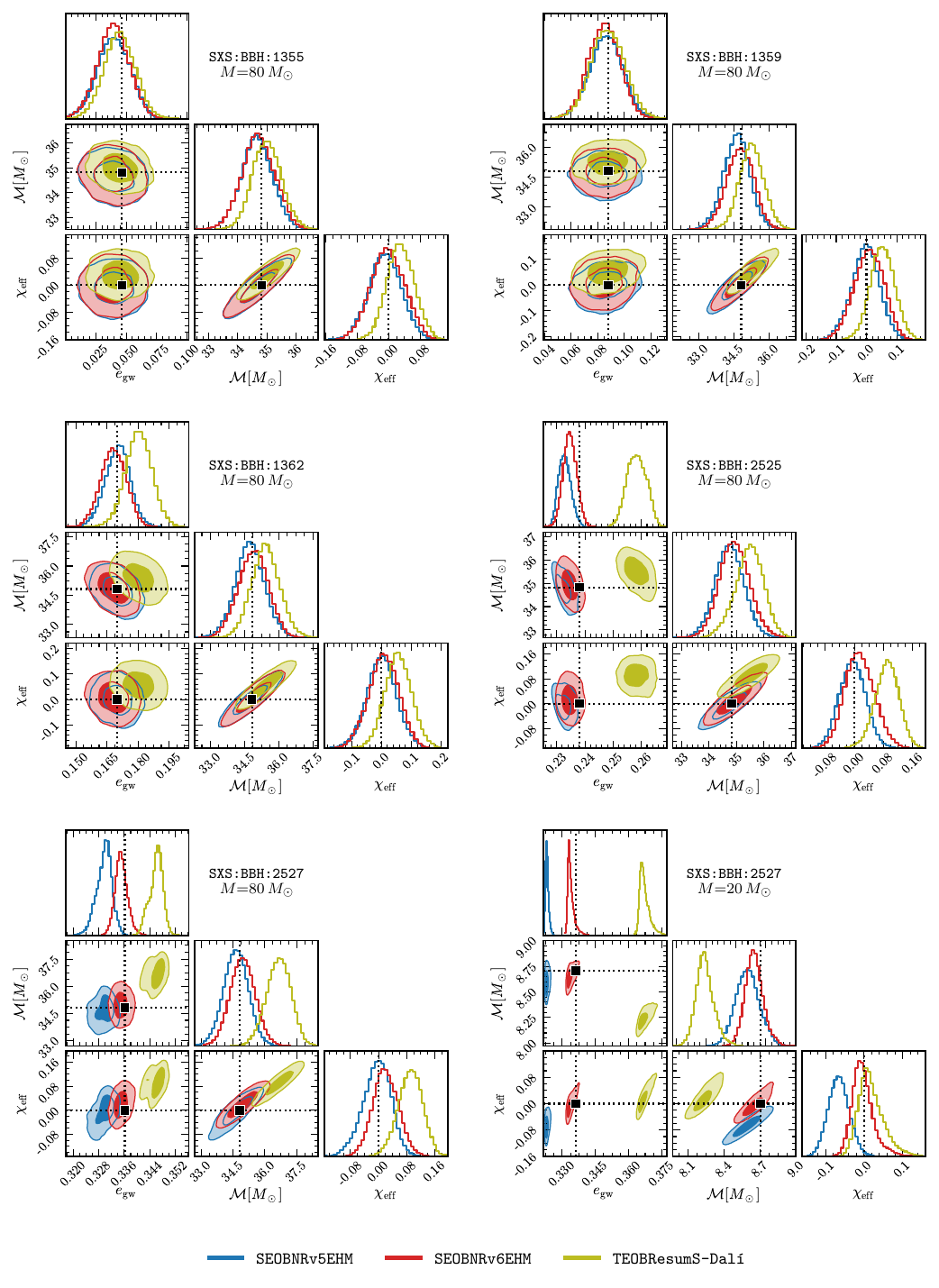}
    \caption{
	Posterior distributions of GW eccentricity $e_{\text{gw}}$, chirp mass $\mathcal{M}$, and effective spin $\chi_{\text{eff}}$ for the six NR synthetic-signal configurations recovered with \texttt{SEOBNRv6EHM}, \texttt{SEOBNRv5EHM}, and \texttt{TEOBResumS-Dal\'i}. The eccentricity is evaluated at a dimensionless orbit-averaged frequency $M \langle f_{\text{ref}} \rangle = 0.01$ (corresponding to $\approx 25\,$Hz for a total mass $80\,M_\odot$). Contours show the $50\%$ and $90\%$ credible regions; the injected values are indicated by the crosshairs.
    }
    \label{fig:NR_inj_corner}
\end{figure*}

\begin{figure*}
    \centering
    \includegraphics[width=\textwidth]{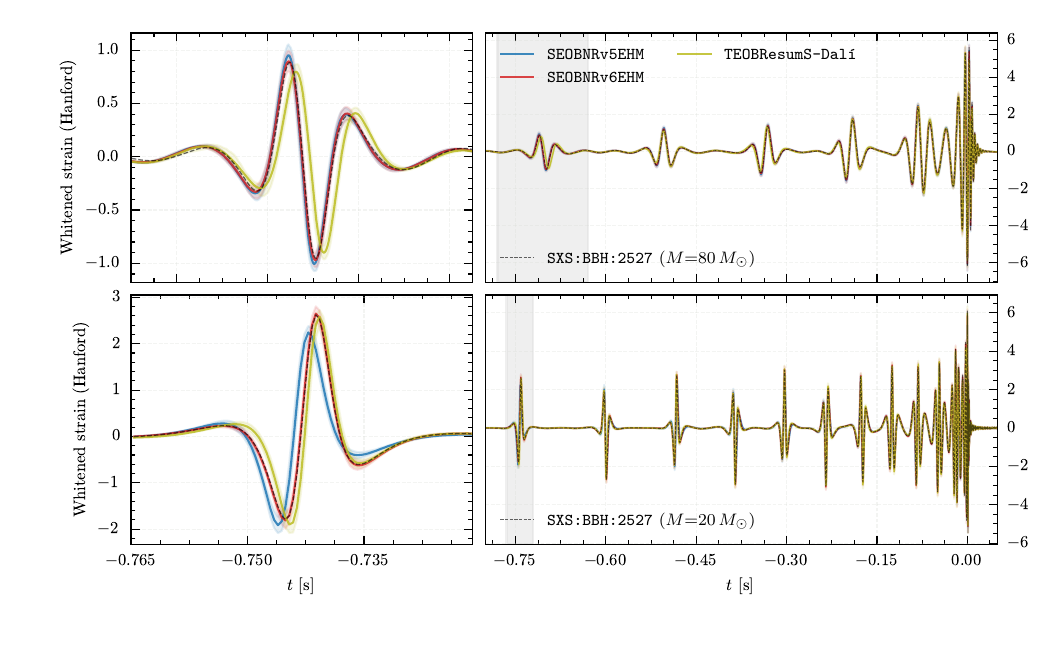}
    \caption{
	Whitened-strain waveform reconstruction for the highest-eccentricity NR configuration (\texttt{SXS:BBH:2527}, injected $e_{\rm gw} = 0.336$ at $M\langle f_{\rm ref} \rangle = 0.01$) in the LIGO Hanford detector, for total mass $80\,M_\odot$ (top) and $20\,M_\odot$ (bottom). The black dashed curves are the injected NR signals; the colored curves show the posterior median of the \texttt{SEOBNRv6EHM}, \texttt{SEOBNRv5EHM}, and \texttt{TEOBResumS-Dal\'i} reconstructions, with the shaded bands indicating the $90\%$ credible intervals across posterior samples. The left panel shows a zoomed-in view of a periastron passage in the early inspiral, indicated by the gray band in the right panel, where differences between the models are most evident.
    }
    \label{fig:NR_inj_waveform}
\end{figure*}

\renewcommand{\arraystretch}{1.5}
\begin{table}[t]
\centering
\small
\begin{tabular*}{\columnwidth}{l@{\extracolsep{\fill}} l c c}
\hline\hline
Configuration & Model & $\rho^{\text{N}}_{\text{mf}}$ & $\log \mathcal{L}$ \\
\hline
\multirow{3}{*}{\makecell[l]{\texttt{2525}, $80\,M_\odot$ \\ ($\rho^{\text{inj}} = 38.3$)}} 
 & \texttt{SEOBNRv6EHM} & $ {37.3}^{+0.1}_{-0.1} $ & $ {690.3}^{+2.7}_{-4.6} $ \\
 & \texttt{SEOBNRv5EHM} & $ {37.3}^{+0.1}_{-0.1} $ & $ {690.3}^{+2.6}_{-4.2} $ \\
 & \texttt{TEOBResumS-Dal\'i} & $ {37.1}^{+0.1}_{-0.1} $ & $ {683.0}^{+2.7}_{-4.6} $ \\
\hline
\multirow{3}{*}{\makecell[l]{\texttt{2527}, $80\,M_\odot$ \\ ($\rho^{\text{inj}} = 40.3$)}} 
 & \texttt{SEOBNRv6EHM} & $ {38.3}^{+0.1}_{-0.1} $ & $ {725.7}^{+2.6}_{-4.5} $ \\
 & \texttt{SEOBNRv5EHM} & $ {38.2}^{+0.1}_{-0.1} $ & $ {722.6}^{+2.9}_{-4.7} $ \\
 & \texttt{TEOBResumS-Dal\'i} & $ {37.4}^{+0.1}_{-0.1} $ & $ {693.3}^{+2.5}_{-4.2} $ \\
\hline
\multirow{3}{*}{\makecell[l]{\texttt{2527}, $20\,M_\odot$ \\ ($\rho^{\text{inj}} = 43.5$)}} 
 & \texttt{SEOBNRv6EHM} & $ {42.7}^{+0.1}_{-0.1} $ & $ {900.4}^{+2.5}_{-4.3} $ \\
 & \texttt{SEOBNRv5EHM} & $ {38.6}^{+0.1}_{-0.1} $ & $ {736.0}^{+2.6}_{-4.4} $ \\
 & \texttt{TEOBResumS-Dal\'i} & $ {40.0}^{+0.1}_{-0.1} $ & $ {790.9}^{+2.5}_{-4.0} $ \\
\hline\hline
\end{tabular*}
\caption{
Network matched-filter SNR $\rho^{\text{N}}_{\text{mf}}$ and log-likelihood
$\log \mathcal{L}$ (median and $90\%$ credible interval) for the high-eccentricity NR configurations recovered
with three waveform models. The injected SNR values are given in parentheses
in the first column.
}
\label{tab:pe_systematics}
\end{table}

The multi-model comparison is shown in Figs.~\ref{fig:NR_inj_corner} and~\ref{fig:NR_inj_waveform}, and key diagnostic quantities related to goodness of fit are reported in Table~\ref{tab:pe_systematics}.

Figure~\ref{fig:NR_inj_corner} shows the posterior distributions of GW eccentricity $e_{\text{gw}}$ measured at $M \langle f_{\text{ref}} \rangle = 0.01$, chirp mass $\mathcal{M}$, and effective spin $\chi_{\text{eff}}$, for the six NR configurations recovered with the three waveform models.
For the lower-eccentricity configurations (\texttt{SXS:BBH:1355} and \texttt{SXS:BBH:1359}), all three models recover the injected parameters consistently, with posteriors that largely overlap and cover the true values within the $90\%$ credible region.
For \texttt{SXS:BBH:1362}, a systematic bias begins to appear for \texttt{TEOBResumS-Dal\'i}, with the injected values lying at the edge of the $90\%$ credible region in the two-dimensional marginal distributions involving eccentricity.
More substantial differences emerge for the most challenging configurations.
For \texttt{SXS:BBH:2525} and \texttt{SXS:BBH:2527} at $80\,M_\odot$, \texttt{TEOBResumS-Dal\'i} exhibits significant biases in eccentricity, chirp mass, and effective spin, with the injected values falling entirely outside the $90\%$ credible region in the two-dimensional marginals, while \texttt{SEOBNRv5EHM} shows mild biases, with the injected values lying just outside the $90\%$ credible region. Only \texttt{SEOBNRv6EHM} recovers the injected parameters accurately.
These biases become more severe for the $20\,M_\odot$ configuration of \texttt{SXS:BBH:2527}, where a larger portion of the eccentric inspiral is in band and modeling errors accumulate over many more orbital cycles and periastron passages. In this case, both \texttt{SEOBNRv5EHM} and \texttt{TEOBResumS-Dal\'i} show significant biases, with the injected value lying outside the $90\%$ credible region in all two-dimensional marginals. In contrast, \texttt{SEOBNRv6EHM} continues to recover the injected parameters accurately, demonstrating its improved modeling of high-eccentricity inspirals.

A similar picture emerges from the goodness-of-fit diagnostics in Table~\ref{tab:pe_systematics}.
For \texttt{SXS:BBH:2525}, \texttt{SEOBNRv6EHM} and \texttt{SEOBNRv5EHM} achieve comparable log-likelihoods and recovered SNRs, while \texttt{TEOBResumS-Dal\'i} shows a reduction of $\sim 7$ in log-likelihood and $\sim 0.2$ in SNR.
For the higher-eccentricity configurations, \texttt{SEOBNRv6EHM} achieves the highest log-likelihood and recovered SNR, with the differences growing with eccentricity and fraction of inspiral in band.
Since the analyses use the same data and parameters, differences in the median log-likelihood between models are approximately equal to the log Bayes factor in favor of the higher-likelihood model.
A direct comparison of Bayes factors from the nested-sampler evidences across all models is complicated by the fact that some configurations use different prior bounds; however, for \texttt{SXS:BBH:2527} at $20\,M_\odot$, where all priors are consistent, we verified that the Bayes factors from nested sampling approximately agree with the likelihood differences.
In all cases \texttt{SEOBNRv6EHM} is either preferred or shows a comparable level of preference to both alternatives, with the preference growing with eccentricity and fraction of inspiral in band: for the most challenging configuration (\texttt{SXS:BBH:2527} at $20\,M_\odot$), the differences correspond to $\log_{10}\mathcal{B} \approx 70$ in favor of \texttt{SEOBNRv6EHM} over \texttt{SEOBNRv5EHM}, and $\log_{10}\mathcal{B} \approx 46$ over \texttt{TEOBResumS-Dal\'i}, indicating decisive preference.

A higher likelihood or Bayes factor does not, however, automatically imply more accurate parameter recovery~\cite{Hoy:2022tst}.
For example, for the \texttt{SXS:BBH:2527} injection at $20\,M_\odot$, \texttt{TEOBResumS-Dal\'i} achieves a higher recovered SNR and log-likelihood than \texttt{SEOBNRv5EHM}, yet the recovered parameters are slightly more biased: this is because the model is able to fit the data better by shifting its parameters further away from the injected values.
Only \texttt{SEOBNRv6EHM} achieves both a high likelihood and accurate, unbiased parameter recovery, confirming that its improved waveform accuracy is the source of the gain rather than a better fit to a biased template.

These PE results are consistent with the intrinsic waveform accuracy of each model.
For the \texttt{SXS:BBH:2527} configuration, the mismatch against the NR waveform, optimized over eccentricity and starting frequency, is $\mathcal{M}_{\rm v6} \sim 2\%$ for \texttt{SEOBNRv6EHM}, $\mathcal{M}_{\rm v5} \sim 20\%$ for \texttt{SEOBNRv5EHM}, and $\mathcal{M}_{\rm TEOB} \sim 30\%$ for \texttt{TEOBResumS-Dal\'i}~\cite{Gamboa:2026jht}.
These values can be compared against the indistinguishability criterion $\mathcal{M} \lesssim 1/(2\rho^2)$~\cite{Lindblom:2008cm, Chatziioannou:2017tdw} (or, more precisely, $\mathcal{M} \lesssim \mathcal{Q}(1{-}p;\, 1)/(2\rho^2)$, where $\mathcal{Q}(1{-}p;\, 1)$ denotes the $(1{-}p)$-quantile of the $\chi^2$ distribution with one degree of freedom, for a single parameter of interest at confidence level $p$~\cite{Thompson:2025hhc}): for the injected SNRs of $\rho \approx 40$ and $p = 0.9$, this gives a threshold of $\mathcal{M} \lesssim 8 \times 10^{-4}$.
All three models exceed this threshold, but the criterion provides only a sufficient condition for indistinguishability, not a necessary one: mismatches above the threshold do not guarantee parameter biases.
Indeed, \texttt{SEOBNRv6EHM} recovers largely unbiased parameters despite its mismatch of ${\sim}\,2\%$.
In contrast, the substantially larger mismatches of \texttt{SEOBNRv5EHM} and \texttt{TEOBResumS-Dal\'i} are consistent with the significant biases observed in the high-eccentricity configurations.

The waveform reconstructions for the two configurations of \texttt{SXS:BBH:2527} in Fig.~\ref{fig:NR_inj_waveform} provide further insight into this mechanism.
When mismatch calculations are performed with fixed intrinsic parameters (varying only eccentricity and starting frequency, as in Ref.~\cite{Gamboa:2026jht}), \texttt{SEOBNRv5EHM} and \texttt{TEOBResumS-Dal\'i} fail to accurately reproduce the early inspiral of the \texttt{SXS:BBH:2527} simulation (see Fig.~13 of Ref.~\cite{Gamboa:2026jht}).
In the PE reconstructions shown here, however, both models are free to shift their intrinsic parameters, and this freedom allows them to produce waveforms that visually track the NR signal more closely than a fixed-parameter comparison would suggest.
This is the source of the parameter biases shown in Fig.~\ref{fig:NR_inj_corner}.
Yet the agreement remains imperfect, as is clearly visible in the zoomed view of the early periastron passages: the reconstructions show discrepancies that manifest in the reduced recovered SNR and non-negligible residuals noted above.
In contrast, \texttt{SEOBNRv6EHM} reconstructs the signal accurately across the full inspiral, with both smaller residuals and unbiased parameter estimates.
These results highlight the importance of waveform accuracy for unbiased eccentricity inference, and demonstrate that \texttt{SEOBNRv6EHM} substantially reduces systematic biases compared to other state-of-the-art models.

\section{Analysis of real gravitational-wave events}
\label{sec:pe_events}

We now apply the \texttt{SEOBNRv6EHM} model to \NEVENTS real GW events spanning the O1--O4 observing runs.
The event selection is driven by two complementary goals.
First, we choose a set of \emph{benchmark events} for which we perform PE with multiple models (\texttt{SEOBNRv5HM}~\cite{Pompili:2023tna}, \texttt{SEOBNRv5EHM}~\cite{Gamboa:2024hli}, and \texttt{SEOBNRv6EHM} in both the QC and eccentric configurations) using identical settings, to directly compare performance and consistency of the new model.
Second, we select events for which previous analyses have reported evidence or mild support for eccentricity, as well as events with extreme parameters (e.g., high mass ratio, high total mass, long duration) to test the robustness of the model. These events are analyzed with \texttt{SEOBNRv6EHM} alone to provide independent eccentricity measurements.

For each event, we use the strain data, calibration uncertainties, and PSDs provided by the Gravitational Wave Open Science Center (GWOSC)~\cite{LIGOScientific:2019lzm}. For events with data quality issues, we use deglitched frames provided in the respective data releases~\cite{ligo_scientific_collaboration_and_virgo_2022_6477076, ligo_scientific_collaboration_and_virgo_2021_5546680, ligo_scientific_collaboration_2025_16857060}.
The sampling rate, segment length, and frequency settings are adapted to each event and closely follow the official LVK analyses.

To assess the evidence for orbital eccentricity, we compute the Bayes factor between the eccentric, aligned-spin (EAS) hypothesis $ \mathcal{H}_{\text{EAS}} $ and the quasi-circular, aligned-spin (QCAS) hypothesis $ \mathcal{H}_{\text{QCAS}} $, defined as
\begin{equation}
\label{eq:bayesfactor}
\mathcal{B}^{\text{EAS}}_{\text{QCAS}} = \frac{\mathcal{Z}(d \mid \mathcal{H}_{\text{EAS}})}{\mathcal{Z}(d \mid \mathcal{H}_{\text{QCAS}})}.
\end{equation}
We estimate the Bayes factor using the Savage--Dickey density ratio~\cite{10.1214/aoms/1177693507}, which computes $ \mathcal{B}^{\text{EAS}}_{\text{QCAS}} $ as the ratio of the posterior to the prior evaluated at the nested-model boundary ($ e = 0 $), requiring only the eccentric run.
The posterior density at $e = 0$ is estimated with a bounded one-dimensional Gaussian kernel density estimator that handles the boundary by reflection, mirroring the kernels about $e = 0$. The uncertainty on the Savage--Dickey estimate is obtained by bootstrap resampling the posterior samples and averaging the result over multiple kernel-density-estimation bandwidths used to estimate the posterior density at $e = 0$. For the four benchmark events of Sec.~\ref{sec:pe_benchmark}, for which the separate QC run allows $\mathcal{B}^{\text{EAS}}_{\text{QCAS}}$ to be computed directly from the nested-sampling evidences, the Savage--Dickey estimate agrees with the evidence ratio within the respective uncertainties.

For events with the highest support for eccentricity, we perform additional runs with the QCP model \texttt{SEOBNRv5PHM}~\cite{Ramos-Buades:2023ehm} (including multipole asymmetries in the co-precessing frame~\cite{Estelles:2025zah}) and compute the Bayes factor $ \mathcal{B}^{\text{EAS}}_{\text{QCP}} $ between the EAS and QCP hypotheses, to assess whether the apparent eccentricity could instead be attributed to spin-precession effects.
Since $\mathcal{B}^{\text{EAS}}_{\text{QCP}}$ compares two different waveform families, it also includes any difference between their QC limits. For the four benchmark events of Sec.~\ref{sec:pe_benchmark}, the Bayes factor between \texttt{SEOBNRv6EHM} ($e=0$) and \texttt{SEOBNRv5HM} ranges from $\log_{10}\mathcal{B} \approx -0.09$ to $0.15$, comparable to its statistical uncertainty ($\pm 0.09$), and the two models yield consistent posteriors, in line with the close agreement between them found in Ref.~\cite{Gamboa:2026jht}. We therefore expect this effect to be subdominant.

\subsection{Benchmark events}
\label{sec:pe_benchmark}

A key advantage of \texttt{SEOBNRv6EHM} is its significantly reduced computational cost compared to the previous \texttt{SEOBNRv5EHM} model~\cite{Gamboa:2024hli}. 
To quantify this improvement in a PE context, we select four benchmark events that span a range of binary types and signal durations, and analyze them with multiple waveform models, including \texttt{SEOBNRv5HM}~\cite{Pompili:2023tna}, \texttt{SEOBNRv5EHM}~\cite{Gamboa:2024hli}, and \texttt{SEOBNRv6EHM} in both the QC ($e=0$) and eccentric configurations.
\begin{itemize}[leftmargin=*]
\item GW150914~\cite{LIGOScientific:2016aoc}: the first detected BBH merger.
\item GW190521~\cite{LIGOScientific:2020iuh}: a massive BBH merger, for which early analyses showed signs of eccentricity~\cite{Romero-Shaw:2020thy, Gayathri:2020coq, Gamba:2021gap}.
\item GW200105\_162426~\cite{LIGOScientific:2021qlt}: an NSBH binary, for which analyses have found signs of eccentricity~\cite{Morras:2025xfu, Planas:2025plq, Kacanja:2025kpr, Jan:2025fps, Clarke:2026cuw}. The longer signal duration makes this event particularly sensitive to the computational cost of the waveform model.
\item GW170817~\cite{LIGOScientific:2017vwq}: the first detected BNS merger. As \texttt{SEOBNRv6EHM} does not include tidal effects, this analysis primarily serves as a test of computational performance and robustness for the longest signals currently relevant for ground-based detectors.
Tides enter the phasing at 5PN relative order and matter only in the last few tens of cycles, whereas eccentricity enters at Newtonian relative order and decays as the binary circularizes, so the two are imprinted on opposite ends of the band. We accordingly expect the eccentricity constraint to be robust against the neglect of tides, and find no spurious support for eccentricity. The inferred $\chi_{\rm eff}$ and luminosity distance agree with the tidal analyses of Ref.~\cite{LIGOScientific:2018hze}, whereas the mass ratio, which correlates with the tidal deformability, has less support near equal masses; mass-ratio and component-mass estimates for this event should therefore be interpreted with care.
\end{itemize}

\begin{table*}
\centering
\small
\input{tab/PE_runtimes.tex}
\caption{
    Analysis settings and run statistics for the four benchmark events analyzed with multiple waveform models.
    For each event, we list the data settings---sampling rate ($\text{srate}$), segment duration, and orbit-averaged initial frequency for waveform generation $\langle f_\text{start} \rangle$---as well as the sampler configuration, either \texttt{Bilby} or \texttt{parallel Bilby}.
    The computing resources are specified as the number of CPU cores per node times the number of nodes. The total computational cost in CPU hours can be obtained by multiplying the wall-clock time by the number of CPU cores employed.
    For each waveform model---\texttt{SEOBNRv5HM}~\cite{Pompili:2023tna}, \texttt{SEOBNRv5EHM}~\cite{Gamboa:2024hli}, and \texttt{SEOBNRv6EHM}~\cite{Gamboa:2026jht} in its quasi-circular ($e=0$) and eccentric configurations---we report the wall-clock run time and the total number of likelihood evaluations ($\mathcal{L}$ eval.).
}
\label{tab:peruntime}
\end{table*}

\begin{figure}
    \centering
    \includegraphics[width=\columnwidth]{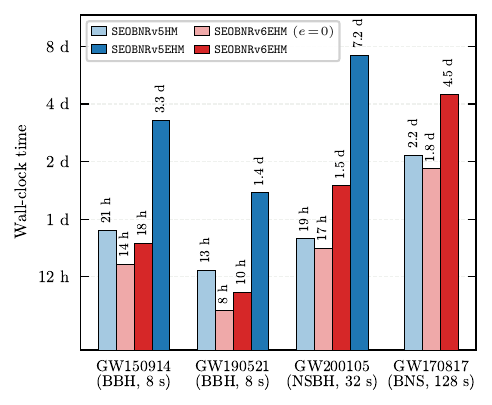}
    \caption{
	Wall-clock run times for the four benchmark events analyzed with \texttt{SEOBNRv5HM}, \texttt{SEOBNRv5EHM}, and \texttt{SEOBNRv6EHM} (in both quasi-circular and eccentric configurations).
	Each group of bars corresponds to one event, with the signal duration and binary type indicated below.
	Time labels are shown above each bar.
	For GW170817, the \texttt{SEOBNRv5EHM} run was not performed as it would be prohibitively expensive.
	The computing resources for each event are listed in Table~\ref{tab:peruntime}.
    }
    \label{fig:runtimes}
\end{figure}

We report the analysis settings, computing resources, wall-clock run times, and number of likelihood evaluations for all benchmark runs in Table~\ref{tab:peruntime}. Note that for GW190521, the starting frequency of the likelihood integration is set to $11\,$Hz, following the LVK analysis~\cite{LIGOScientific:2020iuh}, and we thus start the waveform generation at an orbit-averaged frequency of $5.5\,$Hz. For GW170817, we start both the likelihood and waveform generation at $23\,$Hz, as in the LVK analysis~\cite{LIGOScientific:2018hze}. Following Ref.~\cite{LIGOScientific:2018hze}, we also fix the sky location to the known host galaxy NGC 4993 for GW170817, and use a prior on the spin magnitudes limited to $0.05$ consistent with the low spins expected for BNS systems.

The computational speedup of \texttt{SEOBNRv6EHM} is particularly pronounced for eccentric configurations (see Fig.~\ref{fig:runtimes}): the wall-clock times are reduced by factors of $\sim 3.3$ for GW190521, $\sim 4.4$ for GW150914, and up to $\sim 4.8$ for GW200105\_162426, when compared to \texttt{SEOBNRv5EHM}. The number of likelihood evaluations remains comparable, indicating that the speedup is primarily driven by the reduced cost of waveform generation, rather than by faster sampler convergence or improved posterior structure.

Notably, in its QC limit ($e = 0$) \texttt{SEOBNRv6EHM} is faster than the dedicated QC model \texttt{SEOBNRv5HM} for all four benchmark events, by a factor of $\sim 1.5$ for the shorter-duration BBH events and by $\sim 10$--$20\%$ for the longer-duration NSBH and BNS events, at a comparable number of likelihood evaluations. Thus, the eccentric infrastructure introduces no overhead when eccentricity is not included, making \texttt{SEOBNRv6EHM} equally suitable for QC analyses.

Comparing eccentric and QC configurations of \texttt{SEOBNRv6EHM}, the eccentric runs require comparable time for short-duration signals, and roughly twice the wall-clock time for longer-duration signals, primarily due to increased waveform generation cost.
The likelihood evaluation counts are comparable for short signals, but the longer-duration NSBH event (GW200105\_162426) shows a $\sim 30\%$ increase.
This increase is driven by the more complex posterior structure and additional parameter dimensionality of the eccentric case.

For shorter-duration BBH signals, \texttt{SEOBNRv6EHM} runs complete within a day or less on a single computing node with 64 CPU cores.
For the NSBH event GW200105\_162426 (32-second signal), the eccentric analysis completes in approximately 1.5 days using parallel resources, compared to over 7 days for \texttt{SEOBNRv5EHM}.
The BNS event GW170817 (128-second signal) would be prohibitively expensive with \texttt{SEOBNRv5EHM}, but completes in approximately 4.5 days with \texttt{SEOBNRv6EHM} using parallel resources.
We verify the expected speedup by comparing per-evaluation costs of both models using parameters drawn from the \texttt{SEOBNRv6EHM} posterior, confirming that the improvement for GW170817 is comparable to that for GW200105\_162426 (see Appendix~\ref{app:dali_benchmarks}). Overall, these results demonstrate that \texttt{SEOBNRv6EHM} makes both large-scale catalogs and long-duration eccentric analyses computationally feasible with standard sampling algorithms.

The inferred eccentricity and intrinsic parameters are broadly consistent across all waveform models for the four benchmark events; we defer a detailed discussion of the eccentricity measurements to Sec.~\ref{sec:pe_ecc_events}.

We do not run \texttt{TEOBResumS-Dal\'i} on real events due to its higher computational cost for long-duration signals. However, we present timing comparisons in Appendix~\ref{app:dali_benchmarks}, showing that \texttt{SEOBNRv6EHM} is also faster than \texttt{TEOBResumS-Dal\'i} for all benchmark events, by a factor $\sim 2.5$ for the shortest signal and by more than an order of magnitude for the longer NSBH and BNS signals.

\subsection{Individual event results}
\label{sec:pe_ecc_events}

Beyond the benchmark events of Sec.~\ref{sec:pe_benchmark}, we analyze additional events with the \texttt{SEOBNRv6EHM} model. 
Our selection includes both signals for which previous studies have reported evidence or mild support for eccentricity, and events with extreme parameters (e.g., high mass ratio, high total mass, long duration) that test the robustness of the model across a broad region of parameter space:
\begin{itemize}
\item \textit{Eccentricity candidates from O1--O3.}
We include events with potential eccentricity signatures identified by Romero-Shaw et al.~\cite{Romero-Shaw:2020thy, Romero-Shaw:2022xko}: GW190521, GW190620\_030421, GW191109\_010717, and GW200208\_222617.
The \texttt{SEOBNRv4EHM} analysis of Ref.~\cite{Gupte:2024jfe} confirmed the support for GW200208\_222617 and additionally identified GW200129\_065458 (the most significant candidate) and GW190701\_203306; however, both of the latter are affected by glitches in the LIGO Livingston detector, and the inferred eccentricity varies significantly depending on the glitch-subtraction method~\cite{Gupte:2024jfe, Gupte:2026whi}.
GW190620\_030421 and GW190706\_222641 also show marginal support in Ref.~\cite{Gupte:2024jfe}.
The \texttt{IMRPhenomTEHM} analysis of Ref.~\cite{Planas:2025jny} finds comparable evidence for GW200129\_065458 and GW200208\_222617, as well as mild support for the high-mass events GW190701\_203306 and GW190929\_012149, the latter not found significant in other analyses.
We also include the NSBH event GW200105\_162426, for which multiple independent analyses have reported support for non-zero eccentricity~\cite{Morras:2025xfu, Planas:2025plq, Kacanja:2025kpr, Jan:2025fps}.
\item \textit{Eccentricity candidates from O4a.}
The \texttt{SEOBNRv5EHM} analysis of Ref.~\cite{Gupte:2026whi} identified nine candidates with either $\log_{10}\mathcal{B}^{\text{EAS}}_{\text{QCAS}} \gtrsim 0.2$ or $\log_{10}\mathcal{B}^{\text{EAS}}_{\text{QCP}} \gtrsim 0.2$:
GW230706\_104333, GW230709\_122727, GW230820\_212515, GW231001\_140220, GW231114\_043211, GW231221\_135041, GW231223\_032836, GW231224\_024321, and GW240104\_164932. In all cases the evidence remains modest, with $\log_{10}\mathcal{B}^{\text{EAS}}_{\text{QCP}} \approx 0.62$ for the most significant candidate (GW231001\_140220).
The \texttt{IMRPhenomTEHM} analysis of Ref.~\cite{Xu:2025ajj} found comparable support for a partially overlapping set, including GW230706\_104333, GW231114\_043211, GW231221\_135041, GW231223\_032836, and GW231224\_024321. Ref.~\cite{Xu:2025ajj} additionally found support for GW230712\_090405 and GW231123\_135430 when comparing EAS against QCAS, although not against QCP; neither event shows support even against QCAS with \texttt{SEOBNRv5EHM}~\cite{Gupte:2026whi}.
GW231001\_140220 also shows mild support for eccentricity with \texttt{TEOBResumS-Dal\'i} in Ref.~\cite{Malagon:2026uev}, although the \texttt{SEOBNRv5EHM} model finds no support in this analysis, in contrast to Ref.~\cite{Gupte:2026whi}.
\item \textit{Additional events for robustness testing.}
We include GW190814 and GW191219\_163120~\cite{LIGOScientific:2020zkf, Kagra:2021vkt}, potential NSBH events with the highest mass ratios detected to date, to test the robustness of \texttt{SEOBNRv6EHM} for highly asymmetric binaries.
GW241011\_233834 and GW241110\_124123~\cite{LIGOScientific:2025brd} are asymmetric, high-spin BBH events from O4b, identified as potential products of hierarchical mergers and found consistent with QC orbits with \texttt{SEOBNRv5EHM} and \texttt{TEOBResumS-Dal\'i}~\cite{LIGOScientific:2025brd}.
GW231123\_135430, the most massive BBH event observed to date, is also included~\cite{LIGOScientific:2025rsn}; evidence for the EAS hypothesis over QCAS has been reported with \texttt{TEOBResumS-Dal\'i}~\cite{Jan:2025zcm, Malagon:2026uev} and \texttt{IMRPhenomTEHM}~\cite{Xu:2025ajj}, although in both cases the QCP hypothesis is preferred; with \texttt{SEOBNRv5EHM} the EAS hypothesis is disfavored even relative to QCAS~\cite{Gupte:2026whi}.
We also include GW250114\_082203~\cite{LIGOScientific:2025rid}, the loudest BBH signal observed to date, as a high-SNR benchmark.
The sample is completed by the benchmark events GW150914 and GW170817 discussed in Sec.~\ref{sec:pe_benchmark}.
\end{itemize}

We run all analyses with \texttt{Bilby} on a single computing node with 64 CPU cores, except for GW200105\_162426 and GW170817, which use \texttt{pBilby} as noted in Sec.~\ref{sec:pe_benchmark}. Runtimes range from $\sim 9$ hours for short-duration BBH events, up to a week for GW241011\_233834, a long BBH signal with $32$\,s duration and high SNR.

\setlength{\extrarowheight}{1pt}
\begin{table*}
\centering
\resizebox{\linewidth}{!}{
\input{tab/PE_results.tex}
}
\caption{
    Median and $90\%$ credible intervals for all \NEVENTS gravitational-wave events analyzed with \texttt{SEOBNRv6EHM}, sorted in descending order of the Bayes factor $\log_{10}\mathcal{B}^{\text{EAS}}_{\text{QCAS}}$. The events marked with an asterisk (*) were affected by data-quality issues and have been analyzed with glitch-subtracted frames.
    The reported parameters are: source-frame total mass $M_{\rm src}$, inverse mass ratio $1/q$, effective spin $\chi_{\text{eff}}$, luminosity distance $d_L$, and Keplerian eccentricity $e$ from the \texttt{SEOBNRv6EHM} model evaluated at the orbit-averaged starting frequency of waveform generation $\langle f_{\text{start}} \rangle$ ($10\,$Hz for BBH events, $20\,$Hz for BNS and NSBH events, $5.5\,$Hz for GW190521).
    The column $\log_{10}\mathcal{B}^{\text{EAS}}_{\text{QCAS}}$ reports the Savage--Dickey density ratio Bayes factor comparing the eccentric-and-spin-aligned hypothesis (EAS) against the quasi-circular-and-spin-aligned hypothesis (QCAS); its uncertainty is estimated via bootstrap resampling averaged over multiple kernel-density-estimation bandwidths.
    The column $\log_{10}\mathcal{B}^{\text{EAS}}_{\text{QCP}}$ reports the Bayes factor of \texttt{SEOBNRv6EHM} over \texttt{SEOBNRv5PHM}, comparing EAS against the quasi-circular precessing hypothesis (QCP), for the six most significant events. The Bayes factors and their uncertainties are obtained directly from the nested-sampling evidence estimates. Cell shading indicates the magnitude and sign of each Bayes factor, with green (red) denoting support for (against) the eccentric hypothesis.
}
\label{tab:PE_results_1}
\end{table*}

The results for all \NEVENTS events are reported in Table~\ref{tab:PE_results_1}, which lists the median and $90\%$ credible intervals for the inferred parameters, together with the Bayes factor $\mathcal{B}^{\text{EAS}}_{\text{QCAS}}$ obtained from the Savage--Dickey density ratio, and the Bayes factor $\mathcal{B}^{\text{EAS}}_{\text{QCP}}$ for the six most significant events.
The eccentricity $e$ is the Keplerian eccentricity from the \texttt{SEOBNRv6EHM} model, evaluated at the orbit-averaged starting frequency of waveform generation $\langle f_{\text{start}} \rangle$. This is set to $10\,$Hz for BBH events, $20\,$Hz for BNS and NSBH events (specifically GW170817, GW190814, GW191219\_163120, and GW200105\_162426), and $5.5\,$Hz for GW190521.

\begin{figure*}
    \centering
    \includegraphics[width=0.8\textwidth]{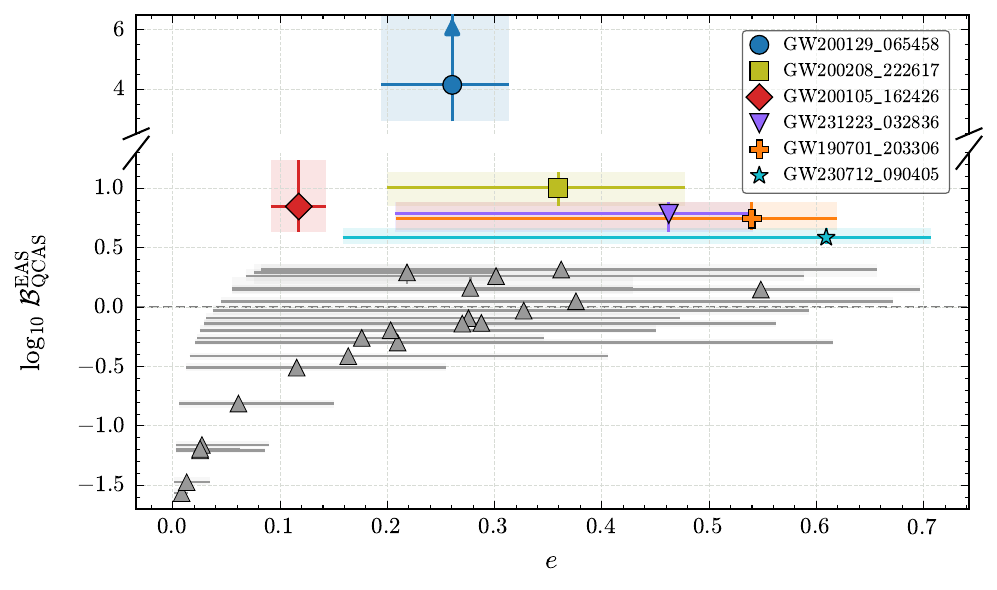}
    \caption{
	$\log_{10}$ Bayes factor favoring the eccentric aligned-spin (EAS) hypothesis over the quasi-circular aligned-spin (QCAS) hypothesis as a function of the median posterior eccentricity $e$ at the orbit-averaged starting frequency $\langle f_{\text{start}} \rangle$ for the \NEVENTS gravitational-wave events analyzed with \texttt{SEOBNRv6EHM}.
	Crosshairs indicate the $90\%$ credible intervals on the eccentricity (horizontal) and the Bayes factor (vertical).
	The Bayes factor uncertainty is estimated via bootstrap resampling of the Savage--Dickey density ratio, averaged over multiple kernel-density-estimation bandwidths.
	Events with $\log_{10}\mathcal{B}^{\rm EAS}_{\rm QCAS} > 0.5$ are shown with distinct markers and listed in the legend in descending order of Bayes factor; the remaining events are shown as gray triangles.
	Note the broken $y$-axis, introduced to accommodate the large Bayes factor of GW200129\_065458; for this event, the upward arrow indicates that the upper bound of the credible interval extends beyond the plotted range due to the near-absence of posterior samples close to $e = 0$.}
    \label{fig:bf_vs_ecc}
\end{figure*}

\begin{figure*}
    \centering
    \includegraphics[width=0.8\textwidth]{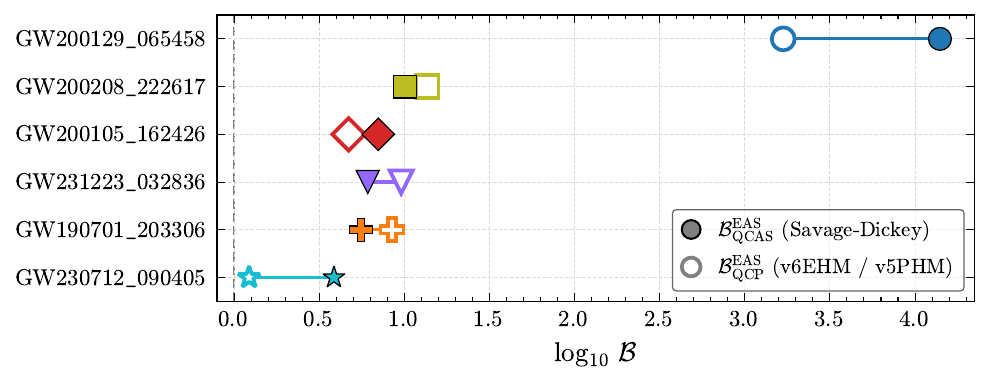}
    \caption{
	Comparison of the $\log_{10}$ Bayes factor favoring the eccentric aligned-spin (EAS) hypothesis over the quasi-circular aligned-spin (QCAS) hypothesis (filled markers) against the Bayes factor favoring EAS over the quasi-circular precessing-spin (QCP) hypothesis (open markers), for the six events in Table~\ref{tab:PE_results_1} with $\log_{10}\mathcal{B}^{\rm EAS}_{\rm QCAS} > 0.5$.
    }
    \label{fig:bf_EAS_QCAS_vs_QCP}
\end{figure*}

The eccentricity measurements and Bayes factors $\mathcal{B}^{\text{EAS}}_{\text{QCAS}}$ are summarized in Fig.~\ref{fig:bf_vs_ecc}.
Most events are consistent with QC orbits, with eccentricity posteriors compatible with zero, and uninformative or negative Bayes factors.
Six events stand out showing mild to moderate support for eccentricity ($\log_{10}\mathcal{B}^{\text{EAS}}_{\text{QCAS}} > 0.5$); in descending order of Bayes factor, these are GW200129\_065458, GW200208\_222617, GW200105\_162426, GW231223\_032836, GW190701\_203306, and GW230712\_090405.
For these six events, we perform additional runs with the QCP model \texttt{SEOBNRv5PHM}~\cite{Ramos-Buades:2023ehm, Estelles:2025zah} and report the Bayes factor $\mathcal{B}^{\text{EAS}}_{\text{QCP}}$ (Fig.~\ref{fig:bf_EAS_QCAS_vs_QCP}).
For GW200129\_065458, GW200208\_222617, and GW200105\_162426, whose eccentricity posteriors are bounded away from the high-eccentricity regime, we additionally compute the waveform-based eccentricity $e_{\rm gw}$~\cite{Shaikh:2023ypz} at the same $\langle f_{\text{start}} \rangle$, finding that its median and $90\%$ credible interval agree with those of $e$ in Table~\ref{tab:PE_results_1} to within $0.01$.

We now discuss the six most significant events individually:
\begin{itemize}
    \item {GW200129\_065458.}
    This high-SNR ($\rho_{\mathrm{mf}}^{\text{N}} \approx 26$) signal from a BBH with source-frame total mass $M_{\rm src} \approx 61\,\solarmass$ shows the highest support for eccentricity in our sample, with $\log_{10}\mathcal{B}^{\text{EAS}}_{\text{QCAS}} \approx 4.2$ and $\log_{10}\mathcal{B}^{\text{EAS}}_{\text{QCP}} \approx 3.2$.
    We recover it as the most significant eccentric candidate, consistent with the findings of Refs.~\cite{Gupte:2024jfe,Planas:2025jny}.
    The Bayes factor decreases noticeably when comparing the EAS hypothesis against the QCP hypothesis rather than the QCAS one, as expected given the evidence for spin precession reported for this event~\cite{Hannam:2021pit}.
    However, as with the evidence for spin precession~\cite{Payne:2022spz}, the inferred eccentricity is significantly affected by the glitch-subtraction method applied to the LIGO Livingston data~\cite{Gupte:2024jfe}.
    We do not re-analyze this event with alternative glitch-subtraction methods, as this has been thoroughly investigated in previous work, but we note that the eccentricity estimate should be interpreted with this caveat in mind.
    \item {GW200208\_222617.}
    With $\rho_{\mathrm{mf}}^{\text{N}} \approx 8$ and ${\sim}\,14$ orbital cycles in band, this event has been identified as the least ambiguous eccentric BBH candidate in GWTC-3~\cite{Romero-Shaw:2025vbc}: it is unaffected by data-quality issues, and its moderate mass ($M_{\rm src} \approx 42\,\solarmass$) makes the eccentricity--precession degeneracy less severe.
    We find $\log_{10}\mathcal{B}^{\text{EAS}}_{\text{QCAS}} \approx 1.0$ and $\log_{10}\mathcal{B}^{\text{EAS}}_{\text{QCP}} \approx 1.13$, consistent with the eccentricity measurements reported in Refs.~\cite{Romero-Shaw:2022xko, Gupte:2024jfe, Planas:2025jny}.
    \item {GW200105\_162426.}
    This NSBH event ($M_{\rm src} \approx 10\,\solarmass$, $\rho_{\mathrm{mf}}^{\text{N}} \approx 14$) shows modest support for eccentricity with $\log_{10}\mathcal{B}^{\text{EAS}}_{\text{QCAS}} \approx 0.85$ and $\log_{10}\mathcal{B}^{\text{EAS}}_{\text{QCP}} \approx 0.67$, broadly consistent with the findings of Refs.~\cite{Morras:2025xfu, Planas:2025plq, Kacanja:2025kpr, Jan:2025fps}.
    The event was originally identified with moderate significance by matched-filter searches that employ non-eccentric templates; its significance increases slightly in a targeted eccentric search~\cite{Phukon:2025cky}.
    \item {GW231223\_032836.}
    This BBH signal ($M_{\rm src} \approx 77\,\solarmass$, $\rho_{\mathrm{mf}}^{\text{N}} \approx 10$) was found with modest significance by both Refs.~\cite{Xu:2025ajj, Gupte:2026whi}, and our results are broadly consistent with these studies, with $\log_{10}\mathcal{B}^{\text{EAS}}_{\text{QCAS}} \approx 0.79$ and $\log_{10}\mathcal{B}^{\text{EAS}}_{\text{QCP}} \approx 0.98$, although with slightly higher significance compared to previous analyses ($\log_{10}\mathcal{B}^{\text{EAS}}_{\text{QCP}} \approx 0.5$ in both studies~\cite{Xu:2025ajj, Gupte:2026whi}).
    \item {GW190701\_203306.}
    This high-mass BBH signal ($M_{\rm src} \approx 96\,\solarmass$, $\rho_{\mathrm{mf}}^{\text{N}} \approx 11$) shows modest support for eccentricity when comparing both against the QCAS and QCP hypotheses, with $\log_{10}\mathcal{B}^{\text{EAS}}_{\text{QCAS}} \approx 0.8$ and $\log_{10}\mathcal{B}^{\text{EAS}}_{\text{QCP}} \approx 0.9$.
    The evidence for eccentricity is reduced compared to the analysis of Ref.~\cite{Gupte:2024jfe} using the \texttt{SEOBNRv4EHM} model, which found $\log_{10}\mathcal{B}^{\text{EAS}}_{\text{QCP}} \approx 1.8$, but is still higher than the \texttt{IMRPhenomTEHM} results of Ref.~\cite{Planas:2025plq} ($\log_{10}\mathcal{B}^{\text{EAS}}_{\text{QCP}} \lesssim 0.2$). All three analyses use the default glitch-subtracted data frames provided by the LVK collaboration. As with GW200129\_065458, this event is also affected by a glitch in the LIGO Livingston detector, and Ref.~\cite{Gupte:2026whi} has shown that the inferred eccentricity is sensitive to the glitch-subtraction method.
    \item {GW230712\_090405.}
    For this BBH signal ($M_{\rm src} \approx 55\,\solarmass$, $\rho_{\mathrm{mf}}^{\text{N}} \approx 9$) we find $\log_{10}\mathcal{B}^{\text{EAS}}_{\text{QCAS}} \approx 0.59$; however, when comparing against the QCP hypothesis, the Bayes factor is significantly reduced, with $\log_{10}\mathcal{B}^{\text{EAS}}_{\text{QCP}} \approx 0.09$, and the evidence for eccentricity is no longer significant.
    This is consistent with the findings of Ref.~\cite{Xu:2025ajj} and can be understood as a consequence of the degeneracy between eccentricity and spin precession~\cite{Romero-shaw:2022fbf, CalderonBustillo:2020xms}. Indeed, GW230712\_090405 is among the most informative events in GWTC-4.0 for the precessing-spin parameter $\chi_{\rm p}$~\cite{LIGOScientific:2025slb}.
\end{itemize}

Other candidates with weaker but still positive support include the high-mass BBH events GW191109\_010717 and GW190620\_030421, both with $\log_{10}\mathcal{B}^{\text{EAS}}_{\text{QCAS}} \approx 0.3$, that also showed some support for eccentricity in previous O3 analyses~\cite{Romero-Shaw:2022xko, Gupte:2024jfe}.
For both systems, the eccentricity posterior exhibits a multimodal structure: one mode lies at low eccentricity ($e_{10\text{Hz}} \approx 0.3$), with some support extending to zero eccentricity, while a second mode peaks at higher eccentricity ($e_{10\text{Hz}} \approx 0.6$). The interpretation of GW191109\_010717 is further complicated by known data-quality issues~\cite{Udall:2024ovp}.
A similar case is the low-mass BBH event GW231224\_024321 ($M_{\rm src} \approx 16.5\,\solarmass$) from O4a, with $\log_{10}\mathcal{B}^{\text{EAS}}_{\text{QCAS}} \approx 0.29$, for which comparable support for eccentricity has been reported in Refs.~\cite{Xu:2025ajj, Gupte:2026whi}. Its posterior likewise shows mild multimodality, with modes located closer together, around $e_{10\text{Hz}} \approx 0.2 - 0.3$, consistent with other analyses.

Several other events that were reported with mild support for eccentricity in previous studies using other waveform models~\cite{Romero-Shaw:2022xko, Gupte:2024jfe, Gupte:2026whi, Planas:2025jny, Xu:2025ajj} do not show comparable significance in our analysis when comparing the EAS and QCAS hypotheses. 
Among these are the high-mass events GW190521, for which we find $\log_{10}\mathcal{B}^{\text{EAS}}_{\text{QCAS}}$ compatible with zero, in agreement with Refs.~\cite{Iglesias:2022xfc, Ramos-Buades:2023yhy, Gamboa:2024hli}, and GW231123\_135430, for which we obtain a slightly negative $\log_{10}\mathcal{B}^{\text{EAS}}_{\text{QCAS}}$, in agreement with \texttt{SEOBNRv5EHM}~\cite{Gupte:2026whi} but in contrast to the support reported with \texttt{IMRPhenomTEHM}~\cite{Xu:2025ajj} and \texttt{TEOBResumS-Dal\'i}~\cite{Jan:2025zcm, Malagon:2026uev}.
This suggests that the eccentricity estimates for these events are sensitive to waveform modeling differences, and underscores the importance of analyzing the events with more accurate waveform models. We also note that, even when comparing analyses employing a uniform prior in eccentricity, Bayes factors depend on the chosen upper bound of the prior, which varies across studies. As a result, comparisons with previous work should be interpreted with this caveat in mind. In this work, we adopt an eccentricity prior extending up to $e = 0.8$, which is typically broader than in other analyses, and may also contribute to the slightly reduced significance observed for some events.

At the opposite end of the sample, the most negative Bayes factors and tightest eccentricity constraints are obtained for the BNS signal GW170817 and for GW250114\_082203, the loudest BBH signal observed to date, both with $\log_{10}\mathcal{B}^{\text{EAS}}_{\text{QCAS}} \approx -1.5$. Our result for GW250114\_082203 is consistent with the eccentric aligned-spin analysis of Ref.~\cite{LIGOScientific:2025rid} and with the eccentric, spin-precessing analysis with \texttt{TEOBResumS-Dal\'i} of Ref.~\cite{Chandra:2025jfc}, both of which also disfavor eccentricity.

\begin{figure*}
    \centering
    \includegraphics[width=0.85\textwidth]{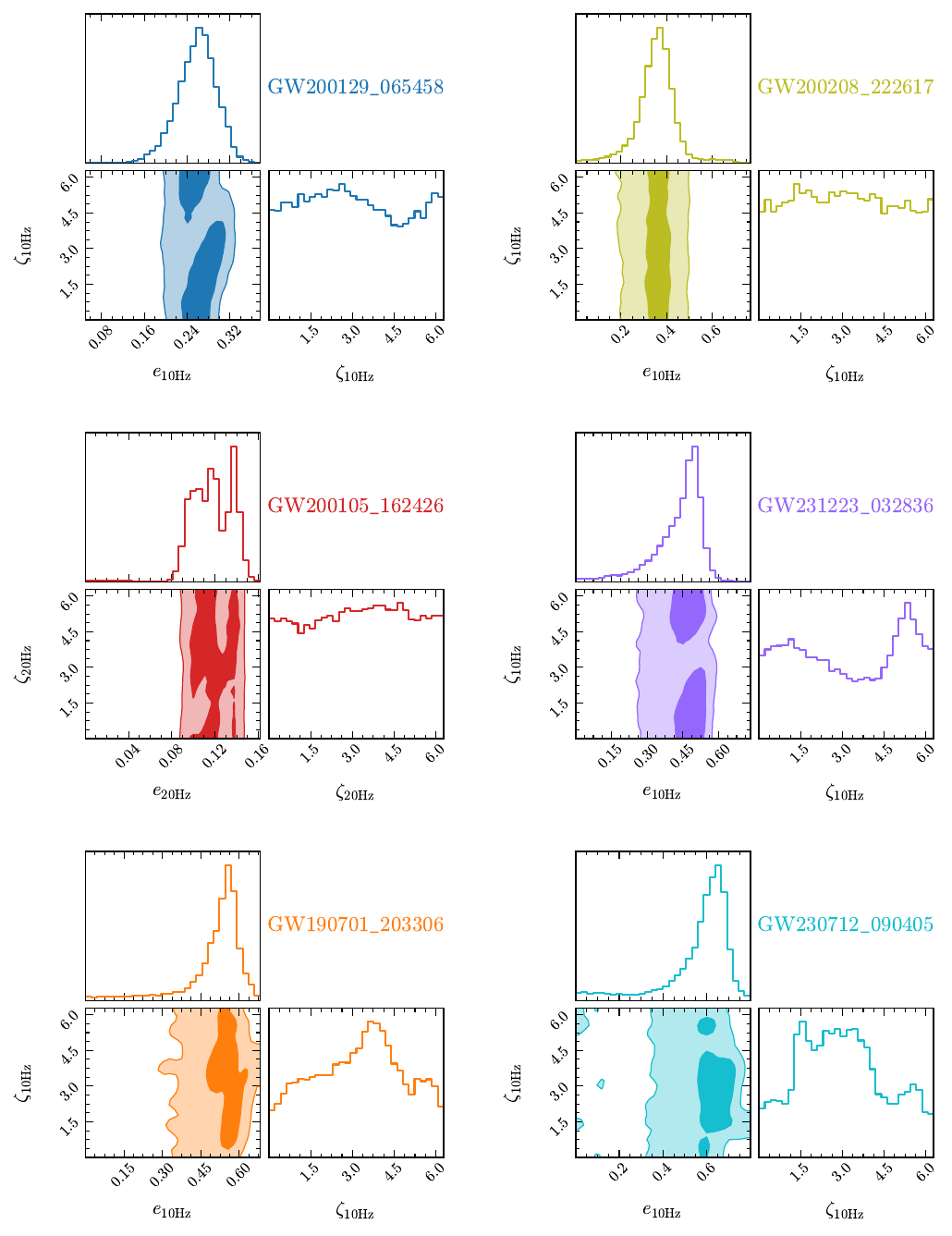}
    \caption{
	Posterior distributions of eccentricity $e$ and relativistic anomaly $\zeta$ at the orbit-averaged starting frequency of waveform generation ($\langle f_{\text{start}} \rangle = 20\,$Hz for GW200105\_162426 and $10\,$Hz for all other events), for the six events with the highest Bayes factor favoring eccentricity.
	Contours show the $50\%$ and $90\%$ credible regions.
	The eccentricity axis starts at $e = 0$ for all events except GW200129\_065458, whose posterior excludes the quasi-circular limit; this absence of posterior support near $e = 0$ underlies the substantially higher Bayes factor recovered for this event.
	Unlike Ref.~\cite{Gupte:2026whi}, we do not overlay the effective prior boundary in the eccentricity--anomaly plane: for \texttt{SEOBNRv6EHM}, waveform generation failures are exceedingly rare (affecting at most $0.6\%$ of the prior volume for a uniform prior up to $e = 0.8$) and show no correlation with high eccentricity or anomaly near periastron, in contrast to the systematic failures reported for \texttt{SEOBNRv5EHM}~\cite{Gupte:2026whi}.
    }
    \label{fig:corner_top_BF}
\end{figure*}

Figure~\ref{fig:corner_top_BF} shows the posterior distributions of eccentricity $e$ and relativistic anomaly $\zeta$ at the orbit-averaged starting frequency of waveform generation for the six events with the highest $\log_{10}\mathcal{B}^{\text{EAS}}_{\text{QCAS}}$.
Since the prior is uniform in both $e$ and $\zeta$ at this frequency, the figure directly shows where the likelihood support for eccentricity is concentrated, without the additional conversion to $e_{\rm gw}$.
Unlike Ref.~\cite{Gupte:2026whi}, we do not overlay the effective prior boundary in the eccentricity--anomaly plane: for \texttt{SEOBNRv6EHM}, waveform generation failures are exceedingly rare, affecting at most $0.6\%$ of the prior volume even with a uniform eccentricity prior up to $e = 0.8$ for the events considered here. The few failures that do occur are concentrated at high total mass and extreme mass ratios, well outside the posterior support of the events analyzed, and show no correlation with high eccentricity or anomaly near periastron---in contrast to the systematic failures reported for \texttt{SEOBNRv5EHM}~\cite{Gupte:2026whi}.

\section{Analysis of high-mass events under an unbound-orbit hypothesis}
\label{sec:pe_unbound}

As discussed in Sec.~\ref{sec:model}, the \texttt{SEOBNRv6EHM} model is valid not only for bound eccentric orbits, but also for generic equatorial orbits, including initially unbound trajectories.
In this section, we exploit this capability to reanalyze a selection of five high-mass events under an unbound-orbit hypothesis: GW231123\_135430, GW190521\_030229, GW191109\_010717, GW190620\_030421, and GW231221\_135041.
For all of these events, the bound-orbit analyses of Sec.~\ref{sec:pe_ecc_events} infer large total masses spanning $M_{\rm src} \approx 70$--$200\,M_\odot$, and broad eccentricity posteriors with substantial support at high values, in some cases with a multimodal structure.
This combination of high mass---hence a small number of cycles in band---and posterior support for high eccentricities suggests that an unbound-orbit configuration without an extended inspiral could provide a comparably good description of the data. For GW190521, this extends the analysis of Ref.~\cite{Gamba:2021gap}, which used the \texttt{TEOBResumS-Dal\'i} model, by including spin effects and higher-order modes.
While this work was being completed, we became aware of a concurrent and independent analysis of Ref.~\cite{Lange:2026eqx}, which uses an updated version of the \texttt{TEOBResumS-Dal\'i} model to study a set of high-mass events under an unbound-orbit hypothesis.

For bound orbits, the binary dynamics is specified by the eccentricity $e$ and relativistic anomaly $\zeta$ at a reference frequency.
For generic orbits, the initial conditions are instead specified by the dimensionless energy $E_0=\mathcal{E}_0/M$ and orbital angular momentum $p_\phi^0  = P_\phi^0/(\mu M)$ at an initial separation $r_0$~\cite{Gamboa:2026jht}. Values $E_0 > 1$ describe initially unbound configurations; depending on the values of $E_0$ and $p_\phi^0$, the resulting orbit falls into one of three qualitatively different morphologies:
\begin{itemize}[leftmargin=*]
\item \emph{Direct capture (plunge):} the binary plunges directly into merger without completing an orbit. The GW signal is a short burst consisting of a plunge and ringdown, with no inspiral.
\item \emph{Dynamical capture:} the first close encounter radiates enough energy to bind the system, which then completes at least one further orbit before merging. Depending on $(E_0, p_\phi^0)$, the post-capture evolution can range from a single periastron passage, to many cycles of an eccentric bound inspiral, in all cases terminating in a standard merger-ringdown.
\item \emph{Scattering:} the binary approaches, interacts gravitationally, but the components separate again without merging. The GW signal is a transient burst with a decaying tail, and no ringdown.
\end{itemize}
These morphologies are continuously connected: the boundaries between direct captures, dynamical captures, and scattering encounters in the $(E_0, p_\phi^0)$ plane depend on the mass ratio and spins~\cite{East:2012xq, Gold:2012tk, Nagar:2020xsk, Kankani:2024may, Albanesi:2024xus}.

A practical subtlety arises when computing the likelihood for unbound waveforms~\cite{Henshaw:2025arb}.
For bound orbits that merge, the ringdown naturally damps the signal to zero, and only the start of the waveform needs to be tapered to suppress spectral leakage in the Fourier transform.
For scattering encounters, however, the binary separates after the close approach, and the GW signal has a slowly decaying tail that does not reach zero due to the GW memory effect.\footnote{
Specifically, scattering encounters display the \emph{linear} memory effect (e.g., \cite{1974SvA....18...17Z, DeVittori:2014psa}), arising from a net change in the time derivatives of the source multipole moments between early and late times.
This effect is included in \texttt{SEOBNRv6EHM} and can be observed, for example, in the right panels of Fig.~15 in Ref.~\cite{Gamboa:2026jht}.
}
If left untapered, this abrupt truncation introduces spectral leakage that contaminates the frequency-domain representation of the signal. To address this, we apply a Planck-taper window to the time-domain waveform, starting from the last peak of the frame-invariant amplitude, for all generic-orbit configurations; for direct captures and dynamical captures that end in merger the taper has negligible effect, but it allows for accurate Fourier transforms for scattering encounters.
The Planck-taper window~\cite{Mckechan:2010kp} is unity at its starting point, with all its derivatives vanishing there, so it departs appreciably from unity only well after the amplitude peak, by which time the ringdown has decayed by orders of magnitude; the ringdown is therefore not suppressed in practice.

We compare the unbound-orbit aligned-spin (hereafter ``unbound'') analyses against three alternative hypotheses: the EAS configuration of \texttt{SEOBNRv6EHM}, and the QCP models \texttt{SEOBNRv5PHM}~\cite{Ramos-Buades:2023ehm, Estelles:2025zah} and \texttt{NRSur7dq4}~\cite{Varma:2019csw}. We use the same data and sampler settings as for the bound-orbit analyses in Sec.~\ref{sec:pe_events}. For the generic-orbit parameters, we adopt uniform priors on the energy $E_0 \in [1.0002, 1.3]$ and angular momentum $p_\phi^0 \in [2.0, 10.0]$ at an initial separation $r_0 = 500\,M$. For GW191109\_010717, we extend the energy prior to $E_0 \in [1.0002, 1.6]$ to avoid railing. For GW190521, this is a broader prior than that used in Ref.~\cite{Gamba:2021gap}, which restricted the analysis to nonspinning binaries with a narrower range of initial conditions. All four models are compared assuming a mass-ratio prior restricted to $1/q \in [1/6, 1.0]$ where \texttt{NRSur7dq4} is applicable, ensuring that the resulting Bayes factors are directly comparable; we have verified that this restriction does not lead to significant prior railing except for GW231123\_135430, for which extend the mass-ratio prior to $1/q \in [1/20, 1.0]$. To keep the resulting Bayes factors consistent with the restricted $1/q \in [1/6, 1.0]$ prior adopted for \texttt{NRSur7dq4}, we correct the \texttt{NRSur7dq4} evidence for the difference in prior volume, assuming negligible \texttt{NRSur7dq4} likelihood support above $1/q = 1/6$ where the model is no longer applicable; this amounts to a correction of $\sim -0.26$, given by the $\log_{10}$ ratio of the two prior volumes for a prior uniform in the component masses, that does not affect any of the conclusions below.
For GW191109\_010717 and GW231123\_135430, whose unbound posteriors are multimodal (see below), we repeat the analyses with an increased number of live points $\texttt{nlive} = 2000$ to ensure that all modes are robustly resolved. For GW191109\_010717 the parallel chains are consistent and the recovered posteriors and Bayes factors are essentially unchanged with respect to the default settings. For GW231123\_135430, instead, we further increase the number of live points to $\texttt{nlive} = 4000$, since the inclusion of the additional generic-orbit parameters $(E_0, p_\phi^0)$ leads to a complex multimodal posterior structure that is particularly challenging for the sampler to resolve.

\begin{figure*}
    \centering
    \includegraphics[width=0.8 \textwidth]{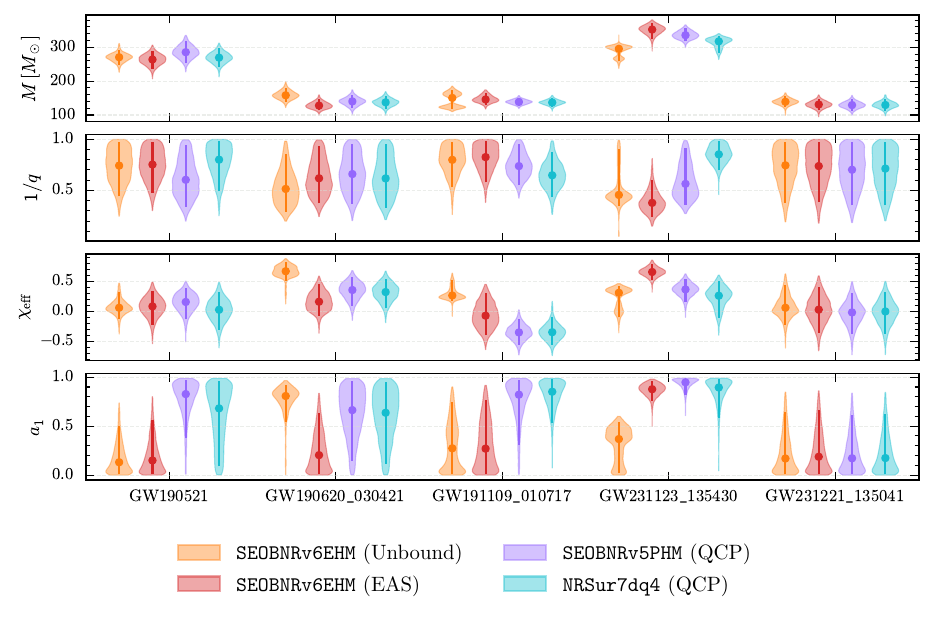}
    \caption{
    Marginal posteriors on the detector-frame total mass $M$, inverse mass ratio $1/q$, effective spin $\chi_\mathrm{eff}$, and primary spin magnitude $a_1$ under the unbound, bound eccentric aligned-spin (EAS), and quasi-circular precessing (QCP) hypotheses for the five high-mass events. For the QCP hypothesis we show results from both \texttt{SEOBNRv5PHM} (with mode asymmetries) and \texttt{NRSur7dq4}. For GW231221\_135041 the inferred parameters are broadly consistent across all four hypotheses; for GW190521 the masses and $\chi_\mathrm{eff}$ are also consistent, and only $a_1$ differs between the aligned-spin and QCP analyses; for GW190620\_030421 and GW191109\_010717 the total mass and mass ratio are also consistent but the spin parameters vary across hypotheses; for GW231123\_135430 all parameters disagree, both between the unbound, EAS, and QCP analyses and between the two QCP models.
    }
    \label{fig:param_shift_unbound_more_events}
\end{figure*}

Figure~\ref{fig:param_shift_unbound_more_events} compares the marginal posteriors on the detector-frame total mass $M$, inverse mass ratio $1/q$, effective spin $\chi_\mathrm{eff}$, and primary spin magnitude $a_1$ under the three hypotheses.

For GW190521 and GW231221\_135041, the masses and the effective spin are broadly consistent across all four hypotheses, with $\chi_{\rm eff}$ compatible with zero in every case.
The primary spin magnitude $a_1$ of GW231221\_135041 is also consistent across hypotheses, with a preference for low values.
For GW190521, instead, the aligned-spin analyses recover small $a_1$, while the QCP analyses favor larger values with significant support for spin precession.

For GW190620\_030421 and GW191109\_010717, the recovered total mass and mass ratio are also broadly consistent, but the inferred spins vary across hypotheses: for GW190620\_030421 the unbound analysis prefers a high effective spin ($\chi_{\rm eff} \approx 0.67$) while the EAS and QCP analyses recover lower values ($\chi_{\rm eff} \approx 0.16$--$0.35$); for GW191109\_010717 the four hypotheses span a wide range, from $\chi_{\rm eff} \approx 0.26$ (unbound) to $\approx -0.35$ (both QCP models). 
The primary spin magnitude $a_1$ of GW190620\_030421 tracks $\chi_{\rm eff}$, with the unbound analysis recovering a high primary spin ($a_1 \approx 0.8$), the EAS analysis a low one ($\approx 0.2$), and the QCP analyses in between. For GW191109\_010717, the two QCP analyses instead have $a_1$ railing towards high values ($\approx 0.8$), driven by a large precessing primary spin, while the aligned-spin analyses prefer lower values.

The most striking shifts are for GW231123\_135430: the EAS analysis recovers both a high effective spin ($\chi_{\rm eff} \approx 0.65$) and a high total mass ($M \approx 350\,M_\odot$), in agreement with a QCAS analysis of the same event using the \texttt{SEOBNRv5\_ROM} model~\cite{Pompili:2023tna}. The unbound analysis instead prefers smaller spins and a substantially lower mass ($M \approx 300\,M_\odot$); the two QCP models additionally disagree on the mass ratio ($1/q \approx 0.56$ for \texttt{SEOBNRv5PHM} while $\approx 0.85$ for \texttt{NRSur7dq4}).\footnote{The \texttt{SEOBNRv5PHM} version used here includes multipole asymmetries in the co-precessing frame~\cite{Estelles:2025zah}, while the LVK analysis of GW231123\_135430~\cite{LIGOScientific:2025rsn} used a version without them. In that case \texttt{SEOBNRv5PHM} recovered more comparable masses, closer to \texttt{NRSur7dq4}. The shift in inferred parameters between the two versions of \texttt{SEOBNRv5PHM} is consistent with mode asymmetries being most relevant for strongly precessing binaries, such as the source of GW231123\_135430. Among the spin-precessing models considered in the LVK analysis, \texttt{SEOBNRv5PHM} with asymmetries gives the highest maximum likelihood and Bayes factor, with $\log_{10}\mathcal{B} \approx 2.9$ relative to \texttt{NRSur7dq4}, $\approx 1.5$ relative to \texttt{SEOBNRv5PHM} without asymmetries, and $\approx 0.6$ relative to \texttt{IMRPhenomXO4a}~\cite{Thompson:2023ase}; we stress again, however, that a higher Bayes factor does not guarantee more accurate parameter recovery~\cite{Hoy:2022tst, Bini:2026kwz}.} The unbound analysis is also the only one to recover a moderate primary spin ($a_1 \approx 0.4$), with the EAS and both QCP analyses railing against the prior boundary at $a_1 \approx 0.9$. This result should be interpreted in light of the substantial waveform-modeling systematics reported for this extremely short signal: the LVK analysis~\cite{LIGOScientific:2025rsn} highlights the sensitivity of the inferred source parameters and Bayesian evidence to the assumed waveform model, and eccentric analyses with different waveform models also yield significantly different parameter estimates and Bayes factors~\cite{Xu:2025ajj, Jan:2025zcm, Malagon:2026uev, Gupte:2026whi}.

For both GW191109\_010717 and GW231123\_135430, the unbound posteriors show bimodalities, in the total mass for GW191109\_010717, and additionally in the effective spin for GW231123\_135430. In both cases the modes are correlated with distinct regions of the initial-condition posterior for $(E_0,\,p_\phi^0)$, indicating genuine degeneracies where different combinations of initial conditions and intrinsic parameters fit the data comparably well.

Across all five events, the posterior on the initial conditions $(E_0,\,p_\phi^0)$ concentrates in the direct-capture region, with negligible support for dynamical-capture or scattering morphologies; the lack of support for scattering is expected given the clear ringdown signature in the observed signals.

\begin{figure*}
    \centering
    \includegraphics[width=0.8\textwidth]{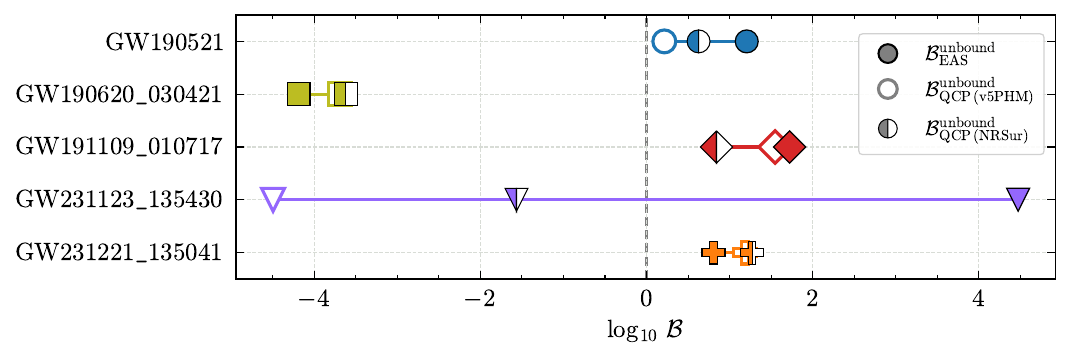}
    \caption{
    Log Bayes factors $\log_{10}\mathcal{B}^{\rm unbound}_{\rm EAS}$ (filled markers) and $\log_{10}\mathcal{B}^{\rm unbound}_{\rm QCP}$ comparing the unbound hypothesis against the bound eccentric aligned-spin (EAS) and quasi-circular precessing-spin (QCP) hypotheses, respectively, for the five high-mass events analyzed under unbound initial conditions. The QCP markers are shown for two precessing waveform models: \texttt{SEOBNRv5PHM} (open markers) and \texttt{NRSur7dq4} (half-filled markers). The unbound model is preferred over EAS for all events except GW190620\_030421. For GW190521, GW191109\_010717, and GW231221\_135041, the unbound configuration is also comparable to or marginally preferred over both QCP models, the two hypotheses being comparable for GW190521; for GW231123\_135430 and GW190620\_030421 the QCP hypothesis is preferred.
    }
    \label{fig:bf_unbound_more_events}
\end{figure*}

{\renewcommand{\arraystretch}{1.5}
\begin{table}
\centering
\small
\input{tab/bf_unbound.tex}
\caption{
    Log Bayes factors $\log_{10}\mathcal{B}^{\rm unbound}_{\rm EAS}$, $\log_{10}\mathcal{B}^{\rm unbound}_{\rm v5PHM}$, and $\log_{10}\mathcal{B}^{\rm unbound}_{\rm NRSur}$ comparing the unbound hypothesis against the bound eccentric aligned-spin (EAS) hypothesis and the quasi-circular precessing-spin (QCP) hypothesis using \texttt{SEOBNRv5PHM} and \texttt{NRSur7dq4}, for the five high-mass events analyzed under unbound initial conditions. Uncertainties are estimated from the nested-sampling evidence error.
}
\label{tab:bf_unbound}
\end{table}}

Figure~\ref{fig:bf_unbound_more_events} and Table~\ref{tab:bf_unbound} summarize the log Bayes factors $\log_{10}\mathcal{B}^{\rm unbound}_{\rm EAS}$ and $\log_{10}\mathcal{B}^{\rm unbound}_{\rm QCP}$ for the five events. The unbound model is preferred over EAS for four of the five: positive evidence is found for GW231123\_135430 ($\log_{10}\mathcal{B}^{\rm unbound}_{\rm EAS} \approx 4.5$), GW191109\_010717 ($\approx 1.7$), GW190521\_030229 ($\approx 1.2$), and GW231221\_135041 ($\approx 0.8$), while GW190620\_030421 strongly prefers the EAS hypothesis ($\approx -4.2$). Since both the unbound and EAS configurations assume an aligned-spin binary, this preference could in principle reflect the unbound waveforms absorbing modulations actually produced by spin precession, rather than a genuine preference for an unbound interpretation. To assess this, we also compare against the QCP hypothesis using two precessing waveform models, \texttt{SEOBNRv5PHM} and \texttt{NRSur7dq4} (open and half-filled markers in Fig.~\ref{fig:bf_unbound_more_events}, respectively). For three of the five events---GW190521 ($\log_{10}\mathcal{B}^{\rm unbound}_{\rm QCP} \approx 0.2$ against \texttt{SEOBNRv5PHM}, $0.6$ against \texttt{NRSur7dq4}), GW191109\_010717 ($\approx 1.6$ and $0.8$), and GW231221\_135041 ($\approx 1.2$ and $1.3$)---the unbound configuration fits the data comparably well to both QCP models, and is marginally preferred for GW191109\_010717 and GW231221\_135041. For GW231123\_135430 the QCP hypothesis is strongly preferred ($\log_{10}\mathcal{B}^{\rm unbound}_{\rm QCP} \approx -4.5$ against \texttt{SEOBNRv5PHM}, $-1.6$ against \texttt{NRSur7dq4}); for GW190620\_030421 the unbound hypothesis is also strongly disfavored against both QCP models ($\log_{10}\mathcal{B}^{\rm unbound}_{\rm QCP} \approx -3.7$).

The preference for the unbound model over both the EAS and QCP hypotheses in three of these five events is a notable result: a no-inspiral, direct-capture configuration provides a comparable or better fit to the data than either a bound eccentric or a QCP configuration with an extended inspiral phase, consistent with the qualitative expectation that short signals from high-mass binaries dominated by the merger--ringdown admit multiple comparable interpretations of the same data~\cite{Romero-shaw:2022fbf, CalderonBustillo:2020xms}.

To test whether this preference could also be produced by a QCP source, in Appendix~\ref{app:unbound_injections} we perform zero-noise injection-recovery analyses of GW190521-like signals. A QCP signal generated with \texttt{NRSur7dq4} is correctly recovered with $\log_{10}\mathcal{B}^{\rm unbound}_{\rm NRSur} \approx -1.0$, but the unbound model recovers it as a direct capture near the separatrix, with an $(E_0, p_\phi^0)$ posterior similar to that of GW190521; conversely, a direct-capture signal generated with the unbound model yields $\log_{10}\mathcal{B}^{\rm unbound}_{\rm NRSur} \approx 1.2$. In both cases the injected hypothesis is correctly preferred: although the two interpretations are significantly degenerate for high-mass signals at the SNR of GW190521, the degeneracy is not complete, and the Bayes factor discriminates between them, albeit only at the level of order ten. At the same time, the direct-capture morphology recovered for the events is not, by itself, diagnostic of an unbound source.

\begin{figure*}
    \centering
    \includegraphics[width=0.95 \textwidth]{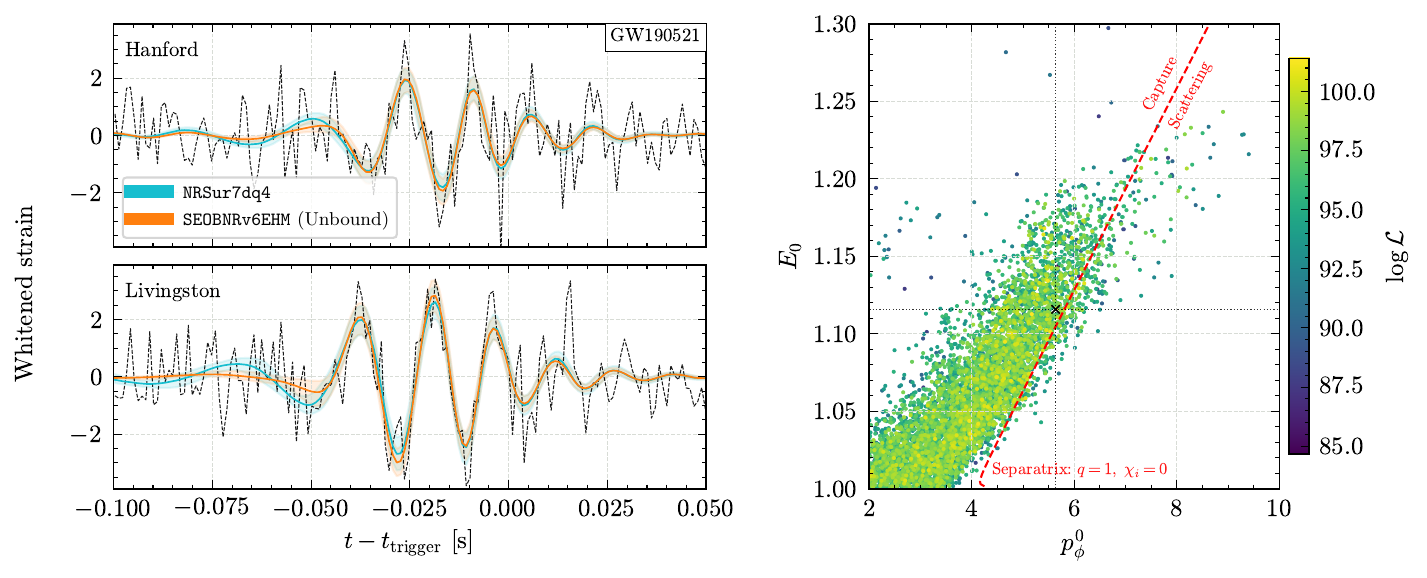}
    \includegraphics[width=0.95 \textwidth]{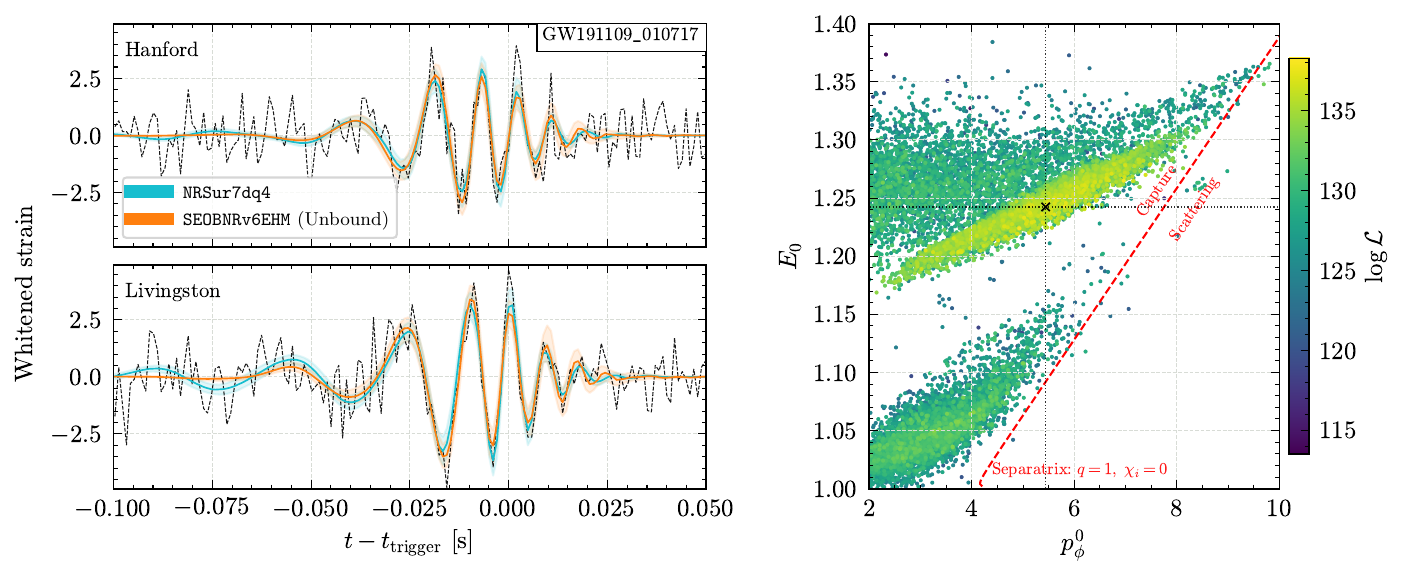}
    \includegraphics[width=0.95 \textwidth]{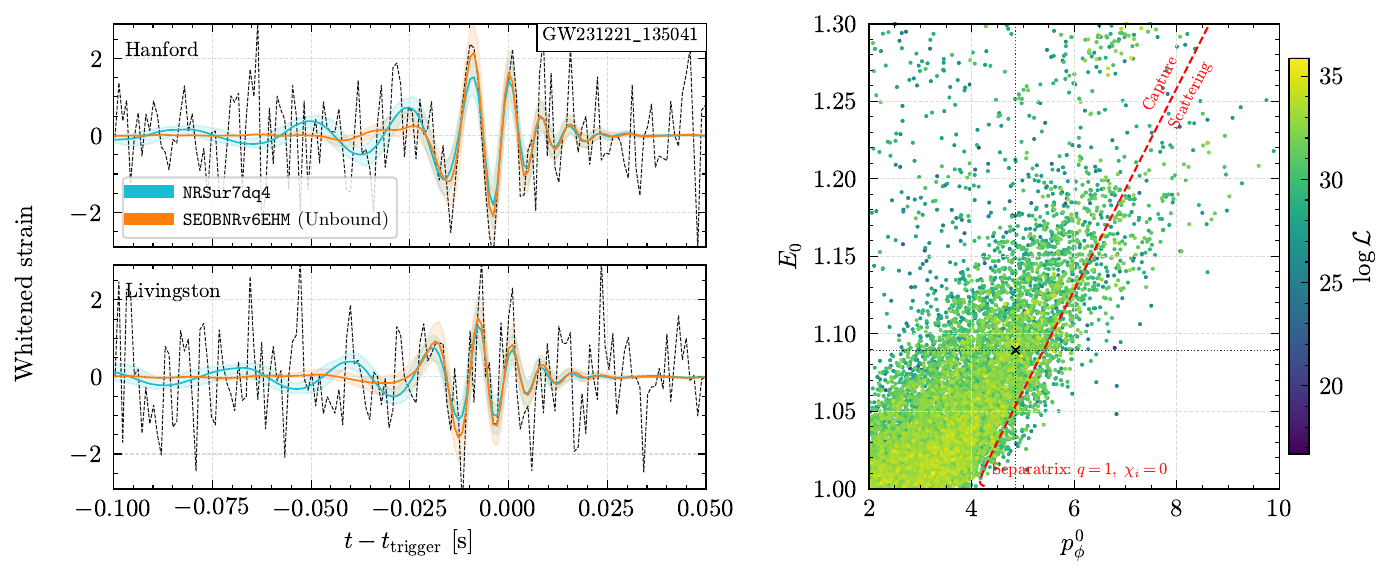}
    \caption{
    Analysis of GW190521, GW191109\_010717, and GW231221\_135041 under the unbound-orbit hypothesis with \texttt{SEOBNRv6EHM}. For each event: \emph{Left:} Whitened waveform reconstruction in the LIGO Hanford (top) and Livingston (bottom) detectors. The median (solid) and $90\%$ credible interval (shaded) of the reconstructed distribution are shown for the unbound \texttt{SEOBNRv6EHM} model and the quasi-circular precessing-spin model \texttt{NRSur7dq4}, compared to the detector data (dashed black). \emph{Right:} Posterior distribution of the initial energy $E_0$ and angular momentum $p_\phi^0$, colored by log-likelihood. The crosshairs mark the maximum-likelihood sample. The red dashed curve shows a fit for the equal-mass, nonspinning separatrix between capture and scattering orbits~\cite{Albanesi:2024xus}.
    }
    \label{fig:unbound_summary_three_events}
\end{figure*}

{\renewcommand{\arraystretch}{2.0}
\begin{table}
\centering
\small
\input{tab/snr_logl_unbound.tex}
\caption{
    Recovered network SNR $\rho_{\mathrm{mf}}^{\text{N}}$ and log-likelihood $\log\mathcal{L}$ (median and $90\%$ credible interval) for GW190521, GW191109\_010717, and GW231221\_135041 under the four model hypotheses: unbound-orbit aligned-spin (Unbound) and bound eccentric aligned-spin (EAS) configurations of \texttt{SEOBNRv6EHM}, and quasi-circular precessing-spin (QCP) analyses with \texttt{SEOBNRv5PHM} and \texttt{NRSur7dq4}.
}
\label{tab:unbound_results}
\end{table}
}

Figure~\ref{fig:unbound_summary_three_events} and Table~\ref{tab:unbound_results} present a more detailed view of the three events---GW190521, GW191109\_010717, and GW231221\_135041---where the unbound model is comparable to or marginally favored over both QCP alternatives. Additinal details on the analysis of GW231123\_135430 are given in Appendix~\ref{app:GW231123_unbound}.

For GW190521, the unbound hypothesis and the higher-likelihood QCP analysis (\texttt{SEOBNRv5PHM}) recover essentially the same median network SNR ($\rho_{\mathrm{mf}}^{\text{N}} \approx 14.2$) and log-likelihoods within $\Delta \log\mathcal{L} \approx 0.2$ of one another, with the bound EAS analysis lower by $\Delta \log\mathcal{L} \approx 4$. For GW191109\_010717 the unbound hypothesis is marginally favored over the higher-likelihood QCP analysis (\texttt{NRSur7dq4}), with comparable SNR ($\rho_{\mathrm{mf}}^{\text{N}} \approx 16.4$) but a $\Delta \log\mathcal{L} \approx 2$ advantage, while the bound EAS analysis is lower by $\Delta \log\mathcal{L} \approx 6$. For GW231221\_135041 the unbound hypothesis exceeds both QCP analyses---which yield almost identical SNR and $\log\mathcal{L}$---by $\Delta\rho_{\mathrm{mf}}^{\text{N}} \approx 0.6$ and $\Delta \log\mathcal{L} \approx 4.4$, with the EAS analysis lying in between (Table~\ref{tab:unbound_results}). 

The whitened waveform reconstructions (left panels of Fig.~\ref{fig:unbound_summary_three_events}) show that the unbound \texttt{SEOBNRv6EHM} and \texttt{NRSur7dq4} models track the data closely during the merger and ringdown; in the preceding cycle, the unbound model exhibits a flatter amplitude profile due to the absence of an inspiral phase, but both reconstructions reproduce the data well within the noise level.
The improvement of the unbound fit over the QCP reconstruction is most apparent for GW231221\_135041, where the \texttt{SEOBNRv6EHM} unbound model matches the first amplitude peak around merger in the Hanford detector more closely than \texttt{NRSur7dq4}, consistent with its larger log likelihood and SNR.

The right panels of Fig.~\ref{fig:unbound_summary_three_events} show, for each event, the posterior distribution in the $(E_0,\, p_\phi^0)$ plane, colored by log-likelihood. For reference, the red dashed curve shows a fit to the equal-mass, nonspinning separatrix between capture and scattering orbits given in Eq.~(8) of Ref.~\cite{Albanesi:2024xus}; the region above this curve is associated with captures, and the region below with scattering encounters. The actual separatrix depends on the mass ratio and spins of each sample, but the curve illustrates the general location of the boundary. For all three events, the posterior is concentrated above this boundary, with the highest-likelihood samples (indicated with crosshairs) corresponding to direct-capture configurations and negligible support for dynamical-capture or scattering morphologies. GW190521 and GW231221\_135041 recover similar initial conditions---$E_0 = 1.06^{+0.10}_{-0.05}$ and $p_\phi^0 = 4.1^{+2.0}_{-1.7}$ for the former, $E_0 = 1.05^{+0.12}_{-0.04}$ and $p_\phi^0 = 3.7^{+2.3}_{-1.5}$ for the latter (median and $90\%$ credible interval)---both favoring direct captures near the separatrix with relatively low energy and angular-momentum. GW191109\_010717, by contrast, is clearly bimodal in $E_0$, with a higher-likelihood high-energy mode ($E_0 \approx 1.2$--$1.3$) separated by a gap from a low-energy mode ($E_0 \approx 1.0$--$1.1$), consistent with the bimodality in $M$ noted in Fig.~\ref{fig:param_shift_unbound_more_events}.

The lower edges of these posteriors rail at the prior boundary $E_0=1.0002$. EAS analyses with an eccentricity prior extending up to $e=0.8$ similarly accumulate posterior support at high $e$, but the high-$e$ tail is bounded by an ``effective prior'' from waveform-generation failures rather than by the data~\cite{Gupte:2026whi}. Pushing the starting frequency to a lower value (e.g., to $\langle f_{\rm start} \rangle = 0.5$ Hz for GW190521, with the likelihood still evaluated above $f_{\rm min}=11$ Hz) lets the bound EAS model reach $e\sim 0.9$ without waveform generation failures, corresponding to initial conditions $E_0  \approx 1 - 10^{-3}$ and $p_\phi^{0}\approx 4$. Across the three events, the maximum-likelihood sample of the unbound analysis is higher than that of the EAS analysis for GW190521 and GW191109\_010717, and lower for GW231221\_135041; in all three cases the highest-likelihood EAS samples lie near the high-$e$ tail of the prior, suggesting that marginally bound configurations could also be preferred if that prior were extended.

The configurations recovered by the unbound analyses correspond to values for the relative velocity at infinity $v_\infty$  well above those of any realistic astrophysical channel: from $E_0 - 1\simeq \tfrac{1}{2}\nu(v_\infty/c)^2$ at leading non-relativistic order, realistic $v_\infty\sim\mathcal{O}(10$--$1000)$ km/s in globular clusters and nuclear star clusters~\cite{Baumgardt2018, Neumayer2020} correspond to $E_0 - 1\sim 10^{-10}$--$10^{-6}$, whereas the lower edge of the unbound prior ($E_0 = 1.0002$) already corresponds to $v_\infty\approx 0.04\,c\sim 10^4$ km/s, and the recovered $E_0 \gtrsim 1.05$ formally implies relativistic $v_\infty$. The recovered configurations are therefore astrophysically unrealistic; at realistic $v_\infty$, marginally unbound and highly eccentric bound orbits cannot be confidently discriminated in band, since the waveform is governed by the strong-field close approach and merger-ringdown, while the asymptotic kinetic energy that separates them is a small perturbation. The preference for the unbound model is more plausibly read as an in-band near-radial plunge---a morphology shared by marginally bound and marginally unbound orbits---than an encounter unbound at infinity, and does not, by itself, support an unbound origin for these events.
Running analyses a with prior on $E_0 - 1\sim 10^{-10}$--$10^{-6}$, closer to parabolic separatrix, is technically challenging because, if the initial conditions are defined at a sufficiently large fixed initial separation, waveforms become increasingly long, and hence each likelihood evaluation increasingly expensive, as $E_0\to1$. Both EAS and unbound analyses therefore push toward the parabolic separatrix $E_0=1$ and stop short of it for technical reasons, with a small region of parameter space near $E_0=1$ unexplored in both runs. A dedicated parameterization of this regime is left to future work.

The direct-capture morphology favored for GW190521 differs from Ref.~\cite{Gamba:2021gap}, which found that the event is best described as a dynamical capture with one or more close encounters before merger. The discrepancy can be attributed to several differences between the two analyses: (i) the \texttt{SEOBNRv6EHM} and \texttt{TEOBResumS-Dal\'i} models differ in their treatment of the EOB dynamics and waveform modes; (ii) our analysis includes aligned spins and higher-order modes beyond the dominant $(2,2)$; (iii) our prior on the initial conditions is broader, encompassing a larger range of energies and angular momenta; (iv) the analyses use different inference libraries (\texttt{Bilby} and \texttt{bajes}~\cite{Breschi:2021wzr}) and settings.

While in agreement with Ref.~\cite{Gamba:2021gap} in finding that an unbound configuration provides a comparable or better fit to the data than a bound eccentric or QCP configuration, our analysis finds weaker evidence for the unbound hypothesis compared to the dynamical-capture interpretation of Ref.~\cite{Gamba:2021gap}, with $\log_{10}\mathcal{B}^{\rm unbound}_{\rm QCP} \approx 0.6$ here against $\approx 3.6$ in Ref.~\cite{Gamba:2021gap}, when compared against \texttt{NRSur7dq4}.
This difference is in part due to the broader priors used here, which include aligned spins---absent from the nonspinning analysis of Ref.~\cite{Gamba:2021gap}---and a wider range of initial conditions, increasing the prior volume of the unbound model.
It is also reflected in the recovered network SNR: the \texttt{TEOBResumS-Dal\'i} dynamical-capture analysis of Ref.~\cite{Gamba:2021gap} reports a slightly higher maximum SNR than our unbound analysis with \texttt{SEOBNRv6EHM} ($\rho_{\mathrm{mf}}^{\text{N}} \approx 15.2$ against $14.5$), which directly translates into a higher Bayes factor.
\footnote{The SNR reported in the original LVK analysis of GW190521 is affected by an error in the likelihood function used in the \texttt{LALInference} and \texttt{Bilby} codes, leading to slightly overestimated SNR values~\cite{Talbot:2025vth}. The \texttt{bajes} library used by Ref.~\cite{Gamba:2021gap} is however not affected by this error. The error has since been fixed in \texttt{Bilby}, and the SNR values reported in this work are also not affected.}
However, a higher SNR does not necessarily indicate a more accurate parameter recovery, as noted in Sec.~\ref{sec:pe_injections}; the accuracy of the waveform model against NR simulations is the more relevant figure of merit.
Comparisons in Ref.~\cite{Gamboa:2026jht} show that \texttt{SEOBNRv6EHM} is generally more accurate than \texttt{TEOBResumS-Dal\'i} in predicting scattering angles and in waveform morphology for a set of NR simulations of scattering encounters and a dynamical capture.
We caution, however, that neither model currently incorporates generic-orbit effects in the merger--ringdown, where both use a QC description; the imprint of the unbound dynamics on the waveform is therefore confined to the pre-merger portion. Ongoing efforts to model the merger--ringdown of generic binaries~\cite{Albanesi:2023bgi, Faggioli:2026alx, Albanesi:2026qtx, Rao:2026lmz} will be important to relax this approximation.
A more detailed comparison between these models in a PE context, as well as extension of the analysis to other events, is left to future work.


\section{Conclusions}
\label{sec:conclusions}

We presented Bayesian PE analyses of eccentric compact binaries using the newly developed \texttt{SEOBNRv6EHM} waveform model~\cite{Gamboa:2026jht}.

We first validated the model through zero-noise injection-recovery analyses of synthetic signals from five equal-mass, nonspinning NR waveforms with increasing eccentricity from the SXS catalog~\cite{Scheel:2025jct}, including a long simulation with the highest eccentricity publicly available.
The simulations span eccentricities from $e_{\text{gw}} \approx 0.05$ to $e_{\text{gw}} \approx 0.34$ at a dimensionless orbit-averaged frequency $M \langle f_{\text{ref}} \rangle = 0.01$.
The eccentricity and other intrinsic parameters are accurately recovered in all six configurations, with the injected values lying within the $90\%$ credible intervals.
By comparing recovery across \texttt{SEOBNRv6EHM}, \texttt{SEOBNRv5EHM}, and \texttt{TEOBResumS-Dal\'i}, we showed that the latter two models yield biased estimates of eccentricity and intrinsic parameters for the most challenging high-eccentricity configurations, while \texttt{SEOBNRv6EHM} significantly reduces these biases.
The biases observed for \texttt{SEOBNRv5EHM} and \texttt{TEOBResumS-Dal\'i} are small in absolute terms but significant relative to the statistical uncertainties at current LVK SNRs. Moreover, waveform residuals from model inaccuracies can have significant impact on tests of general relativity and on global-fit analyses required for the LISA mission, where massive eccentric BBHs are expected at much higher SNR~\cite{LISA:2024hlh}.

We then applied \texttt{SEOBNRv6EHM} to \NEVENTS GW events from the O1--O4 LVK observing runs.
Across the full sample, six events show mild to moderate support for eccentricity over the QCAS hypothesis ($\log_{10}\mathcal{B}^{\text{EAS}}_{\text{QCAS}} > 0.5$), consistent with previous reports in the literature, notably (in decreasing order of Bayes factors) GW200129\_065458, GW200208\_222617, GW200105\_162426, GW231223\_032836,  GW190701\_203306, and GW230712\_090405.
When comparing against the QCP hypothesis, five of these events retain comparable support ($\log_{10}\mathcal{B}^{\text{EAS}}_{\text{QCP}} > 0.5$), while for GW230712\_090405 the evidence for eccentricity is reduced.
The fact that \texttt{SEOBNRv6EHM}---a demonstrably more accurate model---recovers the same candidates as previous analyses suggests that these eccentricity signatures are not artifacts of waveform systematics.
On the other hand, the fact that other eccentric candidates with mild support in previous analyses do not show comparable significance with \texttt{SEOBNRv6EHM} indicates that their inferred eccentricity is sensitive to waveform modeling differences.

The eccentricity posteriors reported here are obtained under a uniform prior in $e$ that does not reflect the astrophysical population. Since most mergers are expected to have circularized by the time they enter the detector band, the Bayes factors would likely decrease under a more realistic astrophysical prior; for the events with modest Bayes factors, this could reduce the evidence from mildly significant to insignificant.
As a concrete example, the odds ratio $\mathcal{O}^{\text{EAS}}_{\text{QCAS}} = \mathcal{B}^{\text{EAS}}_{\text{QCAS}} \times \mathcal{R}^{\text{EAS}}/\mathcal{R}^{\text{QCAS}}$ combines the Bayes factor with the astrophysical prior odds. Adopting $\mathcal{R}^{\text{EAS}}/\mathcal{R}^{\text{QCAS}} \approx 0.023$ compiled in Ref.~\cite{Gupte:2024jfe}---i.e.\ assuming that around $2\%$ of detected BBHs are measurably eccentric (with $e_{10\,\text{Hz}} > 0.05$)---only GW200129\_065458 would retain a preference for the eccentric hypothesis.
More broadly, single-event posteriors under reference priors should not be directly interpreted as astrophysical measurements; population-informed parameter estimates derived from a hierarchical catalog analysis~\cite{Fishbach:2019ckx, Galaudage:2019jdx, Moore:2021xhn, Mould:2026nle} are ultimately needed to rigorously assess the significance of these candidates.
For the event with the strongest evidence, GW200129\_065458, the Bayes factor is large enough that the prior choice is unlikely to qualitatively change the conclusion; in this case, the dominant caveat is instead the sensitivity of the inferred support for eccentricity to the glitch-subtraction method applied to the LIGO Livingston data~\cite{Gupte:2024jfe}.
The same applies to GW190701\_203306, which is also affected by a glitch in the LIGO Livingston detector~\cite{Gupte:2026whi}; for this event the evidence for eccentricity is reduced compared to some previous analyses~\cite{Gupte:2024jfe}, though still higher than others~\cite{Planas:2025plq}.

Exploiting the applicability of \texttt{SEOBNRv6EHM} to generic planar orbits, we reanalyzed five high-mass events allowing for direct-capture, dynamical-capture, and scattering morphologies. For four of them---including GW231123\_135430---the unbound model gives a comparable or better fit than the bound EAS model, indicating that for short, high-mass signals an unbound-orbit interpretation provides a plausible description of the data. More strikingly, for three of these---GW190521, GW191109\_010717, and GW231221\_135041---a direct-capture configuration is even comparable to, or marginally preferred over, the QCP hypothesis ($\log_{10}\mathcal{B}^{\rm unbound}_{\rm QCP} \approx 0.2$--$1.6$, depending on the event and reference precessing model). To our knowledge, this is the first time that GW events beyond GW190521 show support for an unbound-orbit interpretation. 
Several caveats temper these results, however.
Direct-capture configurations are astrophysically rare: single-single capture rates in nuclear star clusters are estimated at $0.005$--$0.02~{\rm Gpc}^{-3}~{\rm yr}^{-1}$~\cite{Tsang:2013mca}, compared to the LVK-inferred local BBH merger rate of $14$--$26~{\rm Gpc}^{-3}~{\rm yr}^{-1}$~\cite{LIGOScientific:2025pvj}, giving prior odds against the unbound hypothesis of order $10^{-3}$--$10^{-4}$ that can largely absorb Bayes factors of the size reported here. More fundamentally, as discussed in Sec.~\ref{sec:pe_unbound}, the recovered configurations are astrophysically unrealistic---formally implying relativistic velocities at infinity, far above any realistic dynamical formation channel---and cannot be confidently discriminated from highly eccentric bound orbits. The results are most plausibly interpreted as near-radial plunges rather than encounters unbound at infinity, and do not, by themselves, support an unbound origin for these events. A prior on the initial conditions informed by astrophysically relevant velocities ($v_\infty\sim \mathcal{O}(10 - 1000)$ km/s) could reduce the support for unbound configurations; sampling that near-parabolic regime is, however, currently computationally challenging, and a dedicated parameterization of that regime is left to future work.

Furthermore, both GW191109\_010717 and GW231221\_135041 were analyzed on glitch-subtracted data. Residual glitch artifacts have been shown to significantly affect the spin inference of GW191109\_010717~\cite{Udall:2024ovp}, and could be especially consequential for these short-duration signals. Quantifying both effects, alongside dedicated NR validation in the unbound regime, is needed before any of these candidates can be confidently interpreted as having an unbound origin. For GW190521 specifically, we extended the analysis of Ref.~\cite{Gamba:2021gap} by including spin effects and higher-order modes.
The evidence is weaker than reported in Ref.~\cite{Gamba:2021gap}, which found a dynamical capture to be the most preferred configuration; this is in part a consequence of the larger prior volume due to the inclusion of spins, though we also recover a slightly lower SNR than reported there. A more detailed comparison of \texttt{SEOBNRv6EHM} with \texttt{TEOBResumS-Dal\'i} in unbound-orbit scenarios, validation of both models with NR injections, and analysis of other events are left to future work.

From a computational standpoint, \texttt{SEOBNRv6EHM} is three to five times faster than its predecessor \texttt{SEOBNRv5EHM} in PE analyses, with the most significant improvements for longer-duration signals; per-waveform benchmarks in Ref.~\cite{Gamboa:2026jht} show similar speedups of $\sim 2-5$ over \texttt{SEOBNRv5EHM} and $\sim 2-8$ over \texttt{TEOBResumS-Dal\'i}.
In its QC limit, the model is even faster than the dedicated QC model \texttt{SEOBNRv5HM}, so the eccentric infrastructure introduces no overhead when eccentricity is not included.
These improvements make large-scale eccentric PE catalogs and long-duration analyses, including NSBH and BNS events, computationally feasible for the first time with a time-domain eccentric EOB model and standard stochastic samplers.

Future improvements to the model will incorporate the joint effects of eccentricity and spin precession within a single model (\texttt{SEOBNRv6EPHM})~\cite{Gamboa:2026ybr}.
This will be crucial to robustly disentangle the signatures of orbital eccentricity from those of spin precession, which can produce similar modulations in GW signals from high-mass binaries~\cite{Romero-shaw:2022fbf, CalderonBustillo:2020xms}.
We plan to validate such a model against eccentric, precessing-spin NR simulations and to reanalyze events in the LVK catalog to provide more robust population-level constraints on orbital eccentricity.


\section*{Acknowledgments}
We thank Nihar Gupte, Coleman Miller, Gonzalo Morr\'as, Peter James Nee, and Matias Zaldarriaga for helpful discussions and for valuable comments on the manuscript.
A.B. would like to thank the Institute for Advanced Study in Princeton for its hospitality during the final stages of this work.
The computational work for this manuscript was carried out on the \texttt{Hypatia} computer cluster at the Max Planck Institute for Gravitational Physics in Potsdam.
L.P. is supported by a UKRI Future Leaders Fellowship (grant number MR/Y018060/1). 
A.B and A.G. are supported in part by the European Research Council (ERC) Horizon Synergy Grant ``Making Sense of the Unexpected in the Gravitational-Wave Sky'' (GWSky-101167314).
This material is based upon work supported by NSF's LIGO Laboratory which
is a major facility fully funded by the National Science Foundation.
This research has made use of data or software obtained from the Gravitational Wave Open Science
Center (gwosc.org), a service of LIGO Laboratory, the
LIGO Scientific Collaboration, the Virgo Collaboration,
and KAGRA. LIGO Laboratory and Advanced LIGO are
funded by the United States National Science Foundation
(NSF) as well as the Science and Technology Facilities
Council (STFC) of the United Kingdom, the Max-Planck-Society (MPS), and the State of Niedersachsen/Germany
for support of the construction of Advanced LIGO and
construction and operation of the GEO600 detector. Additional support for Advanced LIGO was provided by the
Australian Research Council. Virgo is funded, through
the European Gravitational Observatory (EGO), by the
French Centre National de Recherche Scientifique (CNRS),
the Italian Istituto Nazionale di Fisica Nucleare (INFN)
and the Dutch Nikhef, with contributions by institutions
from Belgium, Germany, Greece, Hungary, Ireland, Japan,
Monaco, Poland, Portugal, Spain. KAGRA is supported
by Ministry of Education, Culture, Sports, Science and
Technology (MEXT), Japan Society for the Promotion
of Science (JSPS) in Japan; National Research Foundation (NRF) and Ministry of Science and ICT (MSIT) in
Korea; Academia Sinica (AS) and National Science and
Technology Council (NSTC) in Taiwan.


\section*{Data availability}

Posterior samples for the \NEVENTS events analyzed with \texttt{SEOBNRv6EHM} under the eccentric, aligned-spin hypothesis are available at \url{https://doi.org/10.5281/zenodo.19678261}.

\appendix

\section{Impact of waveform generation starting frequency}
\label{app:f_start_and_t_start}

In this appendix we investigate in detail the impact of the waveform generation starting frequency on the recovered parameters for the two highest-eccentricity NR configurations, \texttt{SXS:BBH:2525} and \texttt{SXS:BBH:2527}, with $e_{\rm gw} \approx 0.24$ and $e_{\rm gw} \approx 0.34$ at $M \langle f_{\rm ref}\rangle = 0.01$, respectively.

For \texttt{SXS:BBH:2525} at $M = 80\,M_\odot$ and \texttt{SXS:BBH:2527} at $M = 20\,M_\odot$, the NR waveform starts at an orbit-averaged $(2,2)$ frequency of approximately $6\,$Hz, so reducing the starting frequency of the template to $\langle f_{\rm start}\rangle = 6\,$Hz is sufficient to capture the full NR signal; we compare this against the default choice of $\langle f_{\rm start}\rangle = 10\,$Hz adopted for the main analyses in Sec.~\ref{sec:pe_injections}.

For \texttt{SXS:BBH:2527} at $M = 80\,M_\odot$, the NR simulation starts at an orbit-averaged frequency of approximately $1.5\,$Hz, so matching the full signal length would require starting waveform generation at $\langle f_{\rm start}\rangle \approx 1.5\,$Hz.
This would significantly increase the computational cost of the analysis: at high eccentricities, the waveform duration at fixed orbit-averaged frequency is much shorter than for a QC binary at the same frequency (since periastron passages radiate energy and angular momentum more efficiently), so a starting frequency of $1.5\,$Hz translates to extremely long waveforms for the near-circular samples that populate the flat eccentricity prior.
As a practical alternative, we supplement the default $\langle f_{\rm start}\rangle = 10\,$Hz run with a backward-in-time integration: for each waveform evaluation, the EOB equations of motion are integrated backward from the nominal orbit-averaged frequency by a fixed duration of $15000~M$, chosen to cover the missing early inspiral cycles; this captures the full signal power without the prohibitive cost of a globally reduced starting frequency. Eccentricity and relativistic anomaly remain defined at the nominal $10\,$Hz reference.
The backward integration is available for the two \texttt{SEOBNR} models but not for \texttt{TEOBResumS-Dal\'i}, which does not currently support this option.

\begin{figure*}
    \centering
    \includegraphics[width=0.95\textwidth]{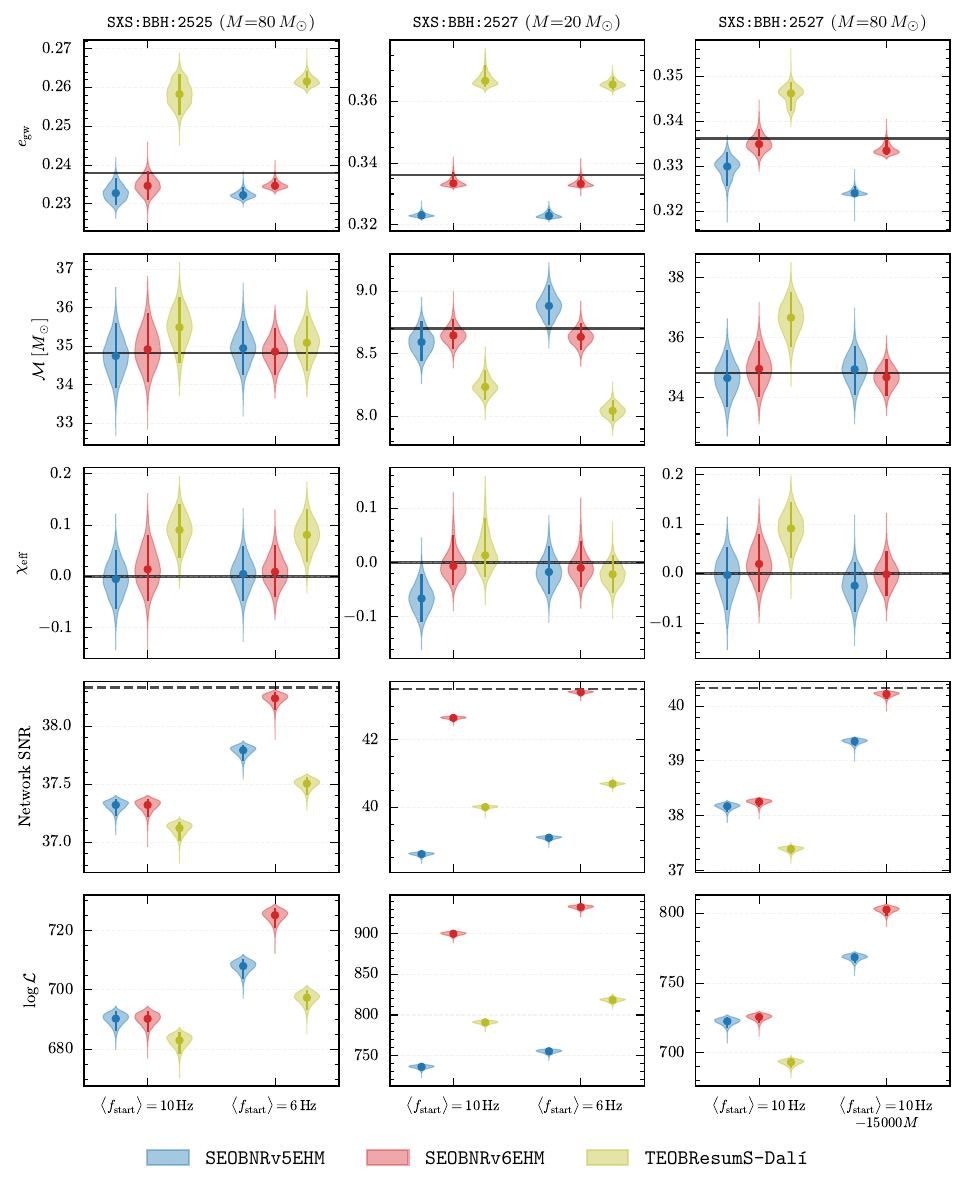}
    \caption{
	Impact of the waveform starting frequency on recovered parameters for three NR injection configurations: \texttt{SXS:BBH:2525} at $M = 80\,M_\odot$ (left column), \texttt{SXS:BBH:2527} at $M = 20\,M_\odot$ (center), and \texttt{SXS:BBH:2527} at $M = 80\,M_\odot$ (right).
	Each violin shows the marginal posterior for the labeled parameter, recovered with \texttt{SEOBNRv5EHM} (blue), \texttt{SEOBNRv6EHM} (red), and \texttt{TEOBResumS-Dal\'i} (yellow); solid horizontal lines mark the injected values, and the dashed line in the SNR row marks the injected network SNR.
	For the first two configurations, we compare $\langle f_{\rm start}\rangle = 10\,$Hz against $6\,$Hz; for the third, we compare $10\,$Hz against $10\,$Hz with backward-in-time integration by $15000\,M$ (available only for the \texttt{SEOBNR} models).
	In all cases, capturing additional early-inspiral signal power increases the recovered SNR and log-likelihood and produces tighter posteriors, most notably on the eccentricity. With a sufficiently low starting frequency, \texttt{SEOBNRv6EHM} recovers nearly the full injected SNR.
    }
    \label{fig:impact_fstart}
\end{figure*}

\begin{figure*}
    \centering
    \includegraphics[width=1.0\textwidth]{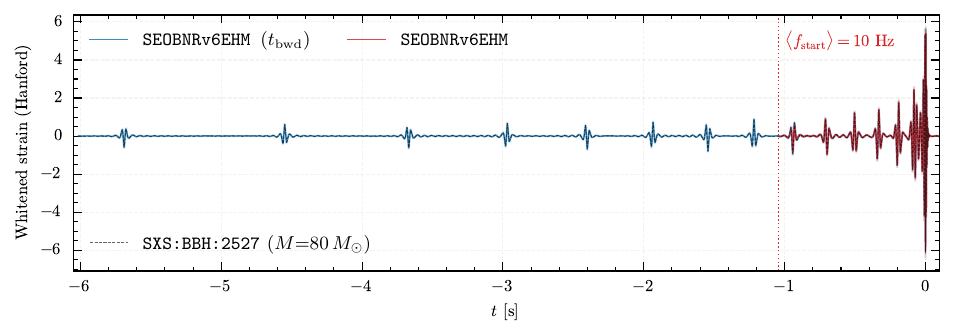}
    \caption{
	Whitened-strain waveform reconstruction in the LIGO Hanford detector for the \texttt{SXS:BBH:2527} NR injection at total mass $M =\nolinebreak 80\,M_\odot$ (injected $e_{\rm gw} = 0.336$ at $M\langle f_{\rm ref} \rangle = 0.01$). The black dashed curve is the injected NR signal; the colored curves show the posterior median of the \texttt{SEOBNRv6EHM} reconstruction with $\langle f_{\rm start}\rangle = 10\,$Hz, without (red) and with (blue) backward-in-time integration ($t_{\rm bwd} = 15000\,M$), with the shaded bands indicating the $90\%$ credible intervals across posterior samples. Without backward integration, waveform generation begins only ${\sim}\,1\,$s before merger, missing the earlier periastron passages visible in the data. Enabling backward integration extends the reconstruction to the earliest periastron bursts at $t \sim -6\,$s.
    }
    \label{fig:wf_recon_tbwd}
\end{figure*}

Figure~\ref{fig:impact_fstart} summarizes the results.
For all three configurations, the default $\langle f_{\rm start}\rangle = 10\,$Hz choice causes the template to miss the earliest periastron passages of the NR signal, reducing the recovered SNR relative to the injected value (see also Table~\ref{tab:peinjections} and the discussion in Sec.~\ref{sec:pe_injections}).
Lowering the starting frequency to $6\,$Hz (for \texttt{SXS:BBH:2525} at $M=80\,M_\odot$ and \texttt{SXS:BBH:2527} at $M=20\,M_\odot$) or supplementing the $10\,$Hz run with backward-in-time integration (for \texttt{SXS:BBH:2527} at $M=80\,M_\odot$) increases the recovered SNR and log-likelihood for all three models.
\texttt{SEOBNRv6EHM} recovers nearly the full injected SNR in all three configurations, while \texttt{SEOBNRv5EHM} and \texttt{TEOBResumS-Dal\'i} remain significantly below the injected value even with the lower starting frequency. Specifically, for \texttt{SXS:BBH:2525} at $M=80\,M_\odot$, the SNR recovered by \texttt{SEOBNRv6EHM} increases from $37.3^{+0.1}_{-0.1}$ to $38.2^{+0.1}_{-0.1}$ (injected SNR $\approx 38.3$) when lowering the starting frequency from $10\,$Hz to $6\,$Hz.
For \texttt{SXS:BBH:2527} at $M=20\,M_\odot$, the recovered SNR increases from $42.7^{+0.1}_{-0.1}$ to $43.4^{+0.1}_{-0.1}$ (injected SNR $\approx 43.5$) when lowering the starting frequency from $10\,$Hz to $6\,$Hz.
For \texttt{SXS:BBH:2527} at $M=80\,M_\odot$, the recovered SNR increases from $38.3^{+0.1}_{-0.1}$ to $40.2^{+0.1}_{-0.1}$ (injected SNR $\approx 40.3$) when supplementing the $10\,$Hz run with the backward-in-time integration.
The log-likelihood for \texttt{SEOBNRv6EHM} increases by $\sim 30$ in the two cases with $\langle f_{\rm start}\rangle = 6\,$Hz and by $\sim 77$ in the case with backward integration; the other two models also improve, but more modestly.

Figure~\ref{fig:wf_recon_tbwd} illustrates this directly by comparing the whitened-strain \texttt{SEOBNRv6EHM} reconstruction with and without backward integration for the \texttt{SXS:BBH:2527} injection at $M=80\,M_\odot$: without backward integration, waveform generation begins only ${\sim}\,1\,$s before merger, missing the earlier periastron passages visible in the injected signal, whereas enabling backward integration extends the reconstruction to the earliest periastron bursts at $t \sim -6\,$s.

For the two cases at $M=80\,M_\odot$, the additional inspiral captured by lowering the starting frequency or by the backward-in-time integration includes a significant number of periastron passages, which are especially informative about eccentricity. As a result, the eccentricity posteriors tighten substantially. For \texttt{SXS:BBH:2525} and \texttt{SXS:BBH:2527} at $M=80\,M_\odot$, the $90\%$ credible interval on the eccentricity shrinks by a factor of $\sim 2$ when lowering the starting frequency from $10\,$Hz to $6\,$Hz or when enabling backward-in-time integration.
For \texttt{SXS:BBH:2527} at $M=20\,M_\odot$, only two periastron cycles are missed in the default configuration, so the impact of lowering the starting frequency is smaller, but still leads to a slight tightening of the posteriors.
Since higher recovered likelihoods and narrower eccentricity posteriors are expected to yield larger Bayes factors favoring eccentricity, ensuring that the analysis captures the full signal power could be important for accurately assessing the significance of eccentricity in real GW events.

For \texttt{SEOBNRv6EHM}, the posteriors become narrower while remaining peaked at similar values across starting-frequency choices, with the injected values for the eccentricity falling just outside the $90\%$ credible interval for the \texttt{SXS:BBH:2525} configuration, while the chirp mass and $\chi_{\rm eff}$ are also more tightly constrained (by up to a factor of $\sim 1.5$) and always remain consistent with the injected values.
For the two configurations at $M=80\,M_\odot$, the biases in \texttt{SEOBNRv5EHM} and \texttt{TEOBResumS-Dal\'i} remain similar in absolute terms to those in the default runs, but become more pronounced relative to the measurement uncertainty, which shrinks accordingly. Only for \texttt{SXS:BBH:2527} at $M=20\,M_\odot$ do the biases change qualitatively: most notably, the chirp mass recovered by \texttt{SEOBNRv5EHM} shifts from being slightly underestimated to overestimated, while $\chi_{\rm eff}$ moves from underestimated to consistent with the injected value. This suggests that the model shifts its intrinsic parameters to compensate for waveform inaccuracies in the early periastron passages that enter the analysis when the starting frequency is lowered from $10\,$Hz to $6\,$Hz.

Overall, these results demonstrate that the choice of starting frequency can significantly affect both the recovered SNR and the eccentricity constraints, particularly for highly eccentric systems where a substantial fraction of the signal power resides in early periastron passages below the default $\langle f_{\rm start}\rangle = 10\,$Hz threshold. 
We note that for real astrophysical signals the impact could be even larger than what is observed here, since the NR waveforms have a finite length and additional periastron passages with instantaneous frequency above $20\,$Hz may occur before the start of the NR simulation, especially at lower total masses (see Fig.~\ref{fig:NR_orbit_averaged_frequencies}).
Related issues with eccentric template initialization have also been identified in the context of early-warning detection, where Ref.~\cite{Sinha:2025vmc} showed that starting waveform generation at the periastron frequency rather than the orbit-averaged frequency improves the recovered SNR and sky localization.

The backward-integration approach is presented here as a proof of principle. In particular, the fixed backward-integration time was chosen based on the known duration of the injected NR signal, which is not available a priori in a real PE run.
In practice, a suitable integration time could be determined from the mass and eccentricity priors, or from preliminary PE analyses, as is done for other analysis settings in the LVK production-PE workflow~\cite{LIGOScientific:2025yae}.
If the power in periastron passages of higher-order $(\ell,m)$ modes is also significant, the starting frequency should be decreased further, by approximately a factor $m/2$; however, for the equal-mass configurations considered here, higher-mode power is subdominant, so the backward-integration time determined from the dominant $(2,2)$ mode is sufficient to recover the full SNR.
We defer a systematic investigation of this approach to future work.

\section{Timing comparisons against \texorpdfstring{\texttt{TEOBResumS-Dal\'i}}{TEOBResumS-Dali}}
\label{app:dali_benchmarks}

\begin{figure}
    \centering
    \includegraphics[width=\columnwidth]{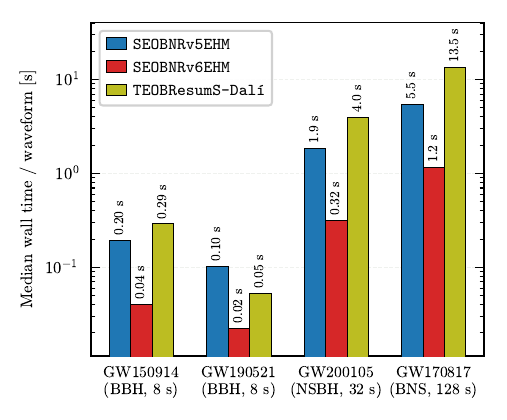}
    \caption{
	Median waveform evaluation time for \texttt{SEOBNRv5EHM}, \texttt{SEOBNRv6EHM}, and \texttt{TEOBResumS-Dal\'i} for the four benchmark events.
    For each event, 100 parameter samples are drawn from the \texttt{SEOBNRv6EHM} posterior and used to evaluate waveforms with each model under identical settings.
    Each group of bars corresponds to one event, with the signal duration and binary type indicated below.
	Time labels are shown above each bar.
	\texttt{SEOBNRv6EHM} is the fastest model for all events, and achieves more than an order-of-magnitude speedup over \texttt{TEOBResumS-Dal\'i} for the long NSBH and BNS signals GW200105\_162426 and GW170817.
    }
    \label{fig:timing_comparison_dali}
\end{figure}

As noted in Sec.~\ref{sec:pe_benchmark}, \texttt{TEOBResumS-Dal\'i} is not included in the real-event analyses due to its higher computational cost for long-duration signals. Nevertheless, to quantify the relative performance of all three models, we compare individual waveform evaluation times on the four benchmark events.
For each event, we draw 100 parameter samples from the \texttt{SEOBNRv6EHM} posterior and evaluate waveforms with \texttt{SEOBNRv5EHM}, \texttt{SEOBNRv6EHM}, and \texttt{TEOBResumS-Dal\'i} under identical settings (frequency range, duration, and sampling rate matching the PE analyses), including conversion to the frequency domain.
While this is not a full PE comparison, the resulting single-waveform timing distributions are consistent with the wall-clock run times reported in Sec.~\ref{sec:pe_events}, with the relative speedups between \texttt{SEOBNRv5EHM} and \texttt{SEOBNRv6EHM} matching those observed in the full PE runs. All timings are performed on an AMD EPYC 7742 64-Core Processor.

The results are shown in Fig.~\ref{fig:timing_comparison_dali}.
Comparing \texttt{SEOBNRv5EHM} and \texttt{TEOBResumS-Dal\'i}, the latter is faster for the short-duration signal GW190521 but slower for all other events, with the gap reaching a factor of $\sim\!2.5$ for the BNS signal GW170817.
Comparing \texttt{SEOBNRv6EHM} and \texttt{TEOBResumS-Dal\'i}, the former is faster for all four events, by a factor $\sim 2.5$ for the short-duration signal GW190521 and by more than an order of magnitude for GW200105\_162426 and GW170817.
These results confirm that \texttt{SEOBNRv6EHM} is the fastest eccentric time-domain EOB model currently available, enabling PE analyses of long-duration signals such as NSBH and BNS events that would be impractical with existing alternatives.

\section{Injection-recovery tests of the unbound hypothesis}
\label{app:unbound_injections}

\begin{figure*}
    \centering
    \includegraphics[width=0.95\textwidth]{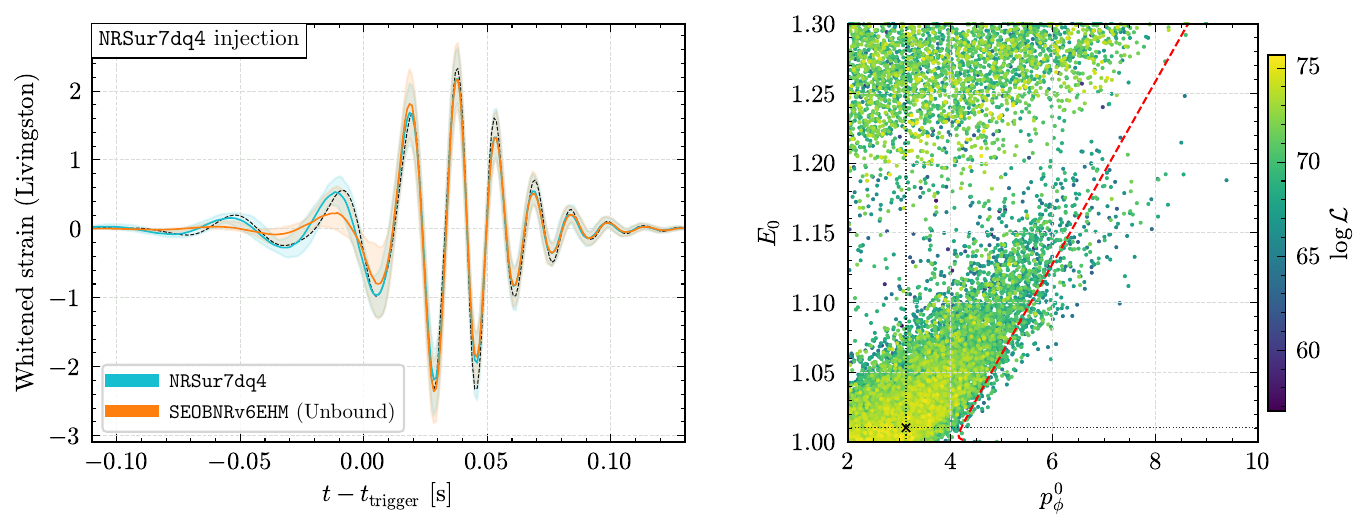}
    \includegraphics[width=0.95\textwidth]{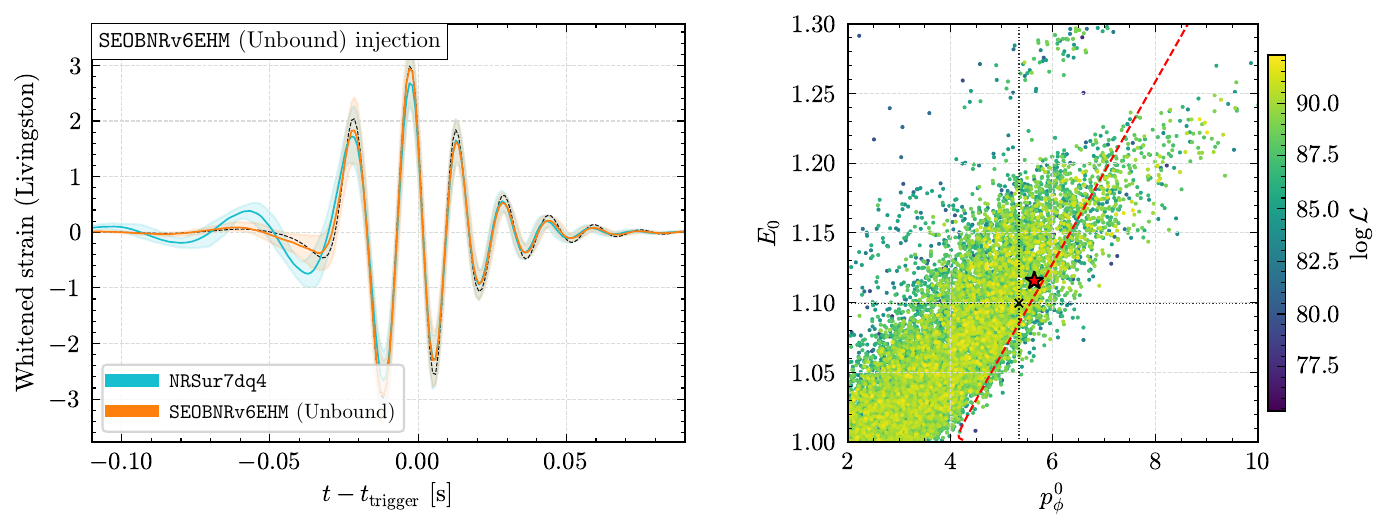}
    \caption{
    Zero-noise injection-recovery analyses of GW190521-like signals. \emph{Top:} quasi-circular, precessing-spin signal generated with \texttt{NRSur7dq4}, using the maximum-likelihood parameters of the \texttt{NRSur7dq4} analysis of GW190521. \emph{Bottom:} unbound, direct-capture signal generated with \texttt{SEOBNRv6EHM}, using the maximum-likelihood parameters of the unbound analysis of GW190521. For each injection: \emph{Left:} Whitened waveform reconstruction in the LIGO Livingston detector. The median (solid) and $90\%$ credible interval (shaded) of the reconstructed distribution are shown for the unbound \texttt{SEOBNRv6EHM} model and the quasi-circular precessing-spin model \texttt{NRSur7dq4}, compared to the injected signal (dashed black). \emph{Right:} Posterior distribution of the initial energy $E_0$ and angular momentum $p_\phi^0$ from the unbound analysis, colored by log-likelihood. The crosshairs mark the maximum-likelihood sample and, for the unbound injection, the red star marks the injected initial conditions. The red dashed curve shows the equal-mass, nonspinning separatrix between capture and scattering orbits~\cite{Albanesi:2024xus}, as in Fig.~\ref{fig:unbound_summary_three_events}.
    }
    \label{fig:unbound_injections}
\end{figure*}

{\renewcommand{\arraystretch}{1.9}
\begin{table*}
\centering
\small
\input{tab/unbound_injection_table.tex}
\caption{
    Injected values and recovered posterior medians with $90\%$ credible intervals for the two GW190521-like zero-noise injections of Fig.~\ref{fig:unbound_injections}, recovered with the quasi-circular precessing-spin model \texttt{NRSur7dq4} (QCP) and with the unbound configuration of \texttt{SEOBNRv6EHM}.
    The reported quantities are: network matched-filter SNR $\rho^{\text{N}}_{\text{mf}}$ (network optimal SNR for the injected signal), log-likelihood $\log\mathcal{L}$, detector-frame total mass $M$, inverse mass ratio $1/q$, effective spin $\chi_{\text{eff}}$, effective precessing spin $\chi_{\rm p}$ (zero for the aligned-spin unbound model), and initial energy $E_0$ and angular momentum $p_\phi^0$ at the initial separation $r_0 = 500\,M$ (defined only for the unbound model).
    The last column reports the Bayes factor $\log_{10}\mathcal{B}^{\rm unbound}_{\rm NRSur}$ between the unbound hypothesis and \texttt{NRSur7dq4}.
}
\label{tab:unbound_injections}
\end{table*}}

The unbound analyses of Sec.~\ref{sec:pe_unbound} find that, for GW190521, GW191109\_010717, and GW231221\_135041, a direct-capture configuration fits the data comparably well to the QCP hypothesis. For signals from high-mass binaries, both configurations produce only a few cycles in band, dominated by the merger--ringdown, and it is not a priori clear whether the unbound model can mimic a QCP source, nor how strongly the data can discriminate between the two hypotheses.
To address both questions, we perform two zero-noise injection-recovery analyses of GW190521-like signals, and recover each of them with the unbound configuration of \texttt{SEOBNRv6EHM} and with the QCP model \texttt{NRSur7dq4}.
First, we inject a QC, spin-precessing signal generated with \texttt{NRSur7dq4}, to test whether the support we find for unbound configurations can be produced by a precessing QC source.
Second, we inject an unbound, direct-capture signal generated with \texttt{SEOBNRv6EHM}, to test whether such a configuration is recovered, and which hypothesis is then preferred.

For the QCP injection, we use the maximum-likelihood parameters of the \texttt{NRSur7dq4} analysis of GW190521 in Sec.~\ref{sec:pe_unbound}. These correspond to a strongly precessing, nearly equal-mass binary, with detector-frame total mass $M \approx 290\,M_\odot$, $1/q \approx 0.94$, a primary spin of magnitude $a_1 \approx 0.97$ lying close to the orbital plane ($\chi_{\rm p} \approx 0.96$, $\chi_{\rm eff} \approx 0.19$).
For the unbound injection, we use the maximum-likelihood parameters of the unbound analysis of GW190521, i.e., $M \approx 292\,M_\odot$, $1/q \approx 0.60$, $\chi_{\rm eff} \approx 0.27$, and initial conditions $E_0 \approx 1.116$, $p_\phi^0 \approx 5.64$ at $r_0 = 500\,M$, which correspond to a direct capture.
Both signals are injected in zero noise in the LIGO Hanford, LIGO Livingston, and Virgo detectors, using the same PSDs and data settings as for the analysis of GW190521.
The results are summarized in Fig.~\ref{fig:unbound_injections} and Table~\ref{tab:unbound_injections}.

For the QCP injection, \texttt{NRSur7dq4} is correctly preferred, with $\log_{10}\mathcal{B}^{\rm unbound}_{\rm NRSur} = -1.02 \pm 0.10$. However, the unbound model fits the precessing signal nearly as well: its recovered SNR is lower by only $\approx 0.1$, and its median log-likelihood by $\Delta\log\mathcal{L} \approx 1.5$.
Both models recover the total mass, mass ratio, and effective spin within the $90\%$ credible intervals, while the precessing spin, which is only weakly constrained at this SNR, is slightly underestimated by \texttt{NRSur7dq4}.
The whitened reconstructions (top left panel of Fig.~\ref{fig:unbound_injections}) show that both models track the injected signal during the merger and ringdown; as for the real events in Sec.~\ref{sec:pe_unbound}, the unbound reconstruction exhibits a flatter amplitude profile in the cycles preceding merger, where the injected signal carries the imprint of the inspiral.
The unbound model recovers the precessing QC signal as a direct capture near the separatrix, with the posterior concentrated at low energies against the lower prior boundary $E_0 = 1.0002$, and a secondary high-energy mode at $E_0 \approx 1.2$--$1.3$ containing ${\approx}\,20\%$ of the samples.
The posterior, with $E_0 = 1.05^{+0.24}_{-0.04}$ and $p_\phi^0 = 3.5^{+2.3}_{-1.3}$, resembles the one found for GW190521 ($E_0 = 1.06^{+0.10}_{-0.05}$ and $p_\phi^0 = 4.1^{+2.0}_{-1.7}$), apart from the secondary high-energy mode, which has negligible support for the event.
The direct-capture morphology recovered for the events in Sec.~\ref{sec:pe_unbound} is therefore not, by itself, diagnostic of an unbound source.

For the unbound injection, the unbound hypothesis is correctly preferred, with $\log_{10}\mathcal{B}^{\rm unbound}_{\rm NRSur} = 1.18 \pm 0.10$, and a higher recovered SNR ($13.6$ against $13.3$) and median log-likelihood ($\Delta\log\mathcal{L} \approx 3.8$) than \texttt{NRSur7dq4}.
The morphology recovered from the unbound analysis correctly corresponds to a direct capture, and the injected initial conditions lie within the $90\%$ credible region ($E_0 = 1.07^{+0.11}_{-0.06}$ and $p_\phi^0 = 4.1^{+2.3}_{-1.7}$), with the maximum-likelihood sample ($E_0 \approx 1.10$, $p_\phi^0 \approx 5.34$) close to the injected values.
The total mass and effective spin are shifted to lower values, with the injected values lying at the upper edge of the $90\%$ credible intervals ($M = 272^{+21}_{-23}\,M_\odot$ and $\chi_{\rm eff} = 0.02^{+0.25}_{-0.23}$).
\texttt{NRSur7dq4} recovers comparable shifts, with the injected values lying just outside the $90\%$ credible intervals, and a broad posterior on the precessing spin ($\chi_{\rm p} = 0.55^{+0.35}_{-0.40}$).
The shifts are consistent with the strong correlation between total mass and effective spin~\cite{Baird:2012cu}, found in both analyses: along this direction the data are only weakly informative, and the posterior is pulled towards the peak of the $\chi_{\rm eff}$ prior at zero. In its reconstruction (bottom left panel of Fig.~\ref{fig:unbound_injections}), \texttt{NRSur7dq4} develops an additional inspiral cycle before merger, which is absent in the injected direct-capture signal.

In both cases, the injected hypothesis is correctly preferred: in zero noise, a GW190521-like QCP source yields $\log_{10}\mathcal{B}^{\rm unbound}_{\rm NRSur} \approx -1.0$, and a GW190521-like unbound source $\approx 1.2$.
Although the two interpretations are significantly degenerate for high-mass signals at the SNR of GW190521, with only a few cycles in band, the degeneracy is not complete: the data retain enough information to discriminate between them, albeit only at the level of Bayes factors of order ten.
The value found for GW190521, $\log_{10}\mathcal{B}^{\rm unbound}_{\rm NRSur} \approx 0.6$ (Table~\ref{tab:bf_unbound}), lies between the two zero-noise values.
These results support the interpretation given in Sec.~\ref{sec:pe_unbound}: the Bayes factors between the unbound and QCP hypotheses are informative, although the preferences they indicate for the events are mild, and the direct-capture morphology alone is not diagnostic of an unbound source.

\section{Unbound analysis of GW231123\_135430}
\label{app:GW231123_unbound}

\begin{figure*}
    \centering
    \includegraphics[width=\textwidth]{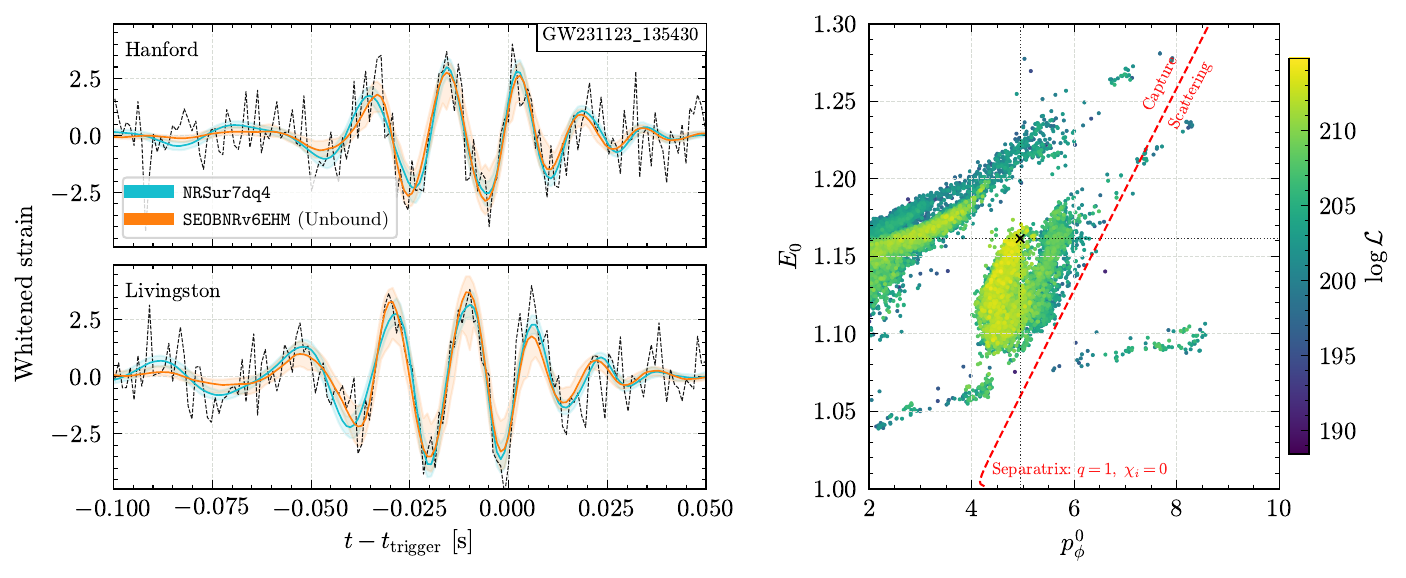}
    \caption{Same as Fig.~\ref{fig:unbound_summary_three_events}, but for GW231123\_135430.}
    \label{fig:unbound_summary_GW231123}
\end{figure*}

Figure~\ref{fig:unbound_summary_GW231123} shows the whitened waveform reconstruction and the $(E_0,\, p_\phi^0)$ posterior for the unbound analysis of GW231123\_135430---the only event in our sample where the unbound hypothesis is preferred over EAS but disfavored against both QCP analyses. The initial-condition posterior is clearly bimodal, with a higher-likelihood mode at $E_0 \approx 1.15$, $p_\phi^0 \approx 4.5$ and a secondary at $p_\phi^0 \approx 3.5$ and slightly higher energy; both lie well above the equal-mass nonspinning separatrix, corresponding to high-energy direct-capture configurations. These two modes map onto the bimodal structure of the $(M,\, \chi_{\rm eff})$ posterior visible in Fig.~\ref{fig:param_shift_unbound_more_events}, providing the most pronounced example in our sample of degeneracies between the initial-condition and  intrinsic-parameters. 

The recovered network SNRs mirror the log-likelihood ordering, with $\rho_{\mathrm{mf}}^{\text{N}} \approx 19.8$ for EAS, $\approx 20.5$ for unbound, and $\approx 20.7$--$20.9$ for QCP; the unbound \texttt{SEOBNRv6EHM} reconstruction reaches a log-likelihood lower by $\Delta\log\mathcal{L} \approx 3$ (\texttt{NRSur7dq4}) and $\approx 8$ (\texttt{SEOBNRv5PHM}). Inspecting the reconstructions, a difference can be seen in the cycle preceding merger in the Livingston detector, where the unbound reconstruction's flatter amplitude profile does not track the data as closely as \texttt{NRSur7dq4}.

\bibliographystyle{JHEP}
\bibliography{references}

\end{document}

%% file: tab/NR_injection_table.tex
\renewcommand{\arraystretch}{1.9}
\begin{tabular*}{\textwidth}{c@{\extracolsep{\fill}} c c c c c c c}

\hline
\hline

Parameter
& \makecell[cc]{Injected \\ value}
& \makecell[cc]{\small \texttt{SXS:BBH:1355}\\$M=80~M_\odot$}
& \makecell[cc]{\small \texttt{SXS:BBH:1359}\\$M=80~M_\odot$}
& \makecell[cc]{\small \texttt{SXS:BBH:1362}\\$M=80~M_\odot$}
& \makecell[cc]{\small \texttt{SXS:BBH:2525}\\$M=80~M_\odot$}
& \makecell[cc]{\small \texttt{SXS:BBH:2527}\\$M=80~M_\odot$}
& \makecell[cc]{\small \texttt{SXS:BBH:2527}\\$M=20~M_\odot$}
\\

\hline
\hline

\centering

$ M/\solarmass $ & $ 80.0~(20.0) $ & $ {80.1}^{+2.1}_{-2.1} $ & $ {80.3}^{+2.0}_{-2.1} $ & $ {80.4}^{+2.4}_{-2.4} $ & $ {80.6}^{+2.2}_{-1.9} $ & $ {80.7}^{+2.3}_{-2.2} $ & $ {19.9}^{+0.6}_{-0.3} $ \\ [0.1cm]

$ \mathcal{M}/\solarmass $ & $ 34.8~(8.7) $ & $ {34.7}^{+0.9}_{-0.9} $ & $ {34.8}^{+0.9}_{-1.0} $ & $ {34.9}^{+1.0}_{-1.1} $ & $ {34.9}^{+0.9}_{-0.8} $ & $ {35.0}^{+0.9}_{-0.9} $ & $ {8.6}^{+0.1}_{-0.1} $ \\ [0.1cm]

$ 1/q $ & $ 1.00 $ & $ {0.89}^{+0.10}_{-0.18} $ & $ {0.89}^{+0.10}_{-0.18} $ & $ {0.90}^{+0.09}_{-0.15} $ & $ {0.87}^{+0.11}_{-0.16} $ & $ {0.87}^{+0.11}_{-0.16} $ & $ {0.86}^{+0.12}_{-0.16} $ \\ [0.1cm]

$ \chi_{\text{eff}} $ & $ 0.00 $ & $ {0.00}^{+0.07}_{-0.07} $ & $ {0.01}^{+0.07}_{-0.08} $ & $ {0.01}^{+0.08}_{-0.08} $ & $ {0.01}^{+0.07}_{-0.06} $ & $ {0.02}^{+0.06}_{-0.06} $ & $ {-0.01}^{+0.06}_{-0.04} $ \\ [0.1cm]

$ \iota $ & $ 0.00 $ & $ {0.35}^{+0.36}_{-0.26} $ & $ {0.36}^{+0.34}_{-0.26} $ & $ {0.36}^{+0.35}_{-0.25} $ & $ {0.33}^{+0.34}_{-0.24} $ & $ {0.33}^{+0.34}_{-0.24} $ & $ {0.41}^{+0.40}_{-0.30} $ \\ [0.1cm]

$ d_L / \mathrm{Mpc} $ & $ 2000~(500) $ & $ {1936}^{+459}_{-454} $ & $ {1921}^{+451}_{-426} $ & $ {1938}^{+447}_{-443} $ & $ {1934}^{+424}_{-428} $ & $ {1937}^{+426}_{-427} $ & $ {482}^{+100}_{-125} $ \\ [0.1cm]

\hline

\multirow{2}{*}{$ e_{\mathrm{gw}} $}
& Injected & $ 0.05 $ & $ 0.09 $ & $ 0.17 $ & $ 0.238 $ & $ 0.336 $ & $ 0.336 $ \\
& Measured & $ {0.04}^{+0.02}_{-0.02} $ & $ {0.08}^{+0.02}_{-0.02} $ & $ {0.17}^{+0.01}_{-0.01} $ & $ {0.235}^{+0.004}_{-0.004} $ & $ {0.335}^{+0.003}_{-0.003} $ & $ {0.333}^{+0.004}_{-0.001} $ \\

\multirow{2}{*}{$ l_{\mathrm{gw}} $}
& Injected & $ 5.01 $ & $ 2.78 $ & $ 2.70 $ & $ 5.13 $ & $ 0.72 $ & $ 0.72 $ \\
& Measured & $ {4.91}^{+1.02}_{-1.34} $ & $ {2.60}^{+0.89}_{-0.90} $ & $ {2.58}^{+0.97}_{-1.01} $ & $ {4.93}^{+0.58}_{-0.69} $ & $ {0.58}^{+5.53}_{-0.48} $ & $ {0.77}^{+0.36}_{-0.49} $ \\

\multirow{2}{*}{$ \rho_{\mathrm{mf}}^{\text{N}} $}
& Injected & $ 36.1 $ & $ 35.9 $ & $ 36.8 $ & $ 38.3 $ & $ 40.3 $ & $ 43.5 $ \\
& Measured & $ {36.0}^{+0.1}_{-0.1} $ & $ {35.8}^{+0.1}_{-0.1} $ & $ {36.7}^{+0.1}_{-0.1} $ & $ {37.3}^{+0.1}_{-0.1} $ & $ {38.3}^{+0.1}_{-0.1} $ & $ {42.7}^{+0.1}_{-0.1} $ \\ [0.1cm]

\hline
\hline

\end{tabular*}

%% file: tab/PE_runtimes.tex
\renewcommand{\arraystretch}{1.3}
\begin{tabular*}{\textwidth}{l@{\extracolsep{\fill}} c c c c c c c c}

\hline
\hline

\makecell[cc]{Event}
& \multicolumn{3}{c}{Data settings}
& \multicolumn{1}{c}{Sampler}
& \makecell[cc]{Computing \\ resources}
& Model
& \multicolumn{2}{c}{Run statistics}
\\

\hline

& \makecell[cc]{srate  \\ (Hz)}
& \makecell[cc]{duration \\(s)}
& \makecell[cc]{$\langle f_\text{start} \rangle$ \\(Hz)}
& 
& cores$\,\times\,$nodes
&
& Run-time
& $\mathcal{L}$ eval. 
\\

\hline
\hline

\multirow{4.1}{*}{ \makecell[ccc]{
GW150914
}}
& \multirow{4.1}{*}{2048}
& \multirow{4.1}{*}{8}
& \multirow{4.1}{*}{10}
& \multirow{4.1}{*}{\texttt{Bilby}}
& \multirow{4.1}{*}{$64 \times 1$}
& \texttt{SEOBNRv5HM} & 21h & $ 6.4 \times 10^7 $ \\
& & & & &
& \texttt{SEOBNRv5EHM} & 3d 7h & $ 6.4 \times 10^7 $ \\
& & & & &
& \texttt{SEOBNRv6EHM} ($ e = 0 $) & 14h & $ 6.3 \times 10^7 $ \\
& & & & &
& \texttt{SEOBNRv6EHM} & 18h & $ 6.4 \times 10^7 $ \\

\hline

\multirow{4.1}{*}{ \makecell[ccc]{
GW190521
}}
& \multirow{4.1}{*}{1024}
& \multirow{4.1}{*}{8}
& \multirow{4.1}{*}{5.5}
& \multirow{4.1}{*}{\texttt{Bilby}}
& \multirow{4.1}{*}{$64 \times 1$}
& \texttt{SEOBNRv5HM} & 13h & $ 4.5 \times 10^7 $ \\
& & & & &
& \texttt{SEOBNRv5EHM} & 1d 9h & $ 4.5 \times 10^7 $ \\
& & & & &
& \texttt{SEOBNRv6EHM} ($ e = 0 $) & 8h & $ 4.5 \times 10^7 $ \\
& & & & &
& \texttt{SEOBNRv6EHM} & 10h & $ 4.7 \times 10^7 $ \\

\hline

\multirow{4.1}{*}{ \makecell[ccc]{
GW200105
}}
& \multirow{4.1}{*}{4096}
& \multirow{4.1}{*}{32}
& \multirow{4.1}{*}{20}
& \multirow{4.1}{*}{\texttt{pBilby}}
& \multirow{4.1}{*}{$32 \times 16$}
& \texttt{SEOBNRv5HM} & 19h & $ 6.6 \times 10^7 $ \\
& & & & &
& \texttt{SEOBNRv5EHM} & 7d 5h & $ 8.8 \times 10^7 $ \\
& & & & &
& \texttt{SEOBNRv6EHM} ($ e = 0 $) & 17h & $ 6.6 \times 10^7 $ \\
& & & & &
& \texttt{SEOBNRv6EHM} & 1d 12h & $ 8.3 \times 10^7 $ \\

\hline

\multirow{3.1}{*}{ \makecell[ccc]{
GW170817
}}
& \multirow{3.1}{*}{4096}
& \multirow{3.1}{*}{128}
& \multirow{3.1}{*}{23}
& \multirow{3.1}{*}{\texttt{pBilby}}
& \multirow{3.1}{*}{$32 \times 16$}
& \texttt{SEOBNRv5HM} & 2d 4h & $ 6.4 \times 10^7 $ \\
& & & & &
& \texttt{SEOBNRv6EHM} ($ e = 0 $) & 1d 20h & $ 6.2 \times 10^7 $ \\
& & & & &
& \texttt{SEOBNRv6EHM} & 4d 12h & $ 6.9 \times 10^7 $ \\

\hline

\end{tabular*}

%% file: tab/PE_results.tex
\begin{tabular*}{\textwidth}{c@{\extracolsep{\fill}} c c c c c c c}

\hline
\hline

Event
& $M_{\rm src}/\solarmass$
& $1/q$
& $\chi_{\text{eff}}$
& $d_L/\text{Mpc}$
& $e$
& $ \log_{10} \mathcal{B}^{\text{EAS}}_{\text{QCAS}}$
& $ \log_{10} \mathcal{B}^{\text{EAS}}_{\text{QCP}}$
\\ [0.05cm]

\hline

GW200129\_065458 (*)
    & $61.1^{+3.2}_{-2.9}$ & $0.80^{+0.17}_{-0.18}$ & $0.05^{+0.08}_{-0.09}$ & $748^{+308}_{-274}$ & $0.26^{+0.05}_{-0.07}$ & \cellcolor[rgb]{0.10,0.55,0.20} $4.15^{+7.72}_{-1.23}$ & \cellcolor[rgb]{0.26,0.63,0.35} $3.23^{+0.12}_{-0.12}$
    \\

\hline

GW200208\_222617
    & $41.5^{+14.9}_{-6.6}$ & $0.58^{+0.37}_{-0.34}$ & $0.06^{+0.29}_{-0.21}$ & $3028^{+2384}_{-1414}$ & $0.36^{+0.12}_{-0.16}$ & \cellcolor[rgb]{0.74,0.85,0.76} $1.00^{+0.14}_{-0.15}$ & \cellcolor[rgb]{0.71,0.84,0.74} $1.13^{+0.10}_{-0.10}$
    \\

\hline

GW200105\_162426 (*)
    & $10.4^{+0.8}_{-0.7}$ & $0.24^{+0.05}_{-0.04}$ & $-0.10^{+0.11}_{-0.12}$ & $264^{+107}_{-108}$ & $0.12^{+0.03}_{-0.03}$ & \cellcolor[rgb]{0.77,0.87,0.79} $0.85^{+0.39}_{-0.21}$ & \cellcolor[rgb]{0.81,0.88,0.82} $0.67^{+0.10}_{-0.10}$
    \\

\hline

GW231223\_032836 (*)
    & $77.4^{+20.1}_{-13.5}$ & $0.73^{+0.24}_{-0.31}$ & $0.04^{+0.31}_{-0.33}$ & $4557^{+3098}_{-2377}$ & $0.46^{+0.08}_{-0.25}$ & \cellcolor[rgb]{0.78,0.87,0.80} $0.79^{+0.10}_{-0.15}$ & \cellcolor[rgb]{0.74,0.85,0.77} $0.98^{+0.09}_{-0.09}$
    \\

\hline

GW190701\_203306 (*)
    & $95.7^{+13.5}_{-11.1}$ & $0.75^{+0.23}_{-0.32}$ & $-0.07^{+0.24}_{-0.29}$ & $2061^{+825}_{-806}$ & $0.54^{+0.08}_{-0.33}$ & \cellcolor[rgb]{0.79,0.88,0.81} $0.75^{+0.14}_{-0.09}$ & \cellcolor[rgb]{0.75,0.86,0.78} $0.93^{+0.10}_{-0.10}$
    \\

\hline

GW230712\_090405
    & $55.0^{+15.8}_{-14.0}$ & $0.53^{+0.40}_{-0.24}$ & $0.05^{+0.37}_{-0.38}$ & $2965^{+2407}_{-1642}$ & $0.61^{+0.10}_{-0.45}$ & \cellcolor[rgb]{0.83,0.89,0.84} $0.59^{+0.08}_{-0.05}$ & \cellcolor[rgb]{0.93,0.94,0.93} $0.09^{+0.10}_{-0.10}$
    \\

\hline

GW191109\_010717 (*)
    & $107.0^{+15.4}_{-13.4}$ & $0.82^{+0.16}_{-0.25}$ & $-0.07^{+0.37}_{-0.33}$ & $2004^{+1404}_{-990}$ & $0.36^{+0.29}_{-0.28}$ & \cellcolor[rgb]{0.88,0.92,0.89} $0.32^{+0.05}_{-0.06}$ & $-$
    \\

\hline

GW231224\_024321
    & $16.5^{+1.3}_{-0.8}$ & $0.77^{+0.20}_{-0.22}$ & $-0.04^{+0.08}_{-0.06}$ & $946^{+283}_{-425}$ & $0.22^{+0.09}_{-0.14}$ & \cellcolor[rgb]{0.89,0.92,0.90} $0.29^{+0.07}_{-0.09}$ & $-$
    \\

\hline

GW190620\_030421
    & $86.2^{+15.4}_{-10.8}$ & $0.62^{+0.32}_{-0.24}$ & $0.16^{+0.29}_{-0.24}$ & $2790^{+1552}_{-1256}$ & $0.30^{+0.29}_{-0.23}$ & \cellcolor[rgb]{0.89,0.92,0.90} $0.26^{+0.06}_{-0.04}$ & $-$
    \\

\hline

GW230706\_104333
    & $27.5^{+4.0}_{-2.8}$ & $0.69^{+0.27}_{-0.24}$ & $0.13^{+0.12}_{-0.14}$ & $1809^{+913}_{-924}$ & $0.28^{+0.15}_{-0.22}$ & \cellcolor[rgb]{0.92,0.93,0.92} $0.16^{+0.09}_{-0.05}$ & $-$
    \\

\hline

GW231221\_135041 (*)
    & $74.1^{+20.2}_{-14.4}$ & $0.73^{+0.24}_{-0.36}$ & $0.02^{+0.37}_{-0.38}$ & $4805^{+3691}_{-2606}$ & $0.55^{+0.15}_{-0.49}$ & \cellcolor[rgb]{0.92,0.94,0.92} $0.15^{+0.04}_{-0.03}$ & $-$
    \\

\hline

GW190521\_030229
    & $149.7^{+19.0}_{-13.1}$ & $0.75^{+0.22}_{-0.27}$ & $0.08^{+0.26}_{-0.31}$ & $4832^{+1636}_{-1763}$ & $0.38^{+0.30}_{-0.33}$ & \cellcolor[rgb]{0.95,0.95,0.95} $0.05^{+0.04}_{-0.03}$ & $-$
    \\

\hline

GW230709\_122727
    & $75.4^{+19.9}_{-13.9}$ & $0.72^{+0.25}_{-0.36}$ & $0.05^{+0.33}_{-0.32}$ & $4409^{+3264}_{-2379}$ & $0.33^{+0.27}_{-0.29}$ & \cellcolor[rgb]{0.95,0.95,0.95} $-0.03^{+0.05}_{-0.03}$ & $-$
    \\

\hline

GW230820\_212515
    & $95.3^{+20.8}_{-15.4}$ & $0.70^{+0.27}_{-0.41}$ & $0.10^{+0.33}_{-0.30}$ & $4233^{+2572}_{-2193}$ & $0.28^{+0.20}_{-0.24}$ & \cellcolor[rgb]{0.94,0.91,0.91} $-0.09^{+0.05}_{-0.03}$ & $-$
    \\

\hline

GW231114\_043211 (*)
    & $35.0^{+5.9}_{-6.5}$ & $0.23^{+0.22}_{-0.06}$ & $0.17^{+0.16}_{-0.20}$ & $1292^{+730}_{-572}$ & $0.29^{+0.05}_{-0.26}$ & \cellcolor[rgb]{0.94,0.90,0.90} $-0.13^{+0.03}_{-0.03}$ & $-$
    \\

\hline

GW231001\_140220
    & $113.9^{+31.5}_{-23.1}$ & $0.53^{+0.36}_{-0.25}$ & $-0.00^{+0.34}_{-0.36}$ & $4451^{+3857}_{-2448}$ & $0.27^{+0.29}_{-0.24}$ & \cellcolor[rgb]{0.94,0.89,0.89} $-0.14^{+0.03}_{-0.03}$ & $-$
    \\

\hline

GW231123\_135430 (*)
    & $204.1^{+33.4}_{-23.5}$ & $0.38^{+0.22}_{-0.14}$ & $0.65^{+0.13}_{-0.13}$ & $4475^{+1959}_{-1775}$ & $0.20^{+0.25}_{-0.18}$ & \cellcolor[rgb]{0.93,0.87,0.87} $-0.19^{+0.04}_{-0.03}$ & $-$
    \\

\hline

GW240104\_164932
    & $71.9^{+11.0}_{-7.9}$ & $0.77^{+0.20}_{-0.26}$ & $0.10^{+0.17}_{-0.18}$ & $1975^{+1039}_{-984}$ & $0.18^{+0.17}_{-0.15}$ & \cellcolor[rgb]{0.92,0.85,0.85} $-0.26^{+0.03}_{-0.05}$ & $-$
    \\

\hline

GW190706\_222641
    & $107.4^{+26.5}_{-17.8}$ & $0.59^{+0.32}_{-0.23}$ & $0.19^{+0.30}_{-0.31}$ & $3539^{+2615}_{-2016}$ & $0.21^{+0.41}_{-0.19}$ & \cellcolor[rgb]{0.92,0.83,0.83} $-0.30^{+0.03}_{-0.02}$ & $-$
    \\

\hline

GW190929\_012149
    & $90.4^{+18.6}_{-16.0}$ & $0.47^{+0.40}_{-0.23}$ & $-0.06^{+0.24}_{-0.31}$ & $3561^{+3148}_{-1719}$ & $0.16^{+0.24}_{-0.15}$ & \cellcolor[rgb]{0.91,0.79,0.79} $-0.41^{+0.02}_{-0.03}$ & $-$
    \\

\hline

GW241110\_124123
    & $24.0^{+3.3}_{-1.9}$ & $0.52^{+0.32}_{-0.21}$ & $-0.34^{+0.23}_{-0.16}$ & $755^{+303}_{-339}$ & $0.12^{+0.14}_{-0.10}$ & \cellcolor[rgb]{0.90,0.75,0.75} $-0.51^{+0.04}_{-0.03}$ & $-$
    \\

\hline

GW150914
    & $63.2^{+3.5}_{-3.0}$ & $0.86^{+0.12}_{-0.19}$ & $-0.09^{+0.14}_{-0.13}$ & $455^{+147}_{-159}$ & $0.06^{+0.09}_{-0.06}$ & \cellcolor[rgb]{0.87,0.63,0.63} $-0.81^{+0.03}_{-0.04}$ & $-$
    \\

\hline

GW191219\_163120 (*)
    & $26.2^{+7.6}_{-3.9}$ & $0.05^{+0.02}_{-0.02}$ & $-0.26^{+0.34}_{-0.25}$ & $643^{+236}_{-270}$ & $0.03^{+0.06}_{-0.02}$ & \cellcolor[rgb]{0.83,0.49,0.49} $-1.16^{+0.03}_{-0.02}$ & $-$
    \\

\hline

GW190814
    & $26.2^{+3.3}_{-2.1}$ & $0.11^{+0.02}_{-0.02}$ & $0.02^{+0.13}_{-0.11}$ & $231^{+46}_{-57}$ & $0.03^{+0.04}_{-0.02}$ & \cellcolor[rgb]{0.83,0.47,0.47} $-1.19^{+0.03}_{-0.03}$ & $-$
    \\

\hline

GW241011\_233834
    & $27.0^{+1.5}_{-1.9}$ & $0.25^{+0.07}_{-0.04}$ & $0.53^{+0.04}_{-0.04}$ & $231^{+45}_{-49}$ & $0.03^{+0.06}_{-0.02}$ & \cellcolor[rgb]{0.83,0.47,0.47} $-1.21^{+0.02}_{-0.02}$ & $-$
    \\

\hline

GW250114\_082203
    & $65.8^{+1.1}_{-1.1}$ & $0.96^{+0.03}_{-0.07}$ & $-0.04^{+0.04}_{-0.04}$ & $403^{+73}_{-69}$ & $0.013^{+0.022}_{-0.012}$ & \cellcolor[rgb]{0.80,0.36,0.36} $-1.47^{+0.04}_{-0.02}$ & $-$
    \\

\hline

GW170817 (*)
    & $2.76^{+0.05}_{-0.03}$ & $0.75^{+0.18}_{-0.11}$ & $0.00^{+0.02}_{-0.01}$ & $37.9^{+8.9}_{-14.3}$ & $0.008^{+0.006}_{-0.007}$ & \cellcolor[rgb]{0.79,0.32,0.32} $-1.57^{+0.05}_{-0.04}$ & $-$
    \\

\hline

\hline

\end{tabular*}

%% file: tab/bf_unbound.tex
\begin{tabular*}{\columnwidth}{c@{\extracolsep{\fill}} c c c}
\hline\hline
Event & $\log_{10}\mathcal{B}^{\mathrm{unbound}}_{\mathrm{EAS}}$ & $\log_{10}\mathcal{B}^{\mathrm{unbound}}_{\mathrm{v5PHM}}$ & $\log_{10}\mathcal{B}^{\mathrm{unbound}}_{\mathrm{NRSur}}$ \\
\hline
GW190521 & $1.21^{+0.10}_{-0.10}$ & $0.21^{+0.10}_{-0.10}$ & $0.63^{+0.10}_{-0.10}$ \\
\hline
GW191109\_010717 & $1.72^{+0.09}_{-0.09}$ & $1.55^{+0.09}_{-0.09}$ & $0.84^{+0.09}_{-0.09}$ \\
\hline
GW231221\_135041 & $0.81^{+0.09}_{-0.09}$ & $1.19^{+0.09}_{-0.09}$ & $1.27^{+0.08}_{-0.08}$ \\
\hline
GW231123\_135430 & $4.47^{+0.09}_{-0.09}$ & $-4.50^{+0.09}_{-0.09}$ & $-1.57^{+0.09}_{-0.09}$ \\
\hline
GW190620\_030421 & $-4.19^{+0.10}_{-0.10}$ & $-3.69^{+0.10}_{-0.10}$ & $-3.62^{+0.10}_{-0.10}$ \\
\hline\hline
\end{tabular*}

%% file: tab/snr_logl_unbound.tex
\begin{tabular*}{\columnwidth}{l@{\extracolsep{\fill}} l c c}
\hline\hline
Event & Model & $\rho^{\text{N}}_{\text{mf}}$ & $\log \mathcal{L}$ \\
\hline
 \multirow{4}{*}{\makecell[l]{GW190521}}
 & \makecell[l]{\texttt{SEOBNRv6EHM} \\ (Unbound)} & $ {14.2}^{+0.1}_{-0.3} $ & $ {96.6}^{+2.9}_{-4.6} $ \\
 
 & \makecell[l]{\texttt{SEOBNRv6EHM} \\ (EAS)} & $ {13.9}^{+0.3}_{-0.3} $ & $ {92.3}^{+4.3}_{-4.8} $ \\
 
 & \makecell[l]{\texttt{SEOBNRv5PHM} \\ (QCP)} & $ {14.2}^{+0.2}_{-0.4} $ & $ {96.4}^{+3.8}_{-5.5} $ \\
 
 & \makecell[l]{\texttt{NRSur7dq4} \\ (QCP)} & $ {14.1}^{+0.3}_{-0.3} $ & $ {94.3}^{+3.8}_{-5.1} $ \\
\hline
 \multirow{4}{*}{\makecell[l]{GW191109\_010717}}
 & \makecell[l]{\texttt{SEOBNRv6EHM} \\ (Unbound)} & $ {16.4}^{+0.3}_{-0.3} $ & $ {129.9}^{+4.7}_{-5.3} $ \\
 
 & \makecell[l]{\texttt{SEOBNRv6EHM} \\ (EAS)} & $ {16.1}^{+0.2}_{-0.3} $ & $ {124.1}^{+3.4}_{-5.0} $ \\
 
 & \makecell[l]{\texttt{SEOBNRv5PHM} \\ (QCP)} & $ {16.3}^{+0.3}_{-0.4} $ & $ {126.1}^{+3.9}_{-6.1} $ \\
 
 & \makecell[l]{\texttt{NRSur7dq4} \\ (QCP)} & $ {16.4}^{+0.3}_{-0.4} $ & $ {127.7}^{+4.0}_{-5.9} $ \\
\hline
 \multirow{4}{*}{\makecell[l]{GW231221\_135041}}
 & \makecell[l]{\texttt{SEOBNRv6EHM} \\ (Unbound)} & $ {8.4}^{+0.2}_{-0.4} $ & $ {30.8}^{+2.6}_{-4.1} $ \\
 
 & \makecell[l]{\texttt{SEOBNRv6EHM} \\ (EAS)} & $ {8.2}^{+0.4}_{-0.6} $ & $ {29.1}^{+4.5}_{-5.4} $ \\
 
 & \makecell[l]{\texttt{SEOBNRv5PHM} \\ (QCP)} & $ {7.8}^{+0.2}_{-0.4} $ & $ {26.4}^{+2.5}_{-4.1} $ \\
 
 & \makecell[l]{\texttt{NRSur7dq4} \\ (QCP)} & $ {7.8}^{+0.2}_{-0.4} $ & $ {26.2}^{+2.5}_{-4.0} $ \\
\hline\hline
\end{tabular*}

%% file: tab/unbound_injection_table.tex
\begin{tabular*}{\textwidth}{c@{\extracolsep{\fill}} c c c c c c c c c c}
\hline\hline
Injection & Model & $\rho^{\text{N}}_{\text{mf}}$ & $\log\mathcal{L}$ & $M/\solarmass$ & $1/q$ & $\chi_{\text{eff}}$ & $\chi_{\mathrm{p}}$ & $E_0$ & $p_\phi^0$ & $\log_{10}\mathcal{B}^{\mathrm{unbound}}_{\mathrm{NRSur}}$ \\
\hline
 \multirow{3}{*}{\makecell[cc]{\texttt{NRSur7dq4} \\ (QCP)}}
 & Injected & $ 12.7 $ & $-$ & $ 289.8 $ & $ 0.94 $ & $ 0.19 $ & $ 0.96 $ & $-$ & $-$ & \multirow{3}{*}{$ -1.02 \pm 0.10 $} \\
\cline{2-10}
 & \makecell[cc]{\texttt{NRSur7dq4} \\ (QCP)} & $ {12.3}^{+0.2}_{-0.3} $ & $ {71.7}^{+3.5}_{-5.1} $ & $ {286.6}^{+28.4}_{-28.7} $ & $ {0.79}^{+0.19}_{-0.37} $ & $ {0.22}^{+0.27}_{-0.34} $ & $ {0.58}^{+0.32}_{-0.37} $ & $-$ & $-$ & \\
 & \makecell[cc]{\texttt{SEOBNRv6EHM} \\ (Unbound)} & $ {12.2}^{+0.2}_{-0.3} $ & $ {70.2}^{+3.5}_{-5.0} $ & $ {291.8}^{+36.0}_{-31.5} $ & $ {0.81}^{+0.17}_{-0.30} $ & $ {0.27}^{+0.35}_{-0.33} $ & $-$ & $ {1.05}^{+0.24}_{-0.04} $ & $ {3.54}^{+2.31}_{-1.34} $ & \\
\hline
 \multirow{3}{*}{\makecell[cc]{\texttt{SEOBNRv6EHM} \\ (Unbound)}}
 & Injected & $ 13.8 $ & $-$ & $ 292.3 $ & $ 0.60 $ & $ 0.27 $ & $ 0.00 $ & $ 1.116 $ & $ 5.64 $ & \multirow{3}{*}{$ 1.18 \pm 0.10 $} \\
\cline{2-10}
 & \makecell[cc]{\texttt{NRSur7dq4} \\ (QCP)} & $ {13.3}^{+0.2}_{-0.3} $ & $ {84.3}^{+3.2}_{-4.8} $ & $ {263.5}^{+27.4}_{-26.6} $ & $ {0.78}^{+0.20}_{-0.30} $ & $ {-0.06}^{+0.29}_{-0.33} $ & $ {0.55}^{+0.35}_{-0.40} $ & $-$ & $-$ & \\
 & \makecell[cc]{\texttt{SEOBNRv6EHM} \\ (Unbound)} & $ {13.6}^{+0.1}_{-0.3} $ & $ {88.1}^{+2.8}_{-4.7} $ & $ {272.1}^{+20.9}_{-22.8} $ & $ {0.79}^{+0.19}_{-0.32} $ & $ {0.02}^{+0.25}_{-0.23} $ & $-$ & $ {1.07}^{+0.11}_{-0.06} $ & $ {4.09}^{+2.34}_{-1.73} $ & \\
\hline\hline
\end{tabular*}